\documentclass[a4paper]{article}

\usepackage[pages=all, color=black, position={current page.south}, placement=bottom, scale=1, opacity=1, vshift=5mm]{background}
\SetBgContents{
}      

\usepackage[margin=1in]{geometry} 

\usepackage{amsmath}
\usepackage{amsthm}
\usepackage{amssymb}

\usepackage[utf8]{inputenc}
\usepackage[colorlinks,citecolor=blue,urlcolor=blue,filecolor=blue,backref=page]{hyperref}
\hypersetup{
	unicode,
	pdfauthor={Author One, Author Two, Author Three},
	pdftitle={A simple article template},
	pdfsubject={A simple article template},
	pdfkeywords={article, template, simple},
	pdfproducer={LaTeX},
	pdfcreator={pdflatex}
}

\usepackage{natbib}
\usepackage{graphicx}

\usepackage{xcolor} 
\usepackage{subfigure} 

\title{Bayesian Framework for Multi-Source Data Integration - Application to Human Extrapolation From Preclinical Studies}
\author{Sandrine Boulet$^{1}$, Moreno Ursino$^{1,2}$, Robin Michelet$^3$, Linda B.S. Aulin$^3$, \\ Charlotte Kloft$^3$, Emmanuelle Comets$^{4}$ and Sarah Zohar$^{1,*}$}

\date{
	$^1$Inserm, Centre de Recherche des Cordeliers, Sorbonne Université, \\
    Université Paris Cité, F-75006 Paris, France; \\
	Inria, HeKA, F-75015 Paris, France\\
	$^2$Unit of Clinical Epidemiology, Assistance Publique-Hôpitaux de Paris, \\
    CHU Robert Debré, Inserm CIC-EC 1426, F-75019 Paris, France\\
	$^3$Department of Clinical Pharmacy \& Biochemistry, Institute of Pharmacy, \\
    Freie Universitaet Berlin, 12169 Berlin, Germany\\
	Inserm, Univ Rennes, EHESP, Irset (Institut de recherche en santé, \\
    environnement et travail) - UMRS 1085, F-35000 Rennes, France\\
	$^4$Inserm, Université Paris Cité, IAME, F-75018 Paris, France\\
	$^{*}$ \textit{email:} sarah.zohar@inserm.fr.\\
}

\begin{document}

\maketitle

\begin{abstract}
In preclinical investigations, e.g. in \textit{in vitro}, \textit{in vivo} and \textit{in silico} studies,  the pharmacokinetic, pharmacodynamic and toxicological characteristics of a drug are evaluated before advancing to first-in-man trial. Usually, each study is analyzed independently and the human dose range does not leverage the knowledge gained from all studies. Taking into account the preclinical data through inferential procedures can be particularly interesting to obtain a more precise and reliable starting dose and dose range.
\\ We propose a Bayesian framework for multi-source data integration from preclinical studies results extrapolated to human, which allow to predict the quantities of interest (e.g. the minimum effective dose, the maximum tolerated dose, etc.) in humans. We build an approach, divided in four main steps, based on a sequential parameter estimation for each study, extrapolation to human, commensurability checking between \textit{posterior} distributions and final information merging to increase the precision of estimation. 
\\The new framework is evaluated via an extensive simulation study, based on a real-life example in oncology inspired from the preclinical development of galunisertib. Our approach allows to better use all the information compared to a standard framework, reducing uncertainty in the predictions and potentially leading to a more efficient dose selection.

\end{abstract}

\section{Introduction}\label{sec:introduction}

At the beginning of the development of a new element, object, compound, etc. (depending on the field), preliminary knowledge on its properties are estimated through several experiments involving small sample sizes. For example, the clinical development of a novel drug molecule is always preceded by numerous preclinical studies. They include \textit{in vitro} studies (studies in subcellular fractions, cell cultures, micro-organisms, organoid models, etc.), \textit{in vivo} studies (animal testing in species such as mouse, rat, dog, monkey, etc.) and \textit{in silico} studies (simulations and synthetic data). 
These preclinical studies generate an abundance of knowledge regarding the safety and efficacy of the compound, including pharmacokinetics/pharmacodynamics (PK/PD) on a cellular, tissue, organ or organism level, which is then used for go/no go decisions and influences the design of clinical trials in humans. However, preclinical trials usually involve small sample sizes, for ethics and budget constraints.

The \cite{us_food_and_drug_administration_2005} provides guidelines regarding the use of preclinical knowledge to compute the Maximum Recommended Starting Dose (MRSD) of the first-in-human (FIH) trial, with healthy volunteers and drug products for which systemic exposure is intended. These guidelines outline an empirical algorithmic approach to compute the FIH dose in four steps; (1) for each animal species used in the \textit{in vivo} studies, the Human Equivalent Dose (HED) is computed based on the No-Observed-Adverse-Effect Level (NOAEL) and on the body surface area; (2) the HED corresponding to the most sensitive animal species is selected and (3) used for the calculation of the MRSD by applying a safety factor accounting for the expected variability coming from animal-to-human toxicity extrapolation; (4) the MRSD is then adjusted based on the predicted pharmacological mechanism. Although this approach is a valuable starting point, it shows several drawbacks. First, there is no precise recommendation for choosing the safety factor and thus ensuring the safety at the starting dose. Second, the dose selection is primarily based on the minimization of toxicity risk, rather than on efficacy. While this approach may be the only option when there is no comparable marker in healthy volunteers, for example because they do not express the target molecule for the tested drug, it is suboptimal when efficacy can be evaluated. 
In this situation, to improve the clinical development process, the \cite{european_medicines_agency_2017} suggests to take under consideration toxicity but also efficacy by calculating the Minimal Anticipated Biological Effect Level (MABEL) based on all \textit{in vitro} and \textit{in vivo} information available.
Other approaches facilitating preclinical to clinical translation are mentioned in \cite{shen_2019} and include PK-driven or PK/PD-driven approaches. Translational PK/PD models extrapolate concentration and drug effect over time rather than dose and account for differences in both PK and PD parameters when moving from animal species to humans.

However, all of these frameworks use only a part of the collected preclinical data (e.g., from the most appropriate animal species based on an empirical estimation), thus ignoring a vast majority of the available data. Additionally, each of the preclinical analyses is conducted independently, without fully using the results already accrued in previous studies. Recently, as part of the European project on Flagellin aerosol therapy as an immunomodulatory adjunct to the antibiotic treatment of drug-resistant bacterial pneumonia (FAIR), \cite{michelet_2021} have proposed to consider the FIH clinical trial as a continuum of serial preclinical studies. The authors propose an approach based on the “learn-predict-confirm” paradigm in which mathematical models are updated at each step and the updated versions are used to optimize the next study \citep{sheiner_1997}. 

This type of approach can be easily placed under a Bayesian framework that can update \textit{posterior} knowledge each time new data becomes available. 
For example, \cite{la_gamba_2019} sequentially introduce knowledge in preclinical investigations within a Bayesian PK/PD setting, using the \textit{posterior} distributions resulting from one trial to build the \textit{prior} distributions for the following trial. 
More widely, the Bayesian approach has also been used to use information obtained from one population for the analysis of another population. For instance, \cite{zheng_2020} use preclinical data from animal to inform the design and \textit{prior} distributions of a phase I clinical trial via a Bayesian decision-theoretic approach and \cite{zheng_2020b} via a meta-analytic approach. Another example is that of \cite{petit_2018} who propose a method allowing extrapolation and bridging of adult data in early-phase dose-finding pediatric studies.

The main issue in these standard Bayesian frameworks for a sequence of preclinical studies, where the \textit{posterior} distributions of the previous study are used to set \textit{prior} distributions for the new study, is that they require the use of the same mathematical models (or sub-models) for each criterion (outcome, marker) between studies. 
Moreover, the process ignores possible differences between species which are not taken into account by extrapolation formulas, and not all previous information can be included in the new study, since, as a general rule in a Bayesian approach, the amount of information of the \textit{prior} should not exceed the information of the current study. Therefore, if at least one study has any parameter not consistent with all other studies, the information chain can be broken, and information gathered before this study may be lost. 

To deal with these challenges and use all relevant information, we propose a Bayesian framework accounting for differences and similarities between preclinical studies at each extrapolation (up-dating) step. More precisely, our approach needs to consider multi-source data (cells, mouse, ...) and be able to predict the quantities of interest (for example, MABEL, NOAEL, Minimal Effective Dose - MED, Maximum Tolerated Dose - MTD, etc.) in humans. This new framework was evaluated via an extensive simulation study. The simulation setting was inspired by the development of an oncology drug, galunisertib, an inhibitor of TGF-$\beta$ signaling which reduces tumor growth, where extrapolation approaches were applied during the preclinical studies based on markers of toxicity and efficacy~\citep{gueorguieva_2014}. 

This manuscript is organized as follow. Methods are presented in Section \ref{sec:methods}. Section \ref{sec:simulation_design} describes the simulation study used to assess the proposed methodology, and the simulation results are given in Section \ref{sec:simulation_results}. Key findings and recommendations are discussed in Section \ref{sec:discussion}.

\section{Methods}\label{sec:methods}

\begin{figure} 
\centerline{
\includegraphics[scale=0.44]{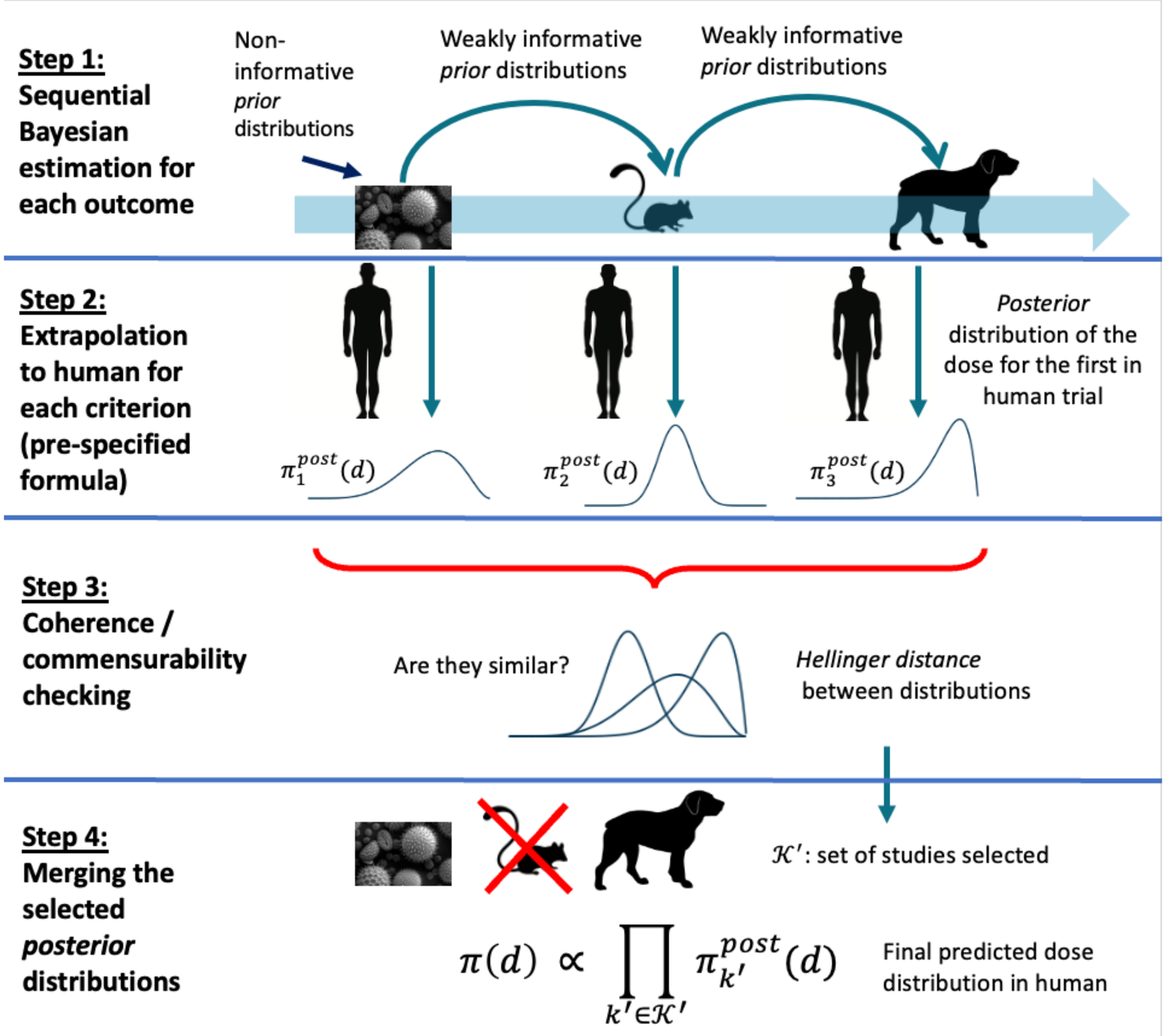}}
\caption{The Bayesian framework in four steps:
(1) Sequential Bayesian estimation, for each outcome, with weakly informative \textit{prior} distributions;
(2) Extrapolation to human for each criterion (via pre-specified formulas);
(3) Coherence/commensurability checking of \textit{posterior} distributions via a divergence-based measure;
(4) Merging the selected \textit{posterior} distributions using an extension of the Bayes formula. 
This figure appears in color in the electronic version of this article.
}
\label{fig:methods}
\end{figure}

We assume that the development plan includes $K$ preclinical studies to be run or, at least, analyzed sequentially before the FIH trial, involving the same outcomes that can be analysed in different ways depending on the study. Our objective is to determine, using appropriate extrapolation-transformation or link functions, the quantities of interest (doses in our case), for example, the MABEL dose or an appropriate range of doses for a FIH dose-ranging trial. We propose a Bayesian framework in four steps (shown in Figure \ref{fig:methods}):
(1) first, we estimate the parameters of the appropriate model for each outcome;
(2) second, we apply extrapolation via pre-specified formulas to obtain parameter distribution in humans for each criterion;
(3) third, we check the coherence/commensurability of \textit{posterior} distributions  via a divergence-based measure;
(4) finally, we merge the selected \textit{posterior} distributions using an extension of the Bayes formula. 
We also investigated an alternative hybrid frequentist-Bayesian approach in Web Appendix A, where the estimation is performed using a frequentist method for faster run-times and a probabilistic sensitivity approach is then used to build parameter distributions. 
In the following, each of the four proposed steps is detailed.

\subsection{First Step: Parameters Estimation}\label{subsec:parameters_estimation}

For $k \in \{1, ..., K\}$, let $\mathbf{y}_k$ denote the outcomes measured in the $k$th study. We suppose that $\mathbf{y}_k$ are related to each other (they could be exactly the same outcome for each study), or that, if different, they can be used to compute the doses of interest, $d_r$, $r \in \{1,\ldots, R\}$ with $R$ the number of doses of interest (for example, MABEL, NOAEL, MED, MTD, etc.).
For each study $k$, Bayesian fixed- or mixed-effect models, $\mathbf{f}_k(\mathbf{y}_k, \boldsymbol{\theta}_k$), (possibly nonlinear) are fitted, via Markov chain Monte Carlo (MCMC) methods, on the dose-outcome relationships (PK/PD indices describing efficacy or toxicity) with weakly-informative \textit{prior} distributions on models parameters $\boldsymbol{\theta}_k$. 
The functions $\mathbf{f}_k(.)$ can differ between studies, as well as the vector $\boldsymbol{\theta}_k$ can have different length or being study specific.
For example, a fixed effect model can be used in a study with only one outcome per subject while a mixed effect model, with the same structure for the population effects but with random effects on a few parameters, can be adopted for longitudinal studies. For the first study, \textit{prior} distributions are chosen based on \textit{prior} knowledge external to this process. When no \textit{prior} knowledge is available, non-informative \textit{prior} distributions can be used. For the following studies, if any component of $\boldsymbol{\theta}_{k-1}$ can be related to an element of $\boldsymbol{\theta}_k$, the \textit{prior} distributions are computed using the estimated \textit{posterior} means of the model parameters of the previous study, $k-1$, (extrapolated between species, if necessary, using similar approaches as described in the next step) and using appropriately large standard deviations. In this way, \textit{prior} distributions act as stabilization tools. 

To summarise, at the end of the process, each study $k \in \{1, ..., K\}$ yields a \textit{posterior} distribution for each element of each $\boldsymbol{\theta}_{k}$.  

\subsection{Second Step: Extrapolation to Human}\label{subsec:extrapolation_to_human}

While in the previous step, extrapolation was used between species/studies, at this step we focus on human extrapolation.
For each study $k \in \{1, ..., K\}$, we extrapolate the preclinical results (the estimated parameters) to humans using predetermined formulas $\mathbf{g}_k(.)$ (transformations), $\boldsymbol{\theta}_{h,k} = \mathbf{g}_k(\boldsymbol{\theta}_{k})$, where $\boldsymbol{\theta}_{h,k}$ represent the vector of extrapolated human parameters from the $k$th study. 
An example of such formulas is the allometric scaling used to extrapolate pharmacokinetic parameters across species by taking into account the size of the organisms \citep*{west_1997}. As stated in the previous section, these formulas could also be used in the previous step to extrapolate parameter values between species/studies. In Bayesian setting, the transformation is applied directly on \textit{posterior} distributions to obtain the \textit{posterior} distributions of the extrapolated parameters (via the random variable transformation theorem). Then, the doses of interest, $d_r = d_{r,k}(\boldsymbol{\theta}_{h,k})$, for human, where $d_{r,k}(.)$ represent the function that links the vector $\boldsymbol{\theta}_{h,k}$ of extrapolated human parameters from the $k$th study to the dose of interest $d_r$, are derived either analytically or from Monte Carlo simulations. Finally, the associated \textit{posterior} distributions of $d_r$, $\pi^{post}_{k}(d_r)$, are deduced (via the random variable transformation theorem or approximated via Monte Carlo methods).   

\subsection{Third Step: Commensurability Checking and \textit{Posterior} Distributions Selection}\label{subsec:commensurability_checking}

In this step, we aim to compare the $K$ predicted human dose distributions, $D_{k,r} \sim \pi^{post}_{k}(d_r)$, for each $r$, and select the ones which seems the most similar to each others. However, \textit{posterior} distributions cannot be directly compared since they bring a different amount of information depending on the corresponding study sample size. One way to make two \textit{posteriors} comparable consists on discounting the likelihood of the study with the highest sample size, as proposed by \cite{ollier_2020}. This method requires the re-analysis of one of the two experiences and the decision on the discounting factor: it could be straightforward in case of fixed effect models, but not really for mixed effects model involving longitudinal studies. Inspired by the effective sample sizes (ESS) notion~ \citep*{morita_2008}, we propose to transform the \textit{posterior} distributions to move to a situation where they could be assumed as Gaussian, and then we standardize them at the highest variance, 
\begin{align}
D^*_{k,r} &= \dfrac{\underset{h \in \mathcal K}{\max} ~ S_{h,r}}{S_{k,r}} \log(D_{k,r}) + \left(1- \dfrac{\underset{h \in \mathcal K}{\max} ~ S_{h,r}}{S_{k,r}}\right) M_{k,r} \label{dose_distr_transf}  \\ 
&\sim \pi^{post}_{k}(d^*_r), \nonumber
\end{align}
for $k \in \{1,... ,K\}$, where $M_{k,r} = E(\log D_{k,r})$ and $S_{k,r}^2 = Var(\log D_{k,r})$ are respectively the mean and the variance of $\log D_{k,r}$.
This transformation allows the distributions to have the same variance across studies but to keep their own expectations. Indeed, the logarithmic transformation of the dose is first used to move to a situation where the distribution could be assumed as Gaussian. According to \cite{morita_2008}, the ESS of a normal distribution, with a normal model and known variance, depends only on the \textit{prior} variance. Thus, the ESS of our transformed distributions are equal (if we consider them as \textit{prior} distribution on a ``Gaussian dose model"). 
Furthermore, for $k \in \{1,... ,K\}$, 
$E(D^*_{k,r}) = M_{k,r}$ 
and
$Var(D^*_{k,r}) = \underset{h \in \mathcal K}{\max} ~ S^2_{h,r}$.

Once the \textit{posteriors} are comparable, the commensurability of the modified distributions $\pi^{post}_{k}(d^*_r)$ between studies $k_1 \in \{1,... ,K\}$ and $k_2 \in \{1,... ,K\}$, $k_2 \ne k_1$, is checked using the Hellinger distance for each $r$,
$H_r(D_{k_1}, D_{k_2}) = \Big[\dfrac{1}{2} \int \big(\sqrt{\pi^{post}_{k_1}(d^*_r)}-\sqrt{\pi^{post}_{k_2}(d^*_r)}\big)^2 \text{d} d^*_r \Big]^{\frac{1}{2}}$~\citep{ollier_2021}. We proposed this distance since it is bounded between 0 and 1, symmetric and it makes easier to define the consistency notion of results between studies.  In fact, the closer the Hellinger distance is to 0, the more consistent the predicted dose distributions in humans are between studies. On the contrary, a distance close to 1 indicates inconsistent results. 

An algorithm, along with a threshold, based on Hellinger distance results should be defined to decide which studies will be selected for the next step. When only a small number of studies are available, ad hoc algorithms can be easily developed, as shown in our example with simple decision rules (see subsection \ref{subsec:implementation}). In more complex settings, clustering methods can be used. We suggest to run simulations of selected/relevant scenarios to optimize the algorithm and to choose a threshold basing on the results accuracy. For each selected scenario, we define the studies that we consider consistent and that we wish to retain for the fourth and final step. Then, via simulation, we compute the Hellinger distances and define a binary variable ``true response" related to each comparison between studies, that is equal to 1 if the two studies are considered similar (theoretically, in the scenario) and 0 otherwise, and, for each possible threshold, a variable ``predicted response" that is equal to 1 if the computed distance is below the threshold and 0 otherwise. 
The value of the threshold is then chosen based on the curve of the accuracy versus the Hellinger distance threshold. 
The accuracy is defined as the proportion of correct predictions (that is, 
true positives and true negatives) among the total number of cases, $\text{acc} = \dfrac{TP+TN}{P+N}, \text{~where~} P = TP+FN \text{~and~} N = TN+FP$, where $TP$, $TN$, $FN$, $FP$, $P$ and $N$ are respectively the numbers of true positives, true negatives, false negatives, false positives, positives and negatives. Examples are given in subsection \ref{subsec:commensurability_checking_res} and in Web Appendix C.

At the end of this step, consistent studies are selected. When no studies are ``clustered together", only the results of the most relevant study are considered for the following step.

\subsection{Fourth Step: Merging the Selected \textit{Posterior} Distributions}\label{subsec:merging_posterior_distributions}

Let $\mathcal K = \{1, 2, \dots, K\}$ be the set of indices of the preclinical studies. Denoting $\mathcal K' \subset \mathcal K $ the subset of studies selected at the previous step, we merge the extrapolated dose distributions $\pi^{post}_{k'}(d_r)$ ($k' \in \mathcal K'$) using an extension of the Bayes formula to obtain the final predicted dose distribution $d_r$ in humans:
\begin{equation}\label{eq:merge}
    \pi(d_r) \propto \prod_{k' \in \mathcal K'} \mathcal \pi^{post}_{k'}(d_r).
\end{equation}
Eq.~\ref{eq:merge} exists when the $\text{Card}(\mathcal K')$ selected distributions share the same support, or at least part of it. It excludes situations where the product term is 0 almost surely on the domain, that is when at least one distribution is 0 almost surely in the part of the domain where the others are different from zero and vice versa. However, this phenomenon should be avoided by the previous steps. Moreover, the second term of eq.~\ref{eq:merge} is Lebesgue integrable if the $\text{Card}(\mathcal K')$ distributions (or at least all but one) are bounded. This condition is sufficient but not necessary, since a product of U-shaped (or J, or inverse J-shaped) beta distribution density functions is still a beta distribution (as shown in Web Appendix A). In case of sequential Bayesian analyses (when the \textit{posterior} of the previous study is used as \textit{prior} for the next one) of the same study design repeated $K$ times, the final \textit{posterior} distribution is given by $\pi^{post}_{K}(\boldsymbol{\theta}) \propto \pi^{\textit{prior}}(\boldsymbol{\theta}) \prod_{k=1}^K \mathcal L(\boldsymbol{\theta}|\mbox{data}_k)$. If $\pi^{\textit{prior}}(\boldsymbol{\theta})$ is highly non-informative and if we write $K-1$ other highly non-informative improper distributions close to each likelihood term, we can see each couple terms of the product (the $k$th likelihood and the associated highly non-informative improper distributions) as a \textit{posterior} distribution. This is why eq.~\ref{eq:merge} can be seen as an extension of the Bayes formula. Because the information from informative \textit{prior} distributions will be counted $\text{Card}(\mathcal K')$ times, we suggest using non-informative \textit{prior} distributions, as the amount of \textit{prior} information can be considered negligible also after the application of eq.~\ref{eq:merge}. Eq.~\ref{eq:merge} then gives the final dose distribution, which takes into account all relevant information and from which inference can be drawn; that is, the expected value (or the median) can be used as point estimator and credible intervals can be extracted. To note, if only the most relevant study was considered at the previous step, eq.~\ref{eq:merge} becomes its \textit{posterior} distribution.

\section{Simulation Settings}\label{sec:simulation_design}

To illustrate and evaluate our approach, we use the preclinical and clinical development of  galunisertib (LY2157299) as a case-study to build different simulation scenarios. In the following section, we describe how to define the MTD using pharmacokinetics results. Therefore, only for the sake of simplicity, $r=1$. However, simulation focusing on MED estimation (along with the MTD, therefore $r=2$) using pharmacokinetics/pharmacodynamics modelling are given in Web Appendix B.

\subsection{Galunisertib Case Study} \label{subsec:case_study}

Galunisertib is a transforming growth factor (TGF)-$\beta$ receptor I kinase inhibitor, developed for patients with glioma (tumor that starts in the glial cells of the brain or the spine) and advanced cancers. It inhibits the transforming growth factor (TGF)-$\beta$ receptor, blocking TGF-$\beta$ signaling which is an important growth regulator in advanced cancer. This pathway is a potential target to simultaneously inhibit tumor growth and neo-angiogenesis. Semi-mechanistic modeling was used during drug development to predict a safe dosing regimen for phase I study and orient dose selection, building a model from data in mouse, rat and dog \citep{bueno_2008, gueorguieva_2014}. Studies in mouse described by \cite{bueno_2008} showed that the inhibition of phosphorylated Smad proteins, a transducer in the signaling cascade, could be used as a surrogate marker for tumor growth inhibition (pSMAD); pSMAD was modelled using an indirect response model depending on the PK of galunisertib. Studies in rat described by \cite{gueorguieva_2014} suggested a toxicity limit on the exposure, measured by the area under the curve (AUC), to prevent cardiovascular side-effects. Finally, allometric scaling combining data from mouse, rat and dog was used to extrapolate PK parameters to humans, assuming the same PD relationship with the biomarker. The predicted dose proved safe and showed clinical benefits in the FIH trial reported by \cite{rodon_2015}.

We used this setting to design the simulation study, simplifying the PK model to a one-compartment model as in \cite*{lestini_2015}. For a given study, let $\mathcal D = (d_1, d_2, ..., d_J)$ be the set of $J$ administered doses.

\subsection{Toxicity Data Simulation}\label{subsec:tox_simu}

In our example, surrogate toxicity data for a given study occurs if the value of a function of the AUC of the drug in blood plasma exceeds a given threshold.

The concentration $C(t)$ at time $t$ after oral administration of the dose $d_j$, $j \in \{1,...,J\}$, at time 0, is first simulated using a first-order absorption linear one compartment model 
\begin{equation}\label{eq:concentration_toxicity_model}
C(t) = \dfrac{d_j}{V} \times \dfrac{k_a}{k_a-\frac{CL}{V}}(e^{-\frac{CL}{V}t}-e^{-k_a t})
\end{equation}
where $k_a$ is the absorption rate constant for oral administration, $CL$ is the clearance of elimination, and $V$ is the volume of distribution. Both the clearance of elimination and the volume of distribution are assumed to follow log-normal distributions within a population: $CL \sim LN(\log(\mu_{CL}),\omega_{CL})$ and $V \sim LN(\log(\mu_{V}),\omega_{V})$.

Then, $\text{AUC} = \int_0^{+\infty}C(t) dt = d / CL$, and binary toxicity is generated for the subject at the given dose if $s(\text{AUC})\ge \tau_T$ where $s(\text{AUC}) = \alpha \text{AUC}$, $\alpha \sim LN(\log(\mu_{\alpha}), \omega_{\alpha})$ and $\tau_T$ is a threshold. In summary, the probability of toxicity for dose $d$ is calculated as 
\begin{align}
p_T(d) &= P(s(\text{AUC})\ge \tau_T | d) \nonumber \\
&= \Phi \Big(\dfrac{\log(d) - \log(\tau_T) - \log(\mu_{CL})}{\sqrt{\omega_{CL}^2+\omega_\alpha^2}} \Big) \label{eq:probability_toxicity_model}
\end{align}
where $\Phi$ is the cumulative function of a standard normal distribution~\citep{ursino_2017}. An unacceptable risk was defined as the probability of exceeding an AUC of $\tau_T$ mg.L$^{-1}$.h being greater than $p_T$. This formula allows in the second step (see subsection \ref{subsec:extrapolation_to_human}) to deduce the distribution of the MTD from the chosen values of the probability of toxicity of this dose $p_T(\text{MTD})$ and the threshold $\tau_T$, as well as the \textit{posterior} distributions of $\mu_{CL}$, $\omega_{CL}$ and $\omega_{\alpha}$. For this paper, we consider that $\alpha$ is constant, and thus $\omega_\alpha = 0$.

In our specific example, the toxicity threshold is related to the total AUC (i.e.  cumulative drug exposure) and therefore directly to the pharmacokinetics. Other toxicity measures could be considered instead, either continuous (high temperature, elevated liver damage marker, maximum concentration, etc.) or of a different nature (occurrence of an adverse event, etc.) which we would then model through the probability of occurrence.

We assume a proportional error of 20\% for the measurements of the concentration of the drug in blood plasma defined as $\tilde{C} \sim LN(\log(C),0.2)$ where $C$ represents the true values of the measurements and $\tilde{C}$ the measured values.

\subsection{Scenarios}\label{subsec:scenarios}

We simulate galunisertib (LY2157299) data corresponding to the dose ranges used in the \textit{vivo} studies. To keep the simulation study simple, we only simulated three sequential studies, with mice, rats and dogs, respectively. Mice ($k=1$), rats ($k=2$) and dogs ($k=K=3$) are assumed to be respectively exposed to oral doses of either 10 ($n = 15$), 30 ($n = 15$), 50 ($n = 15$), 75 ($n = 15$), 100 ($n = 15$), 150 ($n = 15$), 300 ($n = 15$) mg/kg (for mice), 10 ($n = 8$), 30 ($n = 8$), 50 ($n = 8$), 100 ($n = 8$), 300 ($n = 8$) mg/kg (for rats) or 2 ($n=6$), 10 ($n=6$), 30 ($n=6$), 50 ($n=6$), 300 ($n=6$) mg/kg (for dogs). For each of their doses, 3 mice are assumed to be sacrificed at the following times: 15 min, 45 min, 2h, 5h and 15h; while all rats and dogs are respectively sampled at either 15 min, 1h, 2h, 3h40 and 10h (for rats) or 10 min, 1h40, 2h, 5h30 and 15h (for dogs). 

\begin{figure} 
\centerline{
\subfigure[]{\includegraphics[scale=0.25]{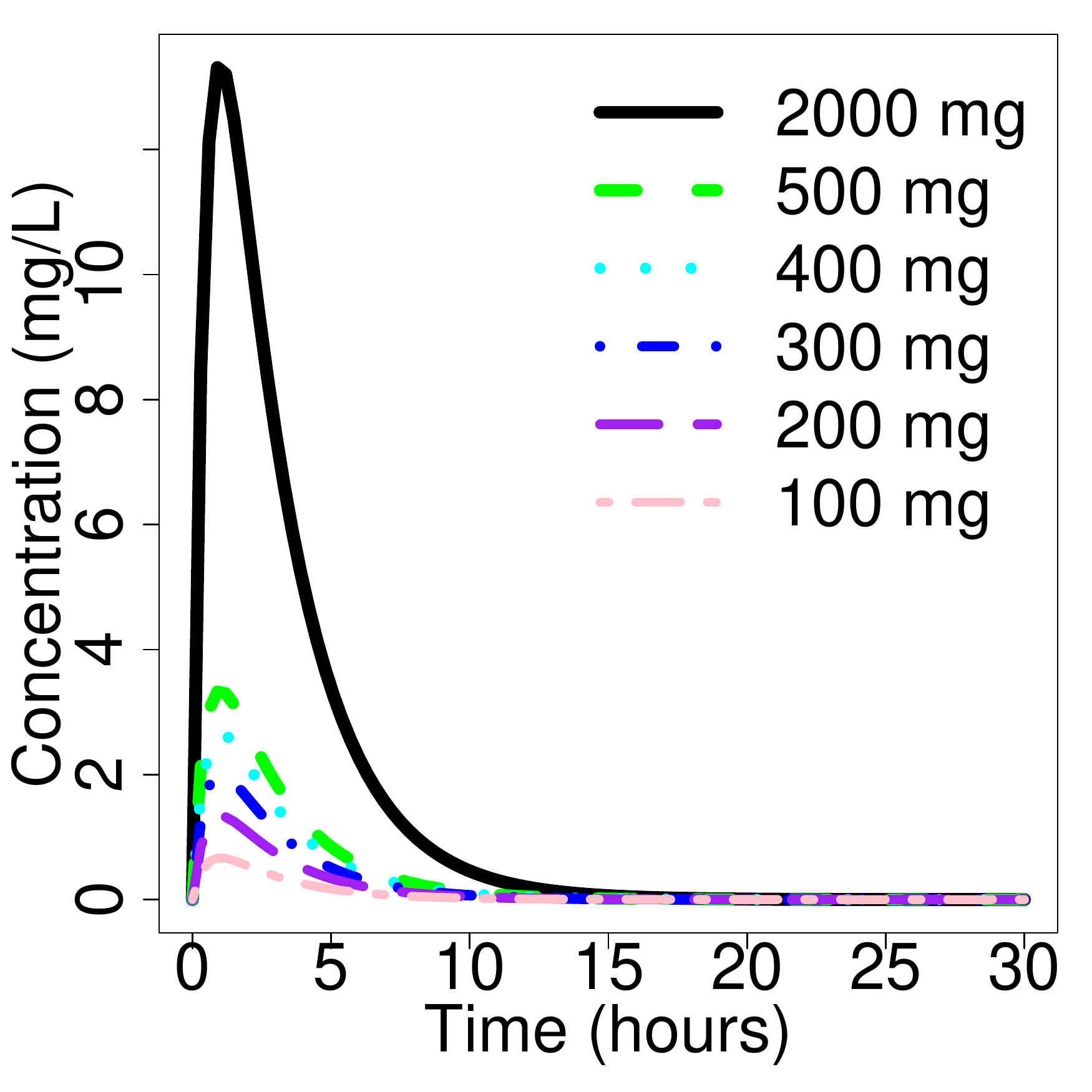}\label{fig:simu_concentration_time_human}}
\subfigure[]{\includegraphics[scale=0.25]{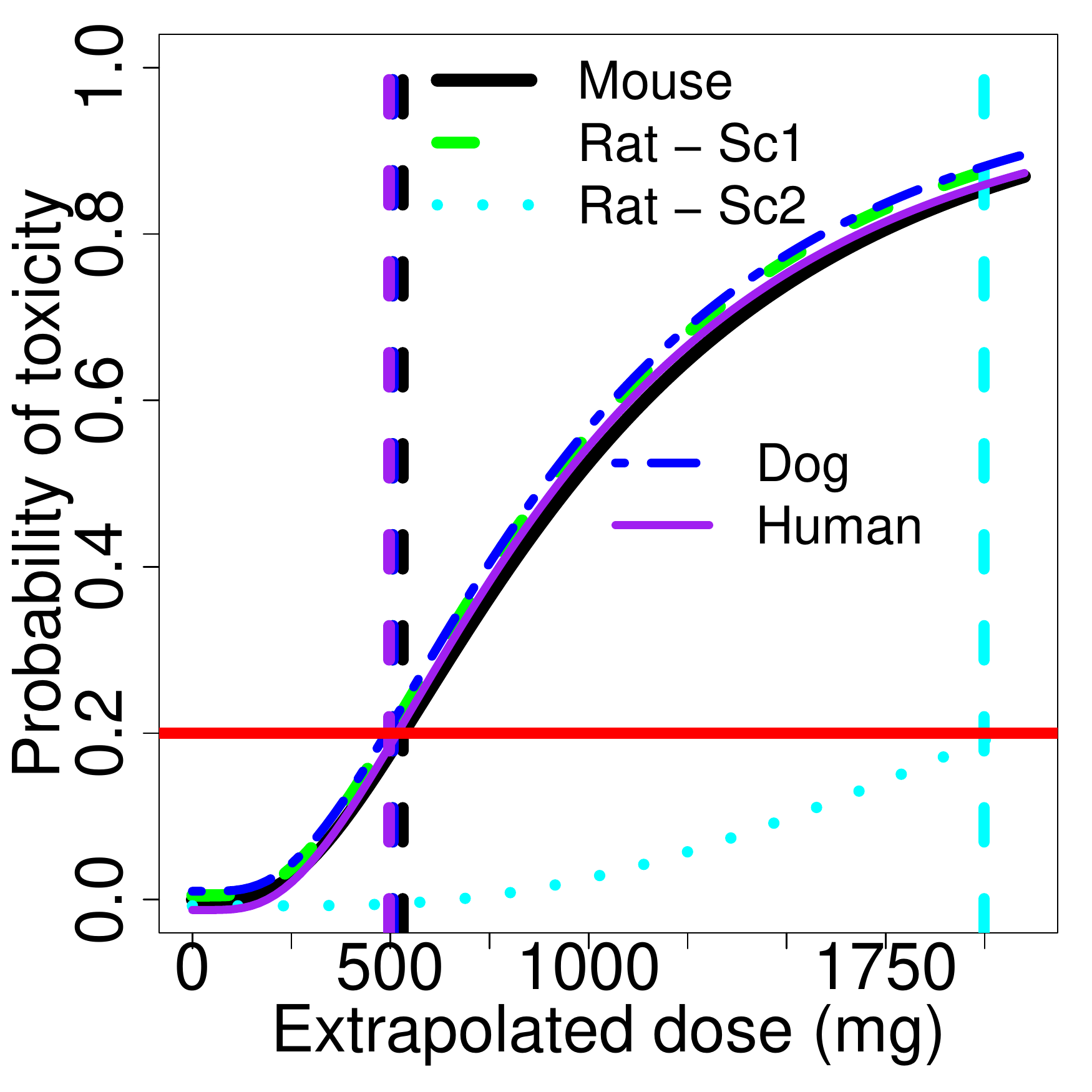}\label{fig:tox_prob_by_extrapolated_dose_sc1and2}}}
\caption{(a) Mean concentration of the drug versus time for several doses for human in scenario 1; (b) Probability of toxicity according to the extrapolated dose for scenarios 1 and 2 and for all species. Red horizontal lines depict the probability of toxicity $p_T$ threshold. The probability of toxicity for a dose should be lower than the threshold $p_T$. This figure appears in color in the electronic version of this article.}
\label{fig:simu_for_PK_model_sc1and2}
\end{figure}

Toxicity data for human in scenario 1 (baseline scenario) are first simulated from models \ref{eq:concentration_toxicity_model} and \ref{eq:probability_toxicity_model}; the parameters were chosen so that the simulated data follow similar distributions to those produced in \cite{lestini_2015} (see Figure \ref{fig:simu_for_PK_model_sc1and2}). Then, $\mu_{\text{CL}}$ and $\mu_{\text{V}}$ parameters for dog, rat and mouse in scenario 1 are calculated from human parameters using the following allometric scaling formulas:
\begin{equation}\label{eq:extrapolation_clearance}
\mu_{CL}^{sp2} = \mu_{CL}^{sp1} \times \Big(\dfrac{W^{sp2}}{W^{\text{sp1}}}\Big)^{0.75}
\end{equation}
\begin{equation}\label{eq:extrapolation_volume}
\mu_{V}^{sp2} = \mu_{V}^{{sp1}} \times \Big(\dfrac{W^{sp2}}{W^{{sp1}}}\Big),
\end{equation}
where $sp1$ and $sp2$ are respectively the human and the animal species considered, and $W$ is the weight of the human or animal species. Note that these formulas hold for conversions between all animal species. All the other parameters (absorption rate constant $k_a$, variances of the PK parameters) for dog, rat and mouse's scenario 1 are chosen equal to those of human (see Table \ref{tab:simu_parameters_for_PK_model_and_extrapolated_mtd_for_sc1and2}). For scenario 1, using the extrapolated parameters of PK model from dog, rat and mouse to human allows to predict accurately the MTD to be approximately equal to 502 mg. Scenario 2 is based on an inaccurate extrapolation of the parameter clearance $CL$, affecting toxicity (via AUC), for one species. Other two scenarios (scenario 3 and 4) accounting for MED estimation along with the MTD are given in Web Appendix B.
Figure \ref{fig:tox_prob_by_extrapolated_dose_sc1and2} gives the probability of toxicity according to the extrapolated dose for scenarios 1 and 2 and for all species.

\begin{table}
\footnotesize
\caption{Simulation parameters for PK model and approximate extrapolated MTD for all species and scenarios 1 and 2. $W$: Weights;  $k_a$: Absorption rate constant for oral administration; $(\mu_{CL}, \mu_V)$: Exponentials of the respective means of the logarithms of the distributions of the clearance elimination and the volume of distribution;  $(\omega_{CL}, \omega_V)$: Respective standard deviations of the logarithms of the distributions of the clearance elimination and the volume of distribution; $\sigma_C$: Standard deviations of the logarithm of the distribution of the measured concentration of the drug in blood plasma; $\tau_T$: AUC threshold; $p_T$: Probability of toxicity threshold; MTD: Maximum tolerated dose.}\label{tab:simu_parameters_for_PK_model_and_extrapolated_mtd_for_sc1and2}
\begin{center}
\begin{tabular}{lrrrrr}
\hline \hline
\textbf{Parameters}&\textbf{Human}&\textbf{Dog} &\multicolumn{2}{c}{\textbf{Rat}} &\textbf{Mouse} \\ 
&&&\textbf{Sc.1}&\textbf{Sc.2}& \\
\hline 
$W$ (kg) &70&10&0.15&0.15&0.025 \\
$k_a$ (h$^{-1}$) &2&2&2&2&2 \\
$\mu_{CL}$ (L.h$^{-1}$)&40&9.3&0.40&\textbf{1.59}&0.11 \\
$\mu_V$ (L)&100&14&0.21&0.21&0.04 \\
$\omega_{CL}$ &0.7&0.7&0.7&0.7&0.7 \\ 
$\omega_{V}$ &0.7&0.7&0.7&0.7&0.7 \\ 
$\sigma_C$ & 0.2 & 0.2& 0.2& 0.2& 0.2 \\ 
$\tau_T$ (mg.L$^{-1}$.h) &22.6&22.6&22.6&22.6&22.6 \\ 
$p_T$ (\%) &20&20&20&20&20 \\
$\mu_{\alpha}$ &1&1&1&1&1 \\
$\omega_{\alpha}$ &0&0&0&0&0 \\
\hline  
\textbf{Extrapolated} & 502 & 502 & 504 & 2002 & 531 \\
\textbf{MTD (mg)} & & & & & \\
\hline
\end{tabular}
\end{center}
\end{table}

\subsection{Implementation}\label{subsec:implementation}

For each scenario, 500 datasets are simulated. We assume that  equations \ref{eq:concentration_toxicity_model}, \ref{eq:probability_toxicity_model}, \ref{eq:extrapolation_clearance} and \ref{eq:extrapolation_volume} describing toxicity data generation and allometric scaling formulas are known as well as $\tau_T$, $p_T$, $\mu_{\alpha}$, $\omega_{\alpha}$ and $W$. The other model parameters are estimated.
At step 1 of our methodology, for each simulated dataset and for each animal species, we fit the previous mixed-effects model
using Hamiltonian Monte Carlo (\citealp{betancourt_2018, thomas_2021}). For the dog and the rat, random effects (intraindividual variability - IIV) are estimated on $CL$ and 
$V$, therefore $\mathbf{y}_{k,i}  = \mathbf{\tilde{c}}_i$, the vector of concentrations measured of the $i$th subject of study $k$, $k=2,3$, and $\mathbf{f}_k(.): \tilde{c}_{ij} = \log\left(C(t_{ij}, \boldsymbol{\theta}_k = \{ k_{a,k}, \mu_{CL,k}, \mu_{V,k}, \omega_{CL,k}, \omega_{V,k} \})\right) + \epsilon$ with $\epsilon \sim N(0, \sigma_C)$ where $k=2,3$ and $j$ indexes the number of time points for each subject, while for the mouse, random effects are only estimated on $CL$ since there is only one data point per mouse (that is $\mathbf{f}_1(.): \tilde{c}_i = \log\left(C(t_i, \boldsymbol{\theta}_1 = \{ k_{a,1}, \mu_{CL,1}, \mu_{V,1}, \omega_{CL,1}, \})\right)  + \epsilon$ and $\mathbf{y}_1 = \tilde{c}_i$). In this situation, $\boldsymbol{\theta}_1$, $\boldsymbol{\theta}_2$ and $\boldsymbol{\theta}_3=\boldsymbol{\theta}_2$ share several parameters with the same role, therefore extrapolation formulas (eq.~\ref{eq:extrapolation_clearance} and \ref{eq:extrapolation_volume} are adopted as $\mathbf{g}_k$) can be used to set \textit{prior} distributions when stepping from one study to the next one. For the sake of simplicity, in the following we do not use the sub-notation $k$ for the parameters, since it will be clear from the contest to which study the parameters are linked to.

For the next steps of the analysis based on mouse data, we make the assumption that $\omega_{V} = 0.7$ for mouse. As sensitivity analysis, we also consider the assumptions that $\omega_{V} = 0.4$ or $\omega_{V} = 1$ for mouse. In this example, $d_{1,k}(\boldsymbol{\theta}_{h,k})$ is the solution of eq.~\ref{eq:probability_toxicity_model} when solving for $d$ while setting $p_T(d)=0.2$.
We choose to perform $L = 1000$ MCMC runs for steps 2 to 4 of the methodology when approximating $D_k$ and $D^*_k$ distributions.
In step 2, each one of the $L$ values of clearance and volume is extrapolated from animal species to human using the previous formula, while other parameters of the models are kept identical.

In step 3, in practice, the Hellinger distance is approximated using the rectangle method:
\begin{equation*}
\left[0.5 (b-a)/M \times \sum_{m=0}^{M-1} \left(\sqrt{\pi^{post}_{k_1}(d^*_{r,m})}-\sqrt{\pi^{post}_{k_2}(d^*_{r,m})}\right)^2 \right]^{\frac{1}{2}},
\end{equation*}
where $a = \underset{m \in \{0,..., M-1\}}{\min} d^*_{r,m}$, $b = \underset{m \in \{0,..., M-1\}}{\max} d^*_{r,m}$.
To compute the accuracy, as proposed in section~\ref{subsec:commensurability_checking}, for each scenario, the animal species that should be kept for the final step (because they are consistent) have to be defined. 
The extrapolated MTDs to humans from Table \ref{tab:simu_parameters_for_PK_model_and_extrapolated_mtd_for_sc1and2} 
are approximately equal to 502 mg (the true MTD value for humans) for all animal species (mouse, rat and dog) in scenario 1, and only for mouse and dog in scenario 2. Therefore, for the MTD, the binary variable ``true response" is set to 1 for all comparisons between animal species in scenario 1, and only for the comparison between mouse and dog in scenario 2; Otherwise, for all comparisons including rat in scenario 2, it is set to 0. To note, in the case of MTD, all scenarios are considered as relevant for choosing the Hellinger distance threshold. It is not the case for the MED, as described in Web Appendix B.
We used the ad hoc following algorithm. If the three computed Hellinger distances are lower or equal to the selected threshold, the three studies are selected. All three studies are selected also if at least two measures are lower than the threshold. If only a value is lower the threshold, the two corresponding studies are selected. Finally, if all three Hellinger distances exceed the threshold the results of the most relevant animal species are considered for the following step. 
In this work, we assume that the most relevant animal species for galunisertib is the dog.

For step 4, \textit{posterior} distributions can be computed using the kernel density estimator,
$\pi^{post}_{k'}(d_r) = (L h_{k'})^{-1} \sum_{l=1}^{L} \kappa\Big((d_r-d_{{r,k'},l}) / h_{k'}\Big),$
where $\kappa$ is the kernel (the standard normal density in this paper) and $h_{k'} >0$ is the bandwidth.

All analyses are performed using R software version 4.04 with Stan \citep{stan_2021} package \textit{rstan} version 2.21.2. In \textit{rstan}, 3 chains, a burn-in of 3000 and 6000 other iterations are used and, as a convergence criterion, \cite{gelman_1992}'s potential scale reduction factor Rhat = 1.

\section{Simulation Results}\label{sec:simulation_results}

\subsection{Illustration}\label{subsec:illustration}

\begin{figure} 
\centerline{
\subfigure[]{\includegraphics[scale=0.25]{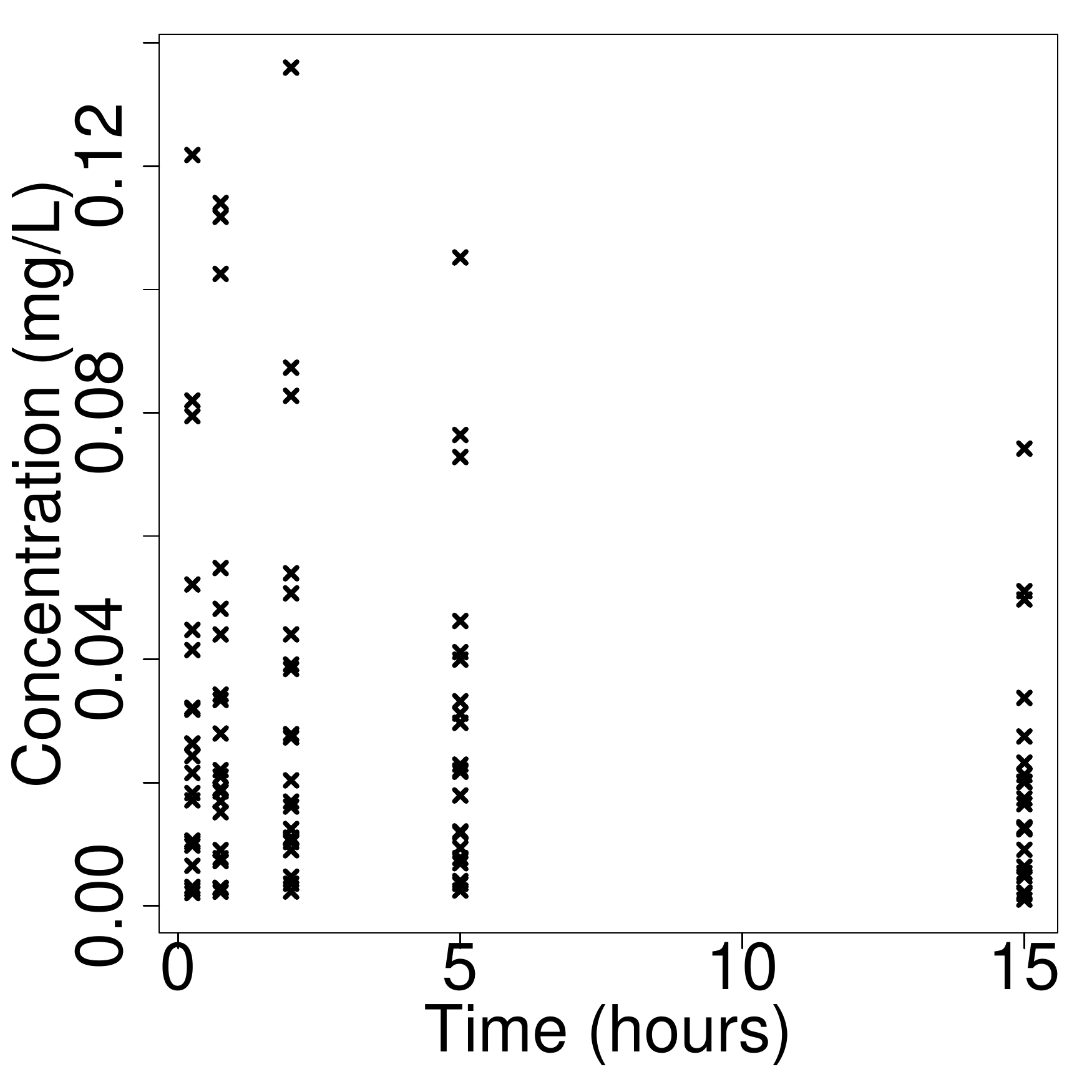}\label{fig:simu_concentration_time_mouse_tr4}}
\subfigure[]{\includegraphics[scale=0.25]{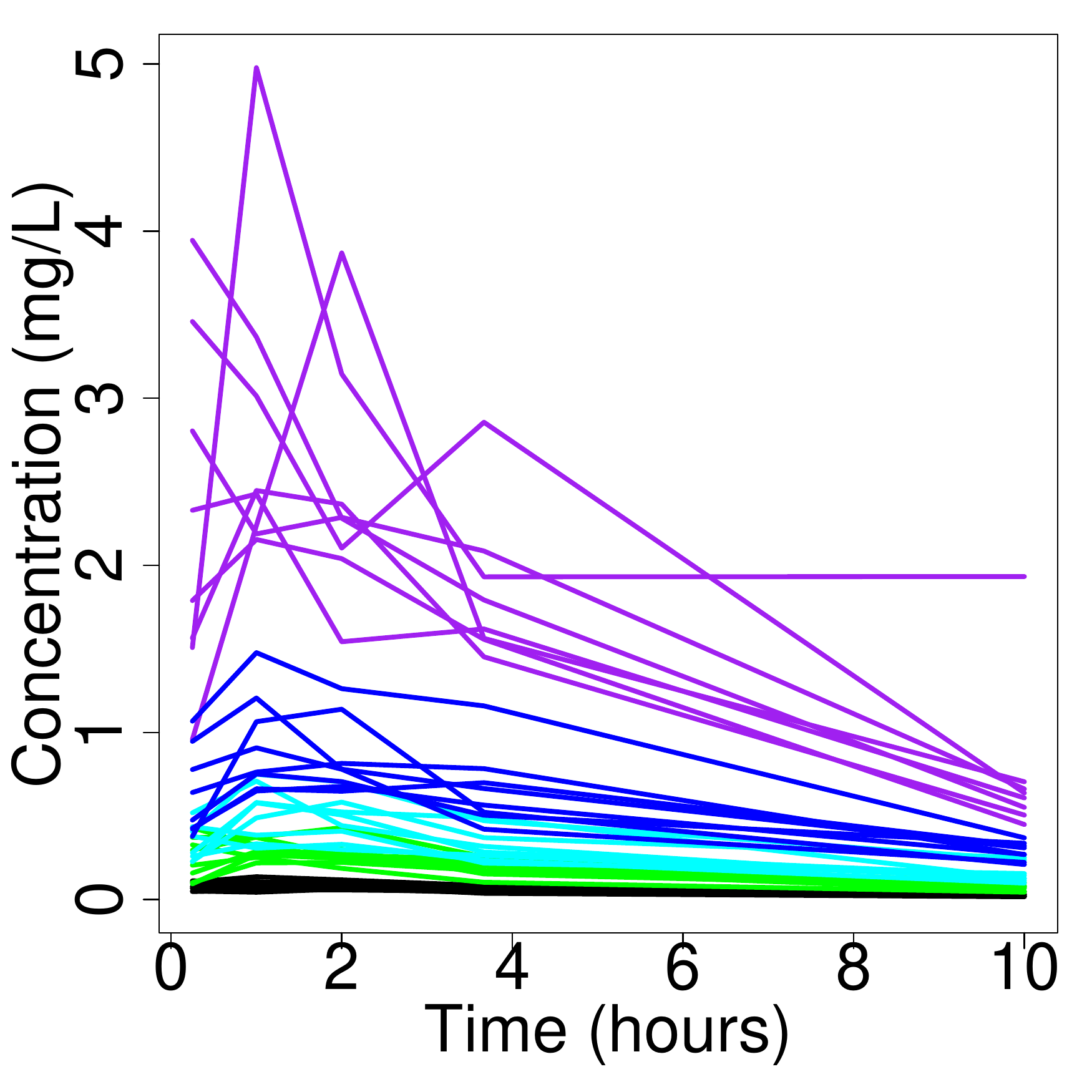}\label{fig:simu_concentration_time_rat_tr4_for_sc1}}}
\centerline{
\subfigure[]{\includegraphics[scale=0.25]{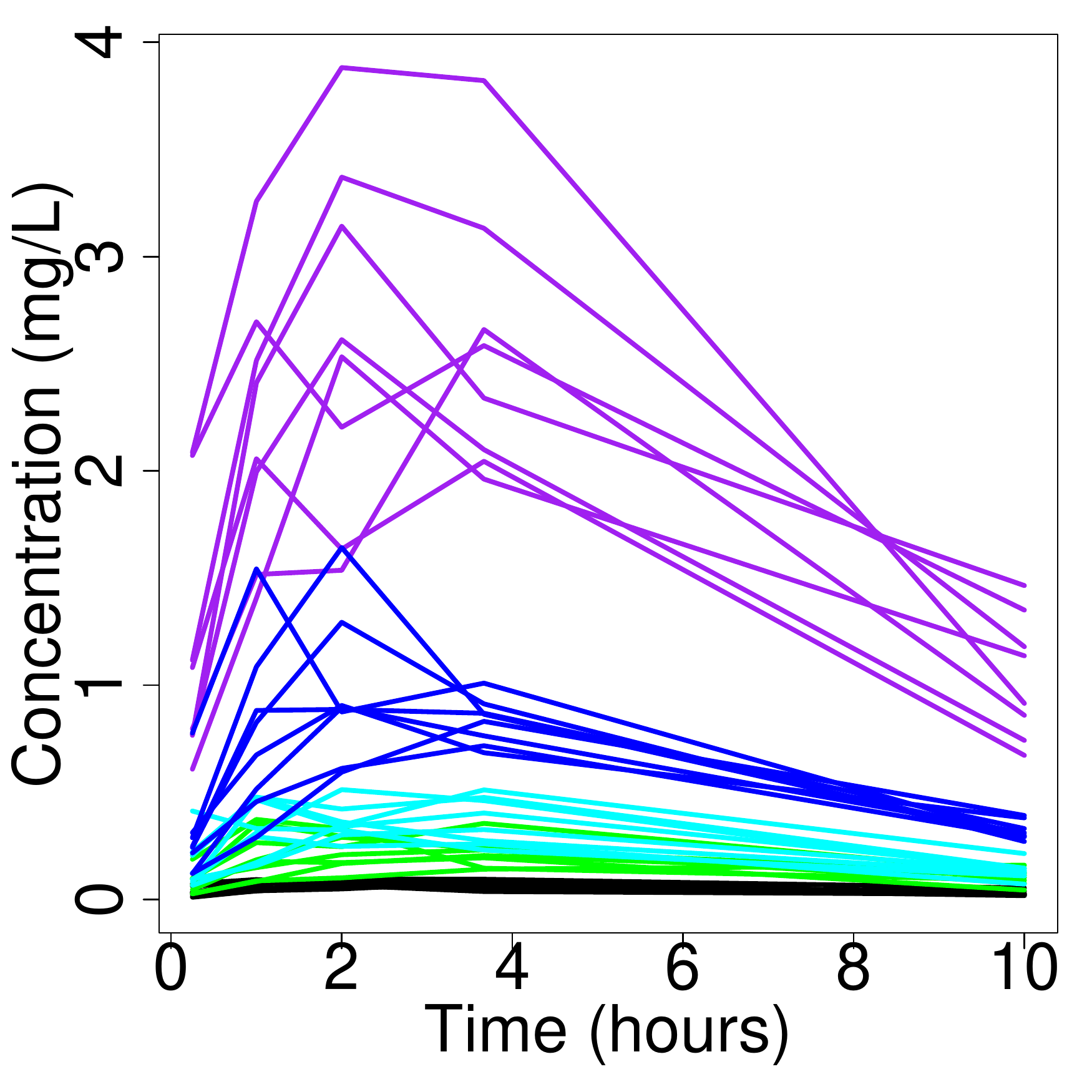}\label{fig:simu_concentration_time_rat_tr4_for_sc2}}
\subfigure[]{\includegraphics[scale=0.25]{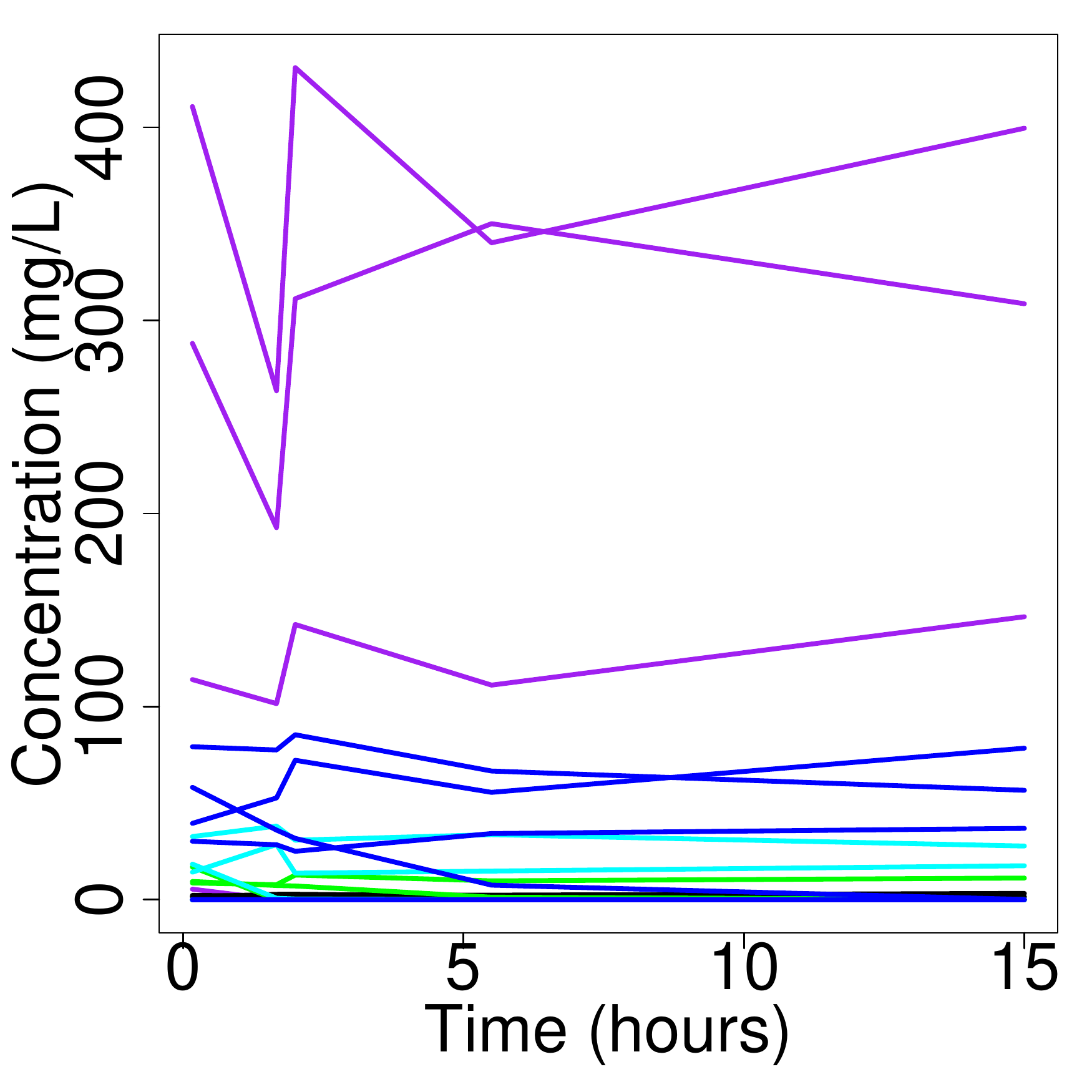}\label{fig:simu_concentration_time_dog_tr4}}} 
\caption{Simulated individual concentration of the drug versus time for one simulated dataset for mouse (a), rat in scenario 1 (b), rat in scenario 2 (c) and dog (d). This figure appears in color in the electronic version of this article.}
\label{fig:simu_concentration_time_tr4_for_sc1and2}
\end{figure}

To illustrate the steps of our methodology, let us first focus on the results from a single simulated dataset. Figure \ref{fig:simu_concentration_time_tr4_for_sc1and2} shows the concentration of the drug versus time for each animal of scenarios 1 and 2 for this dataset. 

In step 1 of our methodology, the Bayesian models for mouse data are first fitted using the non-informative \textit{prior} distributions given in Web Appendix C. The estimated \textit{posterior} means for mouse of $\mu_{CL}$ and $\mu_V$ are equal to 0.0961 L.h$^{-1}$ and 0.0508 L (results shown in Web Appendix C). Using extrapolation formulas \ref{eq:extrapolation_clearance} and \ref{eq:extrapolation_volume} from mouse to rat, we obtain the values of 0.368 L.h$^{-1}$ and 0.305 L that are used as means of the \textit{prior} distributions of the clearance and the volume for rat. Similarly, the estimated \textit{posterior} means for rat of $\mu_{CL}$ and $\mu_V$ are equal to 0.464 L.h$^{-1}$ and 0.216 L for scenario 1 (respectively 1.45 L.h$^{-1}$ and 0.581 L for scenario 2) and we use the extrapolated values of 10.8 L.h$^{-1}$ and 14.4 L (respectively 33.8 L.h$^{-1}$ and 38.7 L) as means of the \textit{prior} distributions of the clearance and the volume for dog.

\begin{figure} 
\centerline{
\subfigure[]{\includegraphics[scale=0.25]{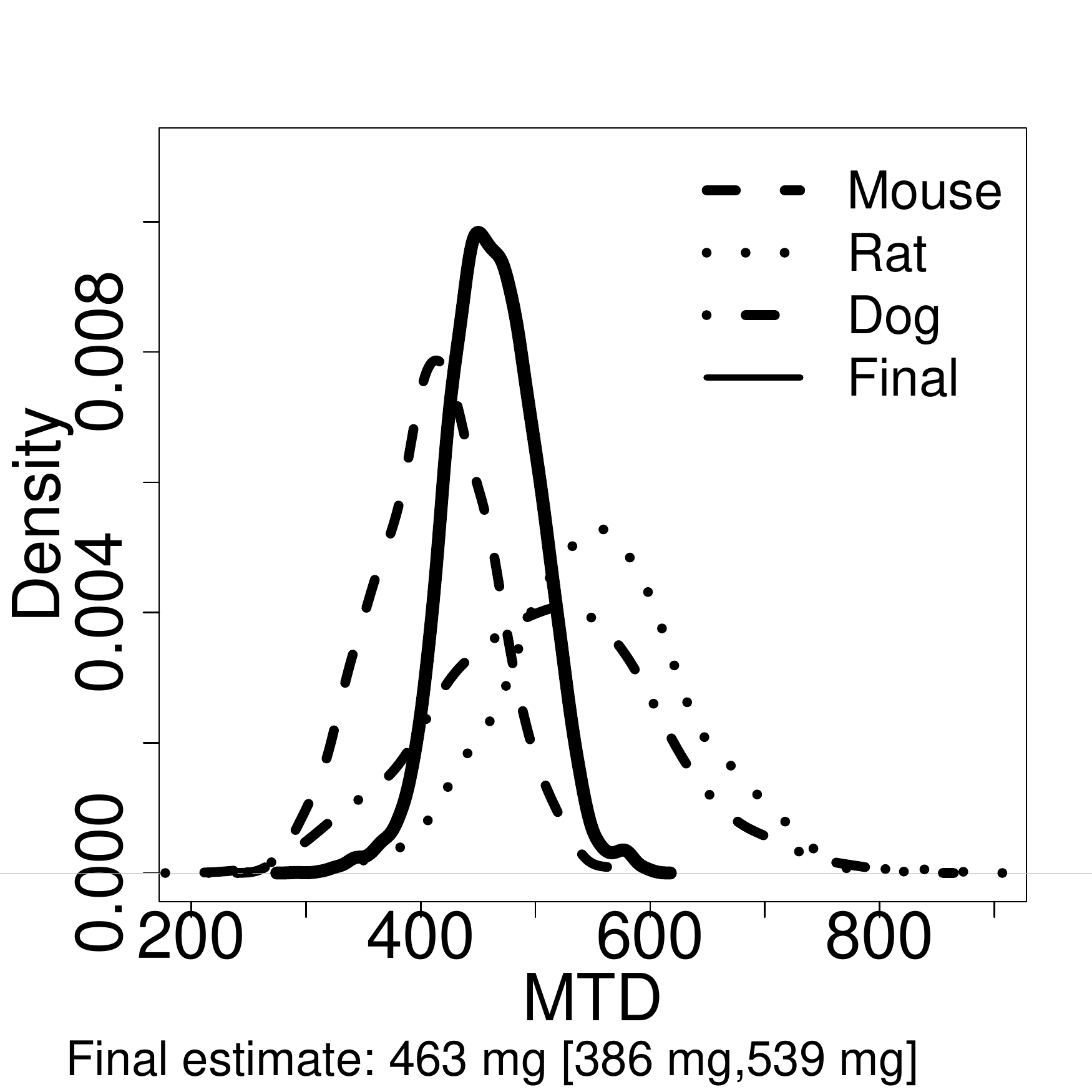}\label{fig:extrapolated_MTD_distributions_Bayesian_tr4_for_sc1}}
\subfigure[]{\includegraphics[scale=0.25]{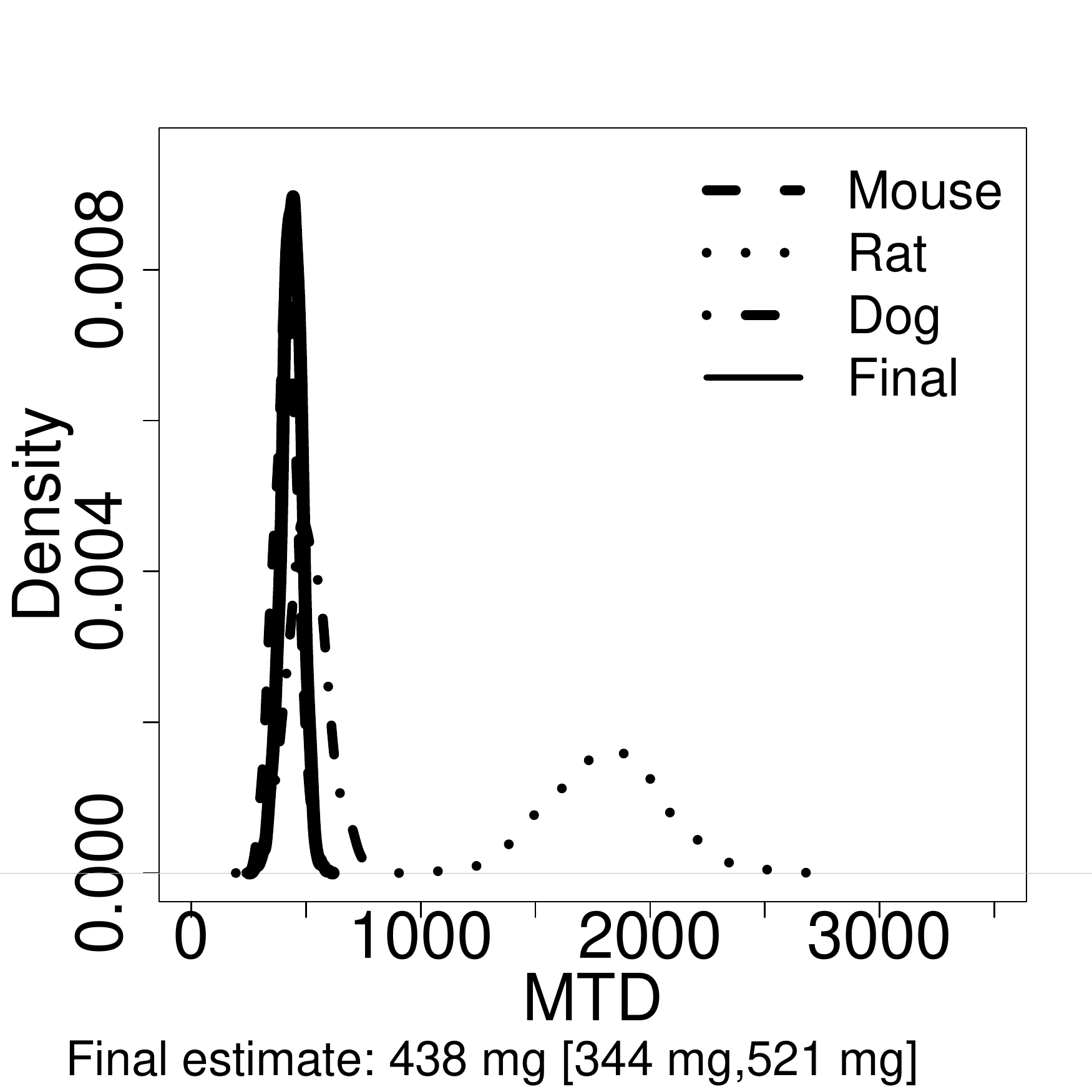}\label{fig:extrapolated_MTD_distributions_Bayesian_tr4_for_sc2}}}
\centerline{\subfigure[]{
\includegraphics[scale=0.25]{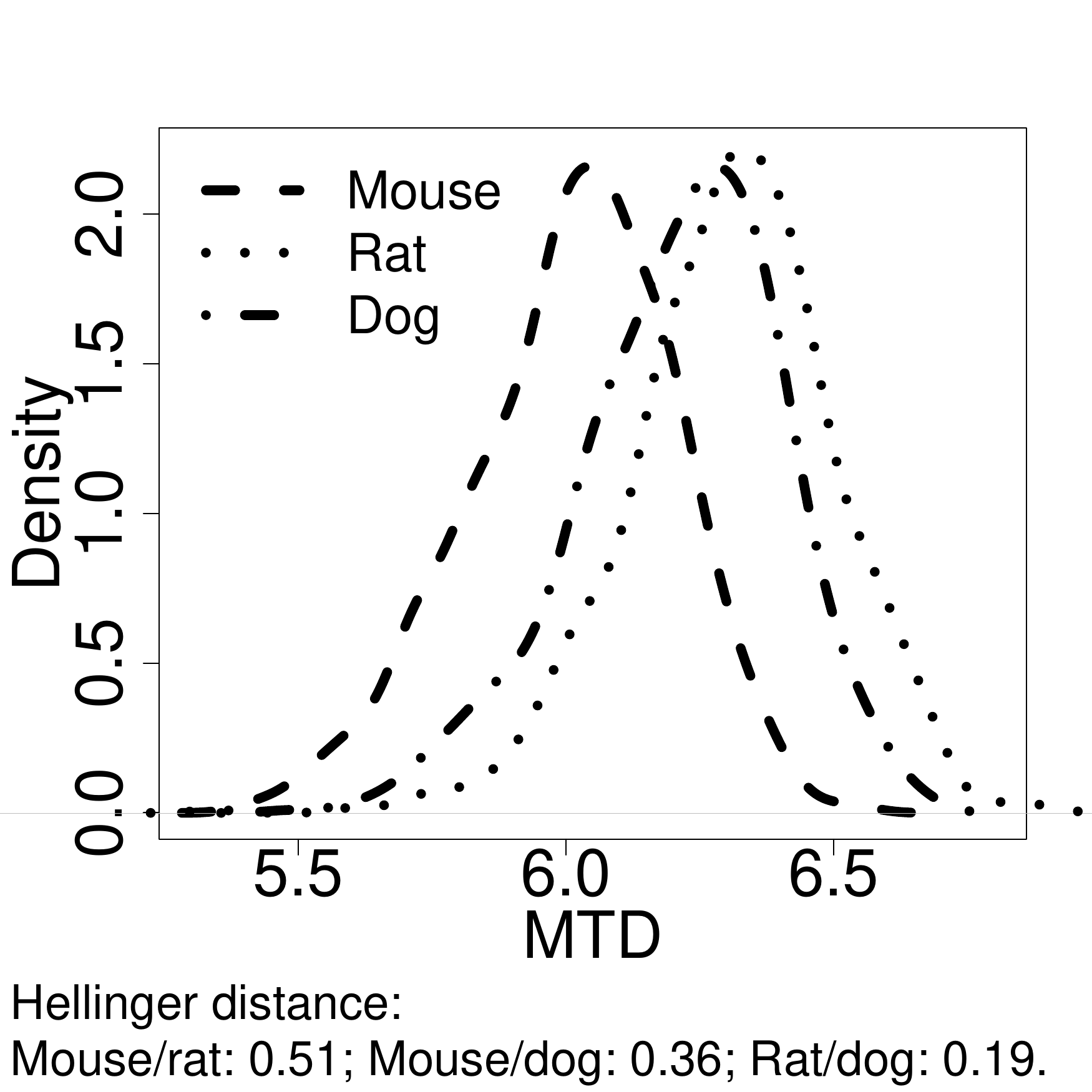}\label{fig:modified_log_extrapolated_MTD_distributions_Bayesian_tr4_for_sc1}}
\subfigure[]{\includegraphics[scale=0.25]{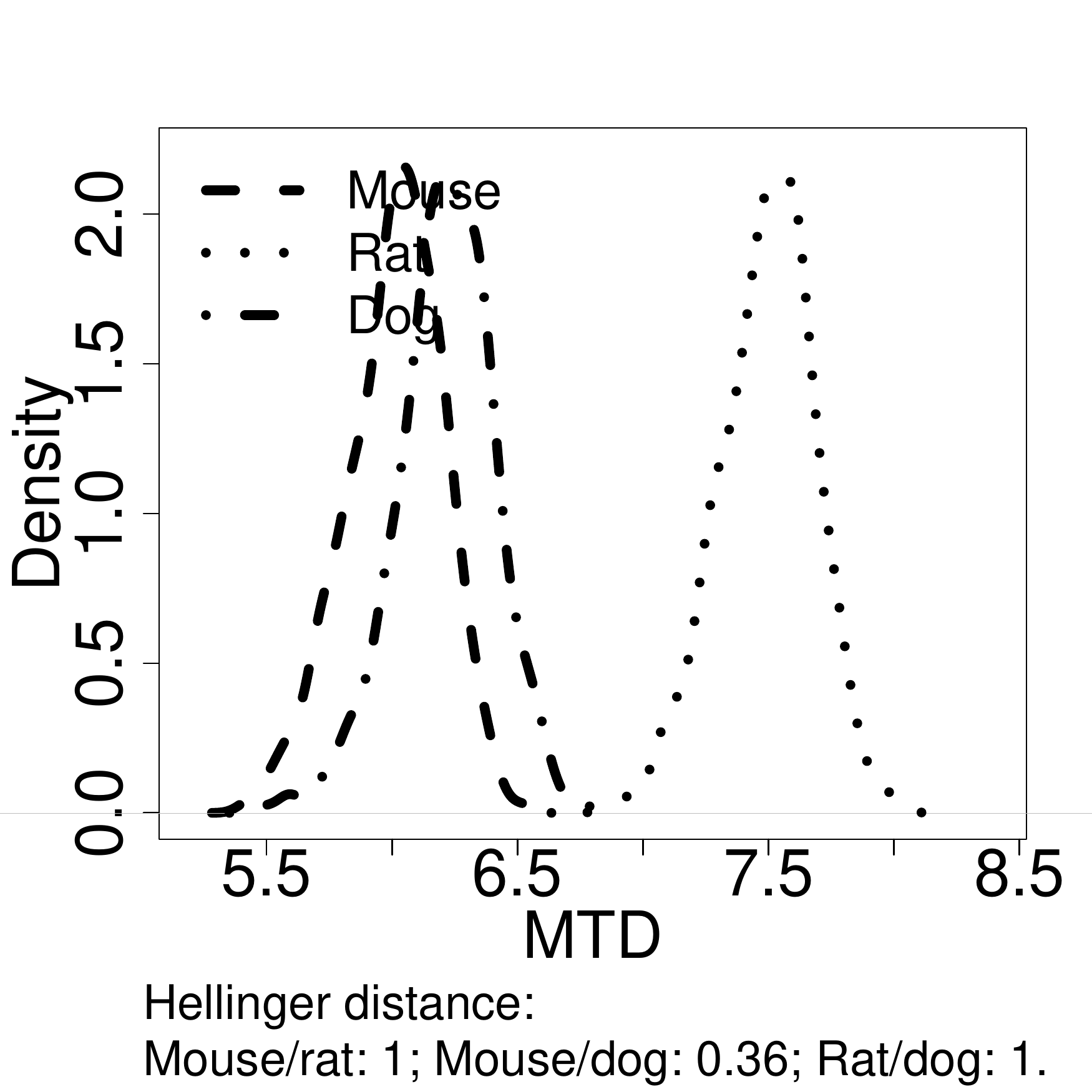}\label{fig:modified_log_extrapolated_MTD_distributions_Bayesian_tr4_for_sc2}}}
\caption{Maximum tolerated dose (MTD) distributions for one simulated dataset in scenarios 1 and 2, for the Bayesian approach. Extrapolated (animals to human) MTD distributions (from Step 2) and final predicted MTD distribution (from Step 4) in the dataset from scenario 1 (a) and  from scenario 2 (b); transformed extrapolated MTD distributions and Hellinger distances (from Step 3) in the dataset from scenario 1 (c) and from scenario 2 (d). The Hellinger distances equal to 1 are due to approximation in computation.}
\label{fig:MTD_distributions_Bayesian_tr4_for_sc1and2}
\end{figure}

The extrapolated MTD distributions from mouse, rat and dog to human resulting from step 2 are drawn in the first line of Figures \ref{fig:MTD_distributions_Bayesian_tr4_for_sc1and2} for these scenarios and each animal species as dotted lines.

In step 3, we transform the distributions from step 2 according to equation \ref{dose_distr_transf} so that the Hellinger distances between animal species do not depend on the difference between sample size in each study. 
The transformed extrapolated MTD distributions and the corresponding Hellinger distances are given in the second line of Figures \ref{fig:MTD_distributions_Bayesian_tr4_for_sc1and2}
for scenarios 1 and 2 and each animal species. For the fourth and final step, we set the threshold for the Hellinger distance at 0.5 for MTD based on the sensitivity analysis described in subsection \ref{subsec:commensurability_checking_res}. 
For scenario 1, Hellinger distances are equal to 0.51 for mouse versus rat, 0.36 for mouse versus dog and 0.19 for rat versus dog. Two of these distances are less than 0.5 so all animal species are selected. For scenario 2, the two distances including rat are greater than 0.5 (mouse versus rat: 1; rat versus dog: 1) but the last distance is less than 0.5 (mouse versus dog: 0.36) so only mouse and dog are selected.  

In step 4, the final predicted MTD distributions are computed by merging the extrapolated dose distributions between the animal species selected in the previous step, and overlayed on the plots in the first line of Figures \ref{fig:MTD_distributions_Bayesian_tr4_for_sc1and2} 
as a solid line. These final distributions have smaller variances than the distributions from each animal species separately. The MTDs are finally estimated to 465 mg and 437 mg for scenarios 1 and 2 respectively, as the \textit{posterior} means of these distributions.

\subsection{Using Hellinger Distance to Check Extrapolability from Animal Species to Human}\label{subsec:commensurability_checking_res}

\begin{figure} 
\centerline{
\subfigure[]{\includegraphics[scale=0.25]{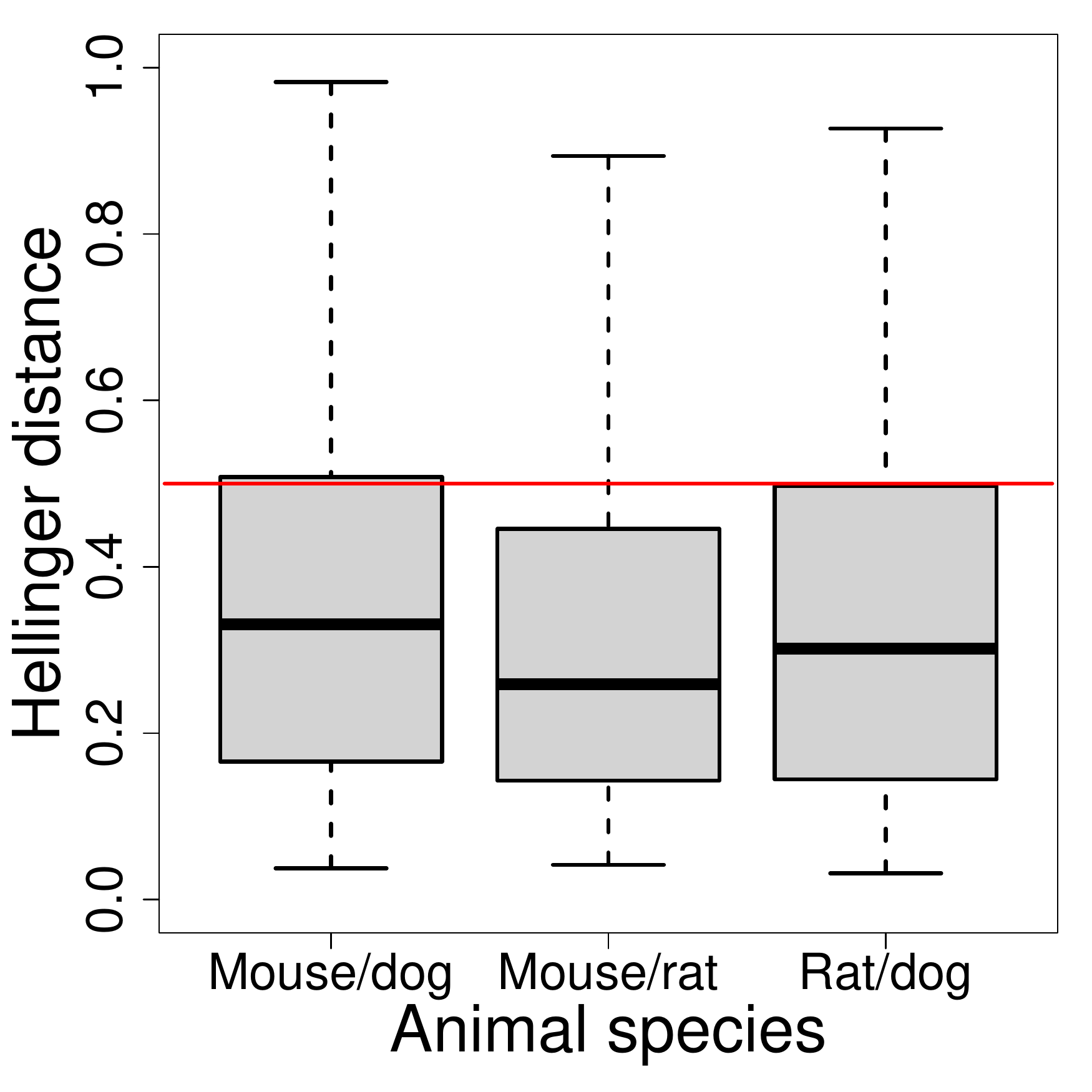}\label{fig:hellinger_dist_MTD_Bayesian_for_sc1}}
\subfigure[]{\includegraphics[scale=0.25]{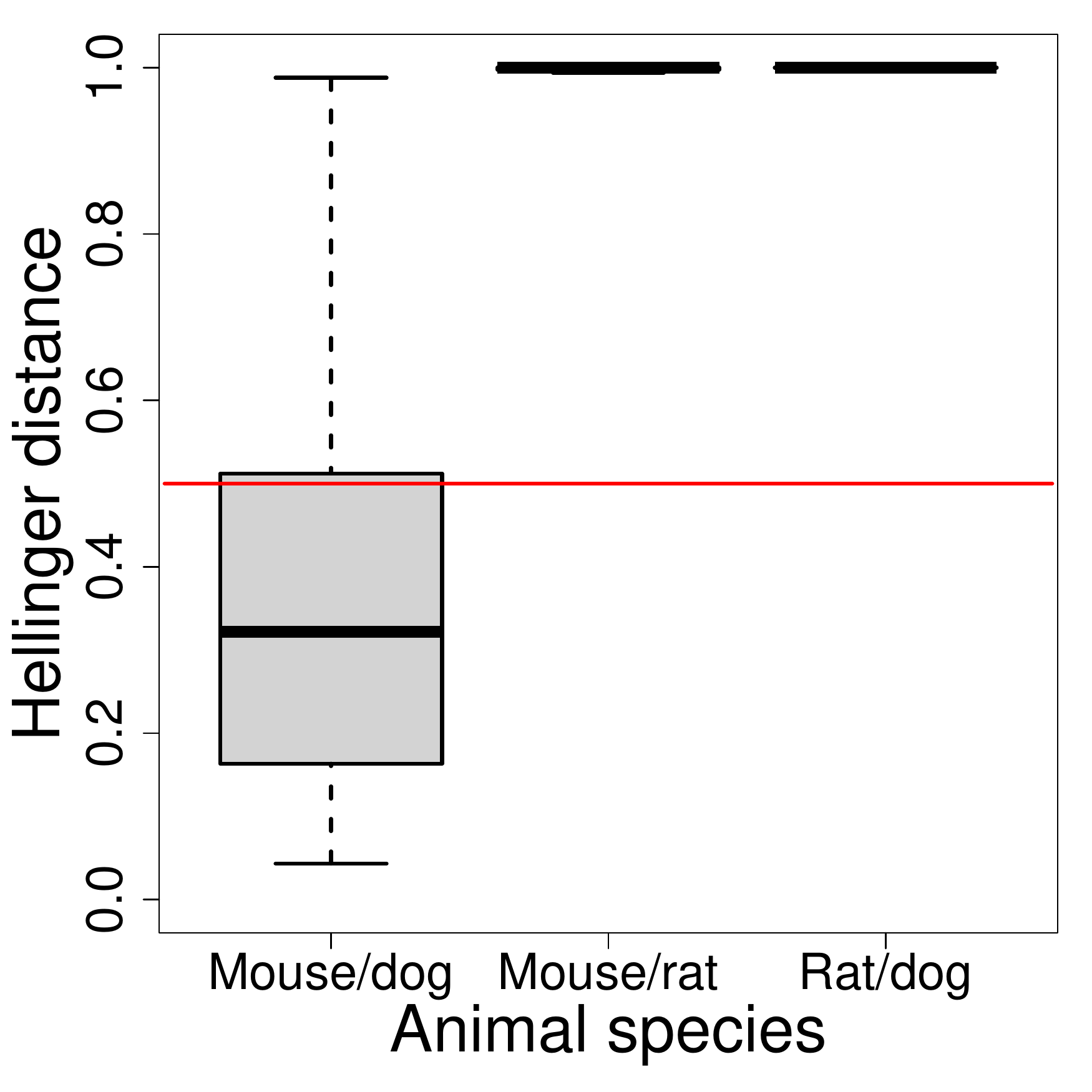}\label{fig:hellinger_dist_MTD_Bayesian_for_sc2}}}
\centerline{
\subfigure[]{\includegraphics[scale=0.25]{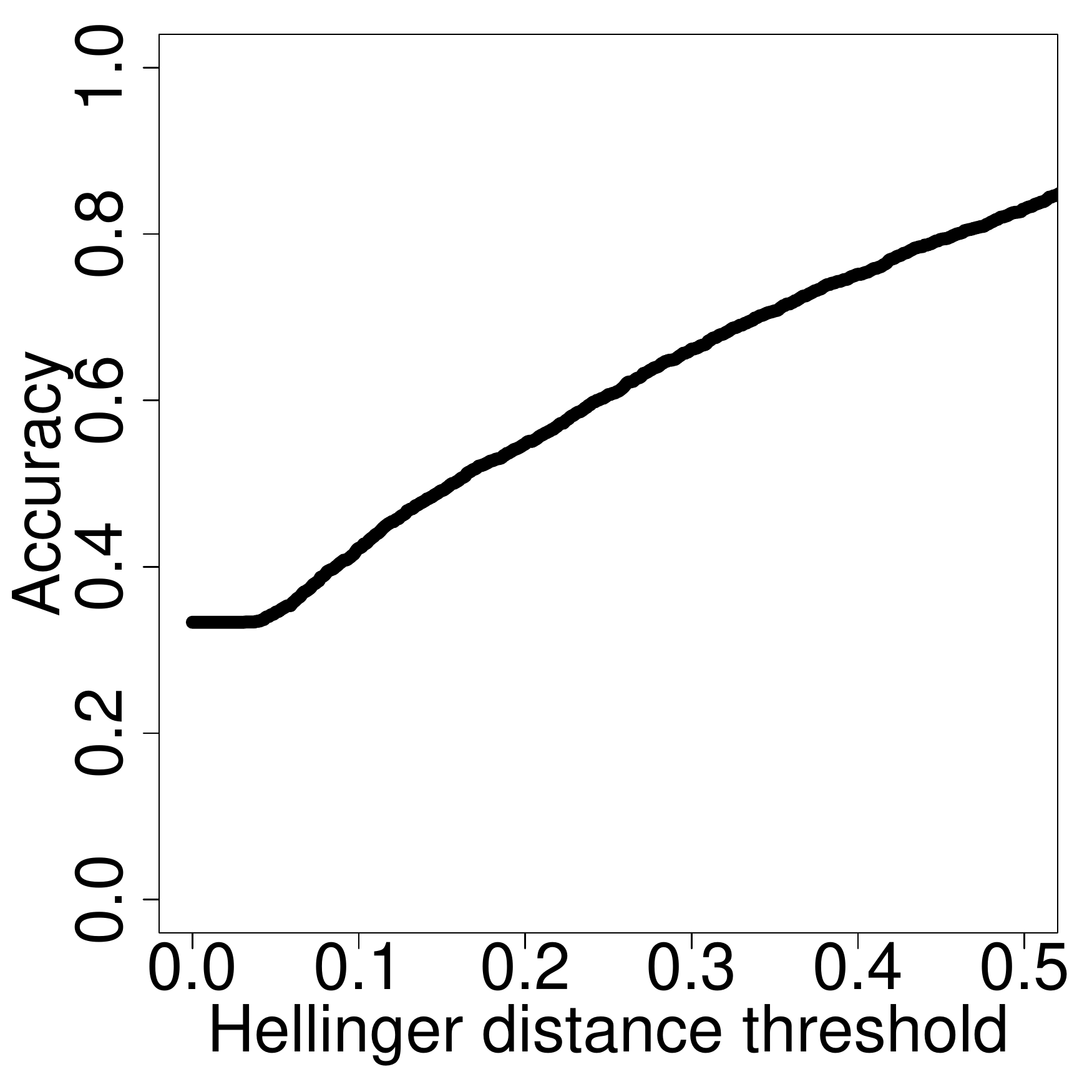}\label{fig:acc_by_thres_MTD_Bayesian_for_sc1and2}}} 
\caption{Hellinger distance of the transformed predicted MTD distributions in humans between animal species for scenario 1 (a) and scenario 2 (b) for the Bayesian approach} over 500 replications (under the assumption that $\omega_{V} = 0.7$ for mouse); Accuracy by Hellinger distance threshold for MTD in scenarios 1 and 2 for the  Bayesian approach over 500 replications (c). MTD: Maximum tolerated dose. This figure appears in color in the electronic version of this article.
\label{fig:hellinger_dist_and_acc_by_thres_MTD_Bayesian_for_sc1and2}
\end{figure}

We then applied our methodology to the scenarios with 500 replications each. Figures \ref{fig:hellinger_dist_MTD_Bayesian_for_sc1} and \ref{fig:hellinger_dist_MTD_Bayesian_for_sc2} show boxplots of the Hellinger distances between two species for the Bayesian approach in the different scenarios, for the MTD. As seen in these figures, the Hellinger distance easily distinguishes animal species between those that show similarities in extrapolated toxicity and those that show dissimilarities.

Indeed, for scenario 1 for which the extrapolation is correct from all animal species to human, 
the Hellinger distances between animal species for the MTD distributions are lower than 0.5 in more than 73\% of cases. This indicates consistent results across all animal species and therefore the possibility of using them all to derive human MTDs.

For scenario 2 that uses an inaccurate extrapolation of toxicity model parameters for rat, the Hellinger distances between mouse and rat, and between rat and dog are close to 1 but again lower for mouse versus dog. This indicates MTD distributions are consistent between mouse and dog, but not for rat.

As shown in Figure \ref{fig:acc_by_thres_MTD_Bayesian_for_sc1and2}, as Hellinger distance threshold increases, so does accuracy for MTD. For MTD, we set the threshold at 0.5. We decided to stop the accuracy evaluation at the threshold of 0.5 since the Hellinger distance is bounded between 0 and 1, and going beyond the median value did not seem relevant to us.

\subsection{Final Estimations of MTD in Humans}\label{subsec:final_estimation}

\begin{figure} 
\centerline{
\subfigure[]{\includegraphics[scale=0.25]{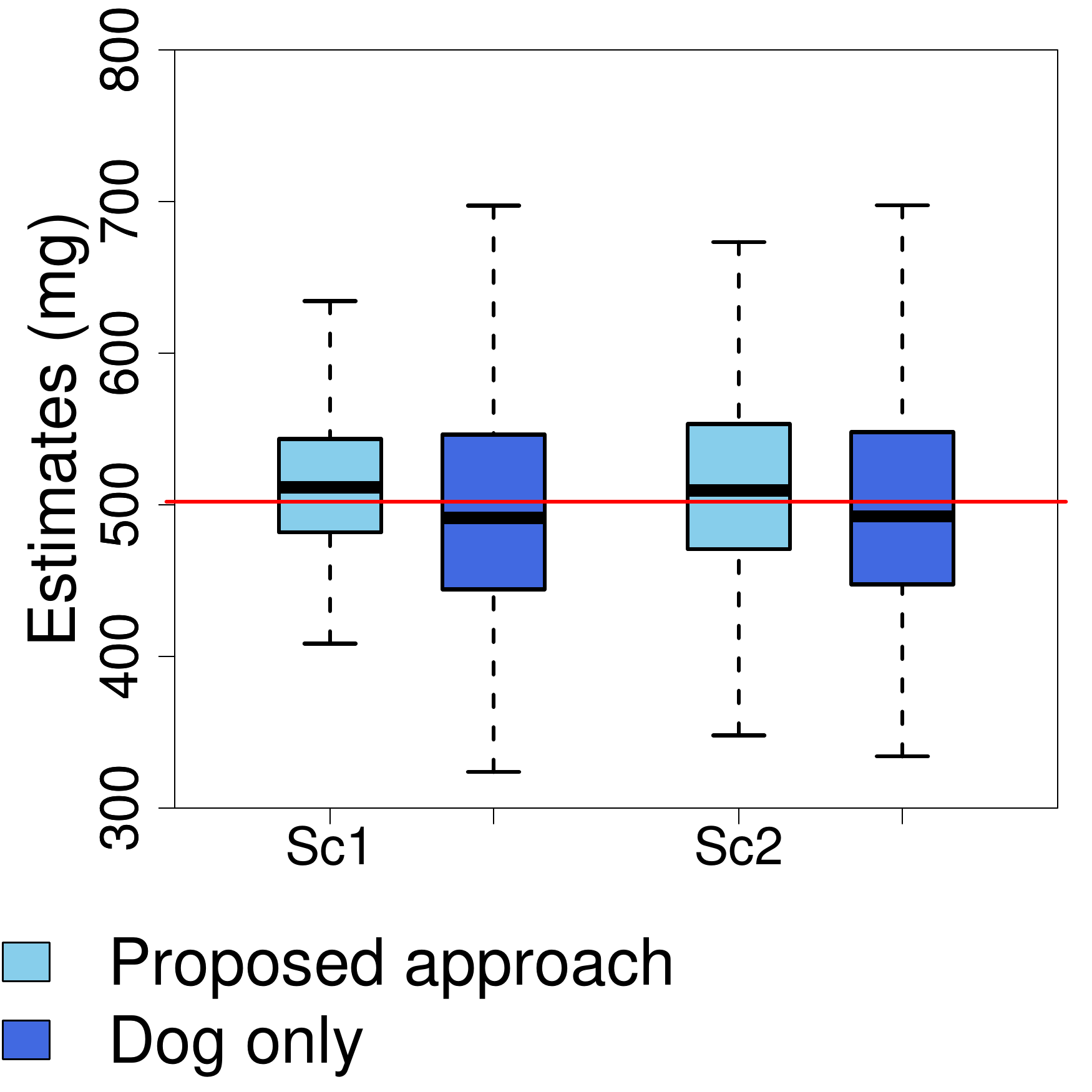}\label{fig:MTD_posterior_mean_Bayesian_for_sc1and2}}
\subfigure[]{\includegraphics[scale=0.25]{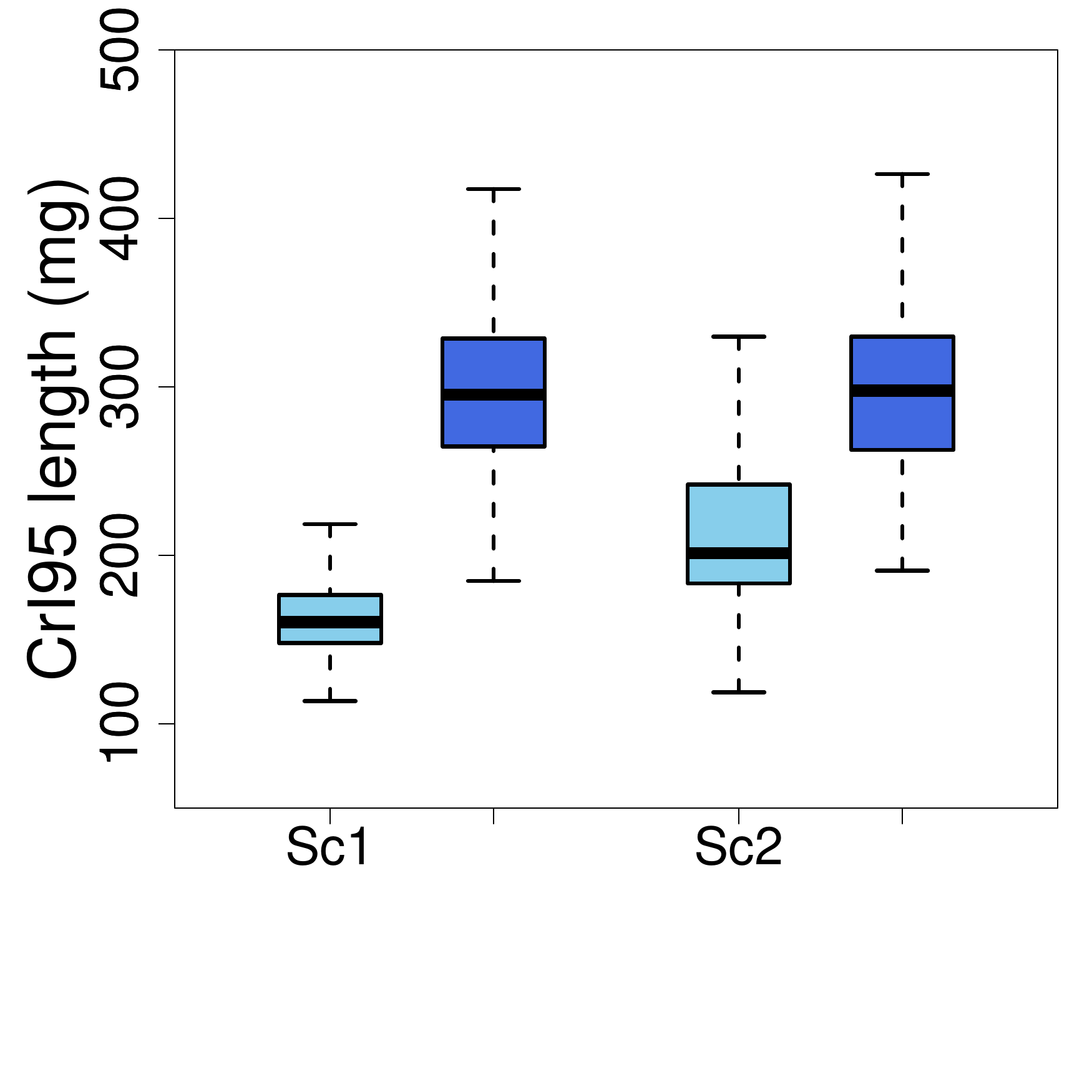}\label{fig:MTD_posterior_IC95_length_Bayesian_for_sc1and2}}}
\caption{Estimated MTD in humans (a) and the length of the 95\% credibility interval (CrI95) (b) for scenarios 1 and 2 for the Bayesian approach and using the standard approach (that is, only dog data) over 500 replications, under the assumption that $\omega_{V} = 0.7$ for mouse. MTD: Maximum tolerated dose. This figure appears in color in the electronic version of this article.}
\label{fig:MTD_posterior_mean_and_IC95_length_Bayesian_for_sc1and2}
\end{figure}

As shown in Figure \ref{fig:MTD_posterior_mean_Bayesian_for_sc1and2}, for scenario 1 for which data from all animal species are often used to estimate the final MTD, the MTD is correctly estimated to 515 mg (standard deviation - sd - equal to 48 mg) by the Bayesian approach (close to the true value of 502 mg). For scenario 2, the MTD is slightly overestimated to 561 mg (sd: 296 mg).

The length of the equal tails 95\% credible interval (CrI95) is considerably greater when only the dog results are used (that is the standard approach) to calculate the MTD than when using the proposed method (see Figure \ref{fig:MTD_posterior_IC95_length_Bayesian_for_sc1and2}). Actually, the mean of the CrI95 lengths is around 300 mg (sd: 47 mg) using only dog results compared to the values of 165 mg and 242 mg (sd equal to 26 mg and 153 mg) for respectively scenarios 1 and 2 using the proposed approach.

The results of the sensitivity analysis for $\omega_{V} = 0.4$ and $\omega_{V} = 1$ for mouse, given in Web Appendix D, do not differ from those obtained for $\omega_{V} = 0.7$ .

\section{Discussion}\label{sec:discussion}

We proposed a Bayesian approach for multisource data integration, that was tailored, in this work, to dose extrapolation from preclinical to clinical research. In particular, steps 3 and 4 require methodological innovations such as a way to normalize the quantity of information of a distribution come from longitudinal data and the modified Bayes formulas. The new framework allows to better use all available information compared to the standard methods, reducing uncertainty in the predictions and potentially leading to a more efficient dose selection. In preclinical setting, this new framework can be seen as a Bayesian generalisation of the \cite{us_food_and_drug_administration_2005} guidelines. Furthermore, the framework is very flexible. Submodels (linear, generalised, mixed-effect, etc.) can be specified according to the study outcomes and could be different between studies.

To evaluate the performance of the approach, we used a case study inspired by the preclinical development of galunisertib. In this example, only data from \textit{in vivo} studies were simulated. However, it is possible to include data from \textit{in vitro} studies as well, by simply considering the appropriate extrapolation formulae to humans, that is working on areas or cells number. In our example, only some estimates of PK fixed effects are extrapolated from animals to humans using allometric formulas. Indeed, to our knowledge, there is no analytical formula to extrapolate the parameters of the PD models, nor to anticipate interindividual variability between species in this setting. However, mouse or rat, for example, which are specially bred for preclinical studies (controlled lines, similar age, same diet, same physical exercises, ...) could be more homogenous than healthy volunteers or patients. In our simulations, a sensitivity analysis with lower variances for these species did not indicate that the uncertainty on the variabilities was an issue, but we could also incorporate expected changes across species in the extrapolation step, for instance to integrate differences in disease course or mechanistic pathways.

In our simulation setting, the theoretical MTD (i.e. the MTD calculated from the true values of the model parameters) extrapolated to humans shown in Table \ref{tab:simu_parameters_for_PK_model_and_extrapolated_mtd_for_sc1and2} 
are not exactly the same between animal species, even for scenario 1, but, in general, very close each other. Therefore, this imperfection of the simulation scenarios should be kept in mind when interpreting the results. However, in real life, the extrapolation of a dose between two species is always subject to a margin of error.

In the third step of our methodology, a rule/algorithm must be defined to check the commensurability between distributions and to select the ones to carry to the next step. In our example, since we have three studies, we proposed a simple algorithm to do it. However, when more than three studies are available, clustering algorithms based on a distance matrix can be investigated to select studies for the last step. Accuracy computation, as done in this work, or ROC (Receiver Operating Characteristic) curves can be adopted to calibrate algorithm thresholds and could be applied directly on distances results (as in this work) or on the final results from the algorithm. Moreover, another distribution distance could be used instead of the Hellinger distance. We suggested it due to its symmetric propriety and, since it is bounded, it is easily interpretable.

To run simulations, we decided to keep the results involving the more numerous cluster, that is, in this case, if at least two species were selected at the coherence step. In reality, if it is well-known that the results on a species are really closed/related to human ones, this species can be selected as default and only species clustered with it will be carried on at the merging step.

Some additional extensions will be investigated. 
It could be interesting to include external sources of information, such as expert opinion or literature data. Such data could be integrated into the weakly informative \textit{prior} distributions of the parameters in step 1, or as additional dose distributions in the final step. External information should be introduced wisely, since if used multiple times it can impact the results of eq.~\ref{eq:merge}. Indeed, when only weakly informative \textit{prior} distribution, carrying on a small and negligible ESS, are adopted, the final impact on eq.~\ref{eq:merge} could be neglected. Otherwise, further studies on how discounting embedded \textit{prior} distribution are needed.
Another perspective would be to handle weighted \textit{posterior} distributions through the Hellinger distance and the merging formula, in order to give more weight to certain results, for example to account for larger sample sizes or on previous knowledge on human similarity. Following this idea, several doses can be proposed regarding the number of \textit{posterior} distribution modes. Then, novel dose-finding methods should be developed to take into account a possible change in the dose panel evaluated in the trial.
In addition, in the methodology, no uncertainty is attributed to the extrapolation approach itself. However, it is possible that a lack of consistency is observed because of inappropriate extrapolation rather than due to mismatch between preclinical data and human. Thus, our framework could be expanded with different extrapolation strategies (allometry, PBPK, etc.) that then will be used at the coherence step, or distribution can be assumed for extrapolation parameter formulas, carrying on this uncertainty during the whole extrapolation process.

Another straightforward extension is using this framework to compute \textit{prior} distributions for human model parameters used in the future (human) dose-finding trial. Indeed, while our work was focused on doses estimation, we can work at parameter level, that is clustering and merging parameter \textit{posterior} distributions. This will give, as results, a distribution for each parameter, that could be used as \textit{prior} distribution on human model parameters. ESS discounting or dynamic borrowing may worth to be studied.

In conclusion, we proposed a new framework to facilitate preclinical to clinical extrapolation in four main steps. Each step could be customized (models, algorithms, extrapolation formulas and hypotheses, etc.) according to the studies the researchers are working on.

\paragraph{Acknowledgements} 

This work is part of the European FAIR project that has received funding from the European Union's Horizon 2020 research and innovation program under grant agreement N$^\text{o}$ 847786. 
Sandrine Boulet and Moreno Ursino made equal contributions and are co-first authors.
Emmanuelle Comets and Sarah Zohar made equal contributions and are co-last authors.

\bibliography{FAIR_art1_main_manuscript_30-Nov-2022_arxiv_format.bib}

\paragraph{Supporting Information} 

Web Appendices, Tables and Figures, referenced in Sections \ref{sec:methods}, \ref{sec:simulation_design} and \ref{sec:simulation_results}, are available with this paper.

\end{document}


\maketitle

\section{Web Appendix A: Complement to the Methods Section}\label{sec:compl_methods}

In this section we first introduce a hybrid approach, where at the first step the models are fitted using traditional frequentist methods.
Then, a theoretical example of the implementation of the merging equation (step 4) is given assuming a binary outcome and one dose.

\subsection{First Step for Hybrid Approach}\label{subsec:hybrid_approach}

The hybrid framework differs from the Bayesian approach only by the first step that is substituted by independent frequentist estimations followed by a probabilistic sensitivity analysis (PSA) approach~\citep{baio_2015}.

For each study, the dose-outcome models (fixed- or mixed-effect models) $\mathbf{f}_k(\mathbf{y}_k, \boldsymbol{\theta}_k$) are fitted following the frequentist paradigm, to obtain parameter estimates. We then assign distributions to each parameter $\boldsymbol{\theta}_k$ using the frequentist point estimates (with standard errors) via the PSA approach, to propagate uncertainty in a decision problem. When using maximum likelihood estimation (or related techniques), the normal approximation of the estimator can be used to associate to each parameter estimated in the model a (marginal) normal distribution with the estimated parameter value as mean and its standard error, computed using the inverse of the observed Fisher information matrix, as standard deviation.  

To summarise, at the end of both the Bayesian and hybrid processes, each study $k \in \{1, ..., K\}$ yields a distribution (\textit{posterior} or associated via PSA approach) for each parameter, that is, each element of each $\boldsymbol{\theta}_{k}$. The following three steps are then applied to the results of both approaches.

\subsection{Beta Distribution Example for the Fourth Step}\label{sec:formula_merge}

We assume we have $K$ sequential studies with a binary outcome and a Bayesian beta-binomial model (on the $\theta$ parameter). Let $x_k$ and $n_k$ represent the number of successes and the total sample size in the $k$-study, $k = 1, \ldots , K$. In a sequential analysis, that is when the \textit{posterior} distribution of $\theta$ of the $(k-1)$-study is used as \textit{prior} for the $k$th study, and if a beta distribution with parameter $a_0$ and $b_0$ is used (it can be seen as the \textit{posterior} of $k=0$), then the \textit{posterior} distribution of the last study is also a beta distribution, 
\begin{equation}\label{eq:beta1}
    \mbox{Beta}\left(a_0 + \sum_{k = 1}^K x_k,  b_0 + \sum_{k = 1}^K (n_k - x_k) \right).
\end{equation}

Applying Eq.2 of the main manuscript to merge the extrapolated dose distributions between the studies selected in the third step of the methodology, and assuming that at each analysis a \textit{prior} $\mbox{Beta}\left(a,  b\right)$ is used, we have:
\begin{align}
    \pi(\theta) \propto & \prod_{k=1}^K \mbox{Beta}\left(a + x_k,  b +  (n_k - x_k) \right) \label{eq:beta2}  \\ 
    \propto  &\prod_{k=1}^K  \theta^{a + x_k -1} (1-\theta)^{b +  n_k - x_k -1 } \nonumber \\ 
     \propto &~ \theta^{Ka - K + \sum_{k=1}^K x_k} (1-\theta)^{Kb - K +  \sum_{k=1}^K (n_k - x_k)} \nonumber \\ 
      = &~ \mbox{Beta}(Ka - K +1 + \sum_{k=1}^K x_k, \nonumber \\ 
     & \quad Kb - K + 1 + \sum_{k=1}^K (n_k - x_k)). \nonumber
\end{align}

Eq.~\ref{eq:beta1} and eq.~\ref{eq:beta2} yield the same results if $a=b=a_0=b_0=1$. Moreover, if $a= (a_0 + K - 1)/K$ and $b= (b_0 + K - 1)/K$, that is if the \textit{prior} information is split among the K studies, both results still coincide.

\section{Web Appendix B: Additional Simulation Settings}\label{sec:additional_simulation_settings}

In this section, we introduce the simulations performed including the MED estimation.

\subsection{Complement to Galunisertib Case Study} \label{subsec:additional_case_study}

To design the simulation study, we simplified the PD model to a direct response model as in~\cite{ursino_2017} to allow for shorter runtimes.

\subsection{Efficacy Data Simulation}\label{subsec:eff_simu}

In the original study, efficacy was measured in terms of percentage of inhibition of the biomarker pSMAD.

Efficacy data for a given study are generated using an $I_{\max}$ model with an effect compartment model to account for the delay between the kinetics of the drug and the observed inhibition:
\begin{equation}\label{eq:efficacy_model}
I(t) = 1-\dfrac{I_{\max} \times C_e(t)}{C_e(t)+IC_{50}}
\end{equation}
where $I_{\max} = 1$ is the maximum inhibition effect; $IC_{50} \sim LN(\log(\mu_{IC_{50} }),\omega_{IC_{50} })$ is the concentration at which 50\% of $I_{\max}$ is reached and $C_e$ is the concentration in the effect compartment given by the following equation:
\begin{align}
C_e(t) = &\dfrac{d k_ak_e}{V} \Big(\dfrac{\exp(-k_at)}{(\frac{CL}{V}-k_a)(k_e-k_a)} + \nonumber \\
& \dfrac{\exp(-\frac{CL}{V}t)}{(k_a-\frac{CL}{V})(k_e-\frac{CL}{V})}+\dfrac{\exp(-k_et)}{(k_a-k_e)(\frac{CL}{V}-k_e)}\Big) \label{eq:concentration_efficacy_model}
\end{align}
where $k_e \sim LN(\text{log}(\mu_{k_e}),\omega_{k_e})$. The MED is defined as the dose that produces a maximum response over $\tau_E$  in more than $p_E$ of the cases. With this model, there is no simple analytical formula to calculate the MED. However, it can be approximated by simulations.
\\
\\ We assume a proportional error of 20\% for the measurements of the percentage of inhibition of pSMAD, defined as $\tilde I \sim LogitN(\text{logit}(I),0.2)$ where $I$ represents  the true values of the measurements and $\tilde I$ the measured values.

\subsection{Additional Scenarios Accounting for Efficacy}\label{subsec:add_scenarios}

\begin{figure}
\centerline{
\subfigure[]{\includegraphics[scale=0.25]{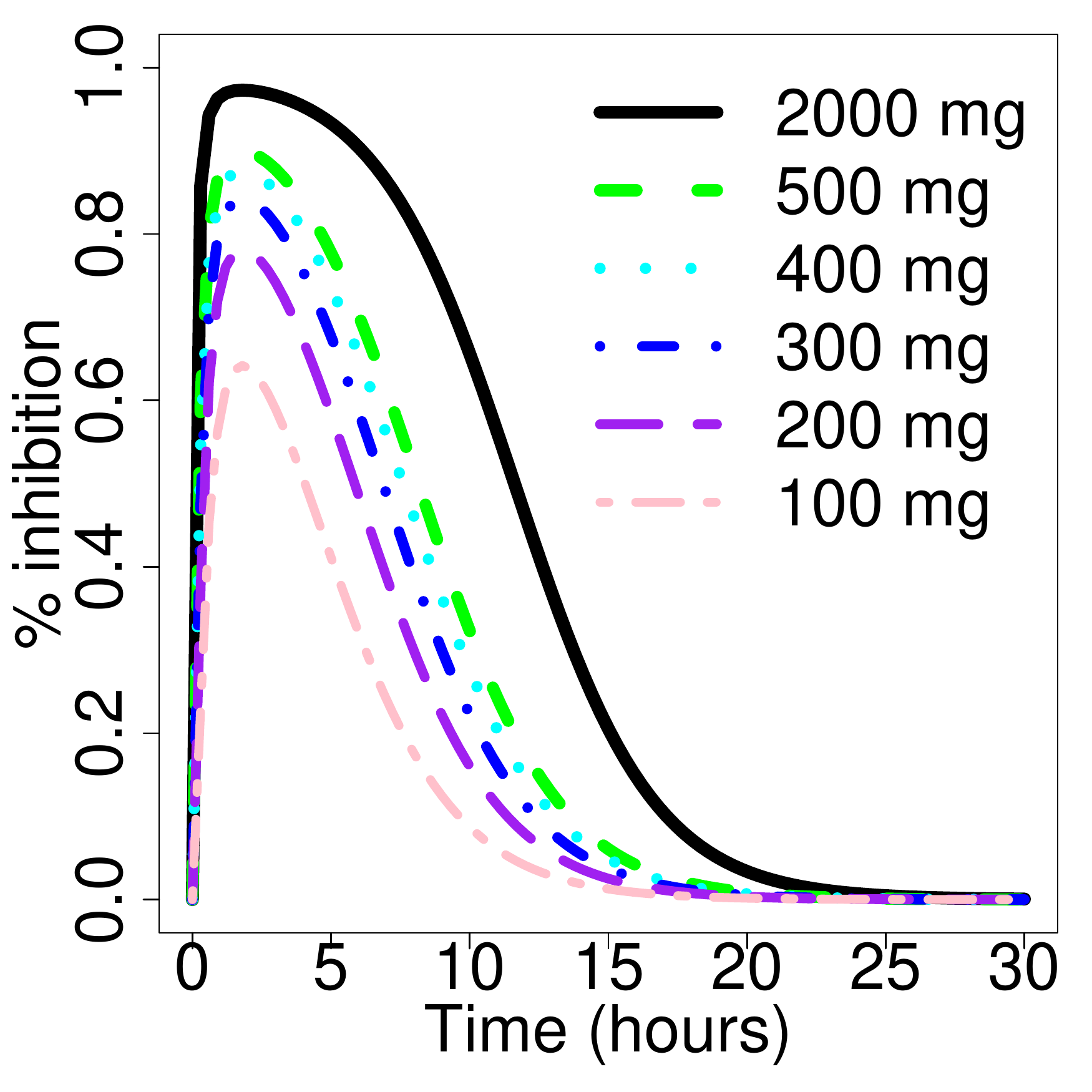}\label{fig:simu_perc_inhib_time_human}}}
\centerline{
\subfigure[]{\includegraphics[scale=0.25]{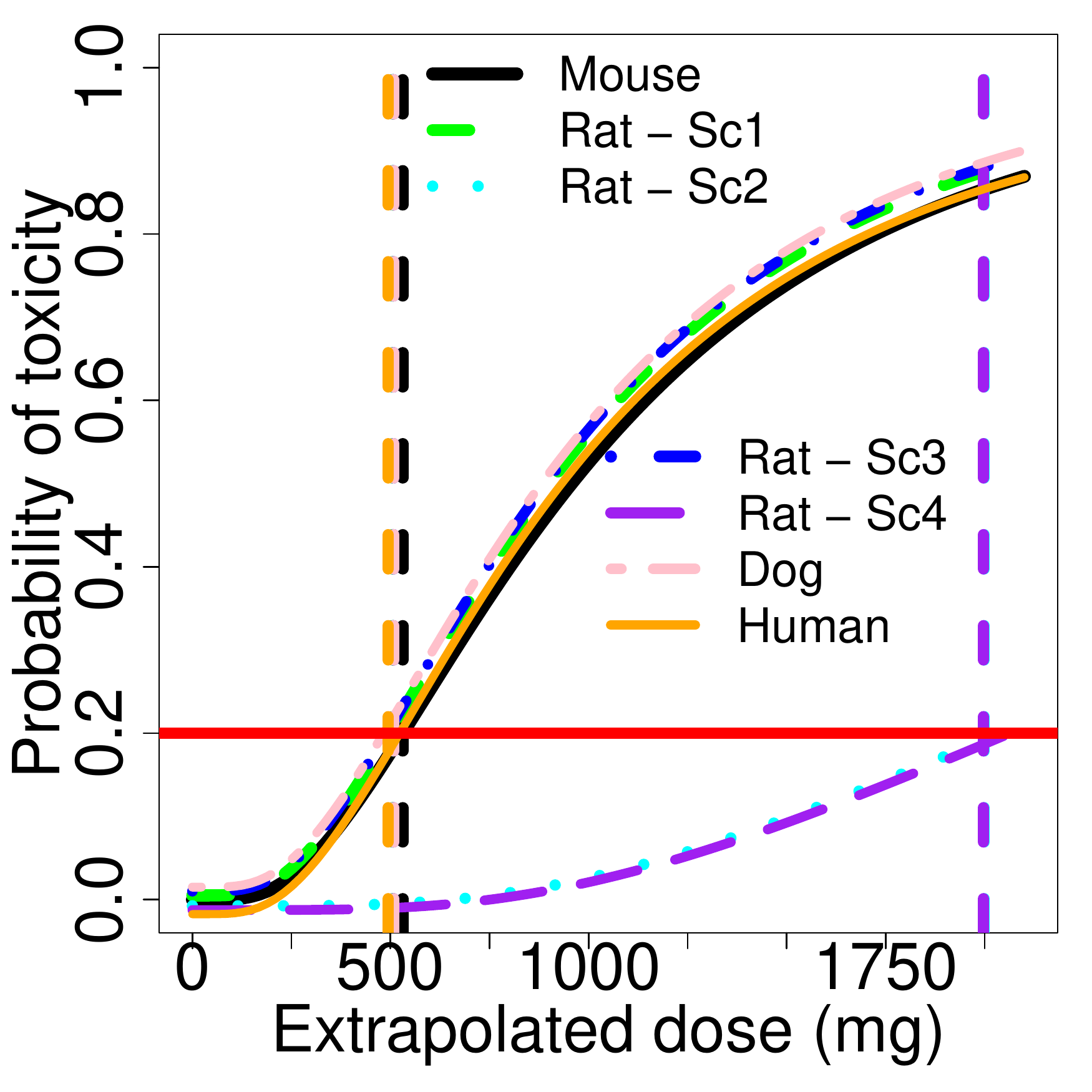}\label{fig:tox_prob_by_extrapolated_dose}}
\subfigure[]{\includegraphics[scale=0.25]{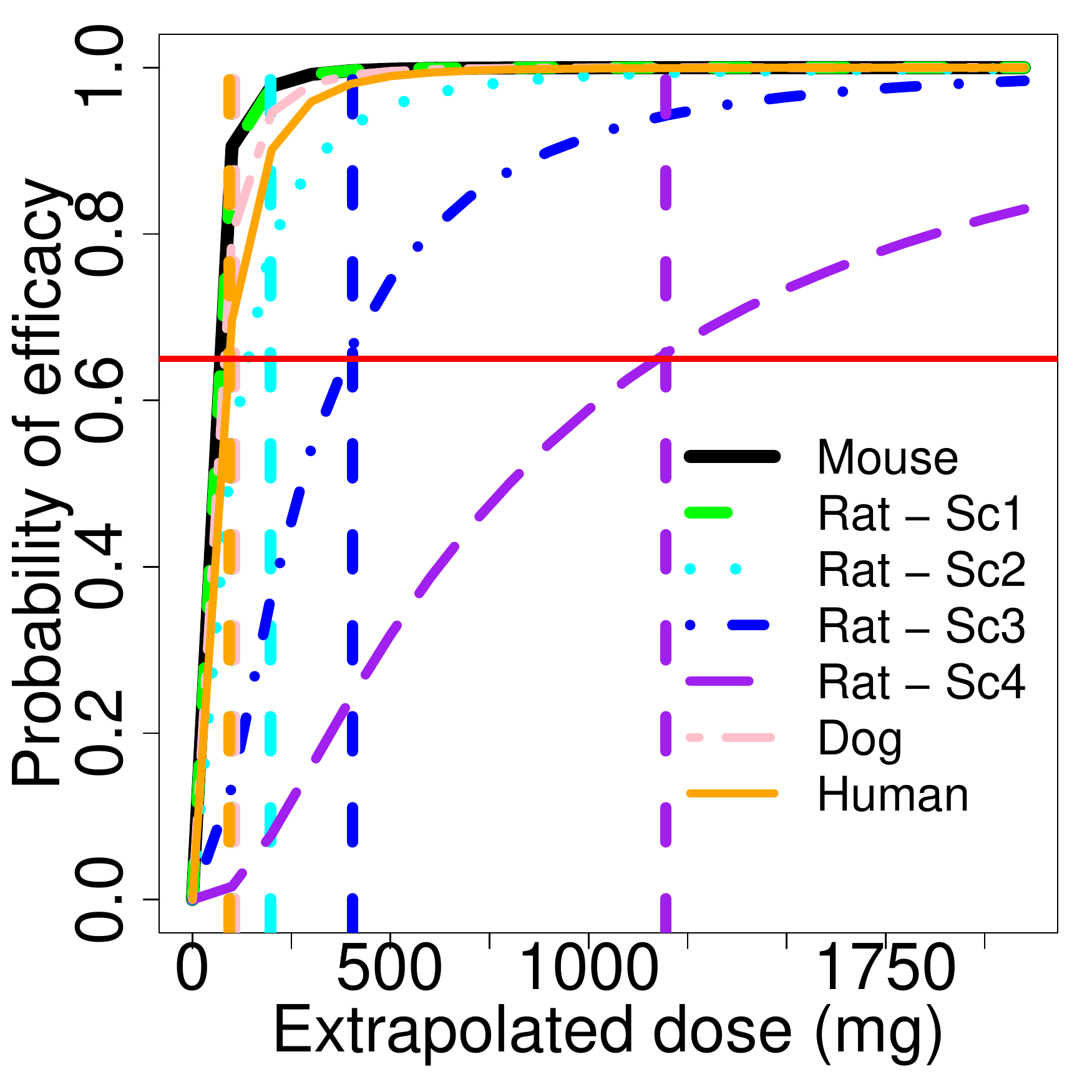}\label{fig:eff_prob_by_extrapolated_dose}}}
\caption{Mean relative inhibition of the drug versus time for several doses for human in scenario 1 (a); Probabilities of toxicity (b) and efficacy (c) according to the extrapolated dose for all scenarios and for all species.
Red horizontal lines respectively depict the probability of toxicity $p_T$ threshold (b) and the probability of efficacy threshold $p_E$ (c). The probability of toxicity for a dose should be lower than the threshold  $p_T$ while the probability of efficacy for a dose should be greater than the threshold $p_E$.}
\label{fig:simu}
\end{figure}

Efficacy data for human in scenario 1 (baseline scenario) are first simulated from models \ref{eq:efficacy_model} and \ref{eq:concentration_efficacy_model}; the parameters were chosen so that the simulated data follow similar distributions to those produced in \cite*{lestini_2015} (see Figures \ref{fig:simu_perc_inhib_time_human} and \ref{fig:eff_prob_by_extrapolated_dose}). Parameters of the PD model for dog, rat and mouse's scenario 1 are chosen equal to those of human (see Table \ref{tab:simu_parameters_and_extrapolated_therapeutic_window}). For scenario 1, using the extrapolated parameters of PK and PD models from dog, rat and mouse to human allows to predict accurately the MED to be approximately equal to 89 mg.

For scenarios 2, 3 and 4, we modified the accuracy of the extrapolation of the toxicity and efficacy models from rat to humans, to assess how our method could detect and correct for the discrepancies (see Table \ref{tab:simu_parameters_and_extrapolated_therapeutic_window}). 
Scenario 2 is based on an inaccurate extrapolation of the parameter clearance $CL$, affecting both toxicity (via AUC) and efficacy (via the concentration in the effect compartment entering the $I_{\max}$ model). For scenario 3, it is the extrapolation of the potency $IC_{50}$ that is imprecise. Scenario 4 combines the modified parameters values of scenarios 2 and 3. 
Figures \ref{fig:tox_prob_by_extrapolated_dose} and \ref{fig:eff_prob_by_extrapolated_dose} respectively give the probability of toxicity and efficacy according to the extrapolated dose for all scenarios and for all species.

\begin{table*}
\caption{Simulation parameters and approximate extrapolated therapeutic window for all species and all scenarios. $W$: Weights;  $k_a$: Absorption rate constant for oral administration; $(\mu_{CL}, \mu_V, \mu_{IC_{50}})$: Exponentials of the respective means of the logarithms of the distributions of the clearance elimination, the volume of distribution and the concentration at which 50\% of $I_{\max}$ is reached;  $(\omega_{CL}, \omega_V, \omega_{IC_{50}})$: Respective standard deviations of the logarithms of the distributions of the clearance elimination, the volume of distribution and the concentration at which 50\% of $I_{\max}$ is reached; $(\sigma_C, \sigma_I)$: Respective standard deviations of the logarithms of the distributions of the measured concentration of the drug in blood plasma and of the measured biomarker pSMAD inhibition; $\tau_T$: AUC threshold; $p_T$: Probability of toxicity threshold; $I_{\max}$: Maximum inhibition effect;  $\tau_E$: Maximum efficacy response threshold; $p_E$: Probability of efficacy threshold; MED: Minimum effective dose; MTD: Maximum tolerated dose.}\label{tab:simu_parameters_and_extrapolated_therapeutic_window}
\begin{center}
\begin{tabular}{l|r|r|rrrr|r}
\hline \hline
\textbf{Parameters}&\textbf{Human}&\textbf{Dog} &\multicolumn{4}{c}{\textbf{Rat}} &\textbf{Mouse} \\  
&&&\textbf{Sc.1}&\textbf{Sc.2}&\textbf{Sc.3}&\textbf{Sc.4}& \\
\hline  \hline
$W$ (kg) &70&10&0.15&0.15&0.15&0.15&0.025 \\
$k_a$ (h$^{-1}$) &2&2&2&2&2&2&2 \\
$\mu_{CL}$ (L.h$^{-1}$)&40&9.3&0.40&\textbf{1.59}&0.40&\textbf{1.59}&0.11 \\
$\mu_V$ (L)&100&14&0.21&0.21&0.21&0.21&0.04 \\
$\omega_{CL}$ &0.7&0.7&0.7&0.7&0.7&0.7&0.7 \\ 
$\omega_{V}$ &0.7&0.7&0.7&0.7&0.7&0.7&0.7 \\ 
$\sigma_C$ & 0.2 & 0.2& 0.2& 0.2& 0.2& 0.2& 0.2 \\ 
$\tau_T$ (mg.L$^{-1}$.h) &22.6&22.6&22.6&22.6&22.6&22.6&22.6 \\ 
$p_T$ (\%) &20&20&20&20&20&20&20 \\
$\mu_{\alpha}$ &1&1&1&1&1&1&1 \\
$\omega_{\alpha}$ &0&0&0&0&0&0&0 \\
\hline
$I_{\max}$ &1&1&1&1&1&1&1 \\
$\mu_{IC_{50}}$ (mg.L$^{-1}$)&0.32&0.32&0.32&0.32&\textbf{2.9}&\textbf{2.9}&0.32 \\
$\mu_{k_e}$ (h$^{-1}$) &1.6&1.6&1.6&1.6&1.6&1.6&1.6 \\
$\omega_{IC_{50}}$ &0.7&0.7&0.7&0.7&0.7&0.7&0.7 \\
$\omega_{k_e}$ &0.7&0.7&0.7&0.7&0.7&0.7&0.7 \\
$\sigma_I$ & 0.2 & 0.2& 0.2& 0.2& 0.2& 0.2& 0.2 \\ 
$\tau_E$ (\%) & 50& 50& 50& 50& 50& 50& 50 \\
$p_E$ (\%) & 65 & 65& 65& 65& 65& 65& 65 \\
\hline  
\textbf{Extrapolated MED (mg)} & 89 & 67 & 43 & 129 & 380 & 1172 & 41 \\
\textbf{Extrapolated MTD (mg)} & 502 & 502 & 504 & 2002 & 504 & 2002 & 531 \\
\hline
\end{tabular}
\end{center}
\end{table*}

\subsection{\textit{Prior} Distributions and Complement to Implementation} \label{subsec:prior_parameter_settings}

Table \ref{tab:parameters_prior_distribution} shows the non-informative \textit{prior} distributions used for all PK/PD parameters and for the variances of intra-individual random effects $\sigma_c$ and $\sigma_I$ in the Bayesian framework. These non-informative \textit{prior} distributions are first used to fit the Bayesian models in step 1 for mouse data. 
Then, the estimated \textit{posterior} means for mouse of $\mu_{CL}$ and $\mu_V$ are extrapolated to rat using the extrapolation formulas. Next, these extrapolated estimates and the estimated \textit{posterior} means for mouse of $k_a$, $\mu_{IC_{50}}$ and $\mu_{k_e}$ are used as means of the \textit{prior} distributions of the model parameters for rat while the other \textit{prior} distributions remain the same as for the mouse. The same procedure is repeated when moving from rat to dog.

\begin{table}
\caption{Non-informative \textit{prior} distributions for each parameter. LN: Log-normal distribution; $k_a$: Absorption rate constant for oral administration; $(\mu_{CL}, \mu_V, \mu_{IC_{50}})$: Exponentials of the respective means of the logarithms of the distributions of the clearance elimination, the volume of distribution and the concentration at which 50\% of $I_{\max}$ is reached;  $(\omega_{CL}, \omega_V, \omega_{IC_{50}})$: Respective standard deviations of the logarithms of the distributions of the clearance elimination, the volume of distribution and the concentration at which 50\% of $I_{\max}$ is reached; $(\sigma_C, \sigma_I)$: Respective standard deviations of the logarithms of the distributions of the measured concentration of the drug in blood plasma and of the measured biomarker pSMAD inhibition.}
\label{tab:parameters_prior_distribution}
\begin{center}
\begin{tabular}{l|r}
\hline \hline
\textbf{Parameters}&\textbf{Non-informative \textit{prior} distribution} \\ \hline
$k_a$ (h$^{-1}$) & LN(-1, 2.5) \\
$\mu_{CL}$ (L.h$^{-1}$)& LN(-1, 2.5) \\ 
$\mu_V$ (L)& LN(-1, 2.5)\\
$\omega_{CL}$ & Half-Student-t(3, 0, 2.8)  \\ 
$\omega_{V}$ & Half-Student-t(3, 0, 2.8)  \\  
\hline
$\mu_{IC_{50}}$ (mg.L$^{-1}$)& LN(0, 2.5) \\
$\mu_{k_e}$ (h$^{-1}$) & LN(0, 2.5) \\
$\omega_{IC_{50}}$ & Half-Student-t(3, 0, 2.8) \\ 
$\omega_{k_e}$ & Half-Student-t(3, 0, 2.8) \\ 
\hline
$\sigma_C$ &  Half-Student-t(3, 0, 2.8)  \\ 
$\sigma_I$ &  Half-Student-t(3, 0, 2.8)  \\ \hline 
\end{tabular}
\end{center}
\end{table}

We assume that the equations \ref{eq:efficacy_model} 
and \ref{eq:concentration_efficacy_model} describing efficacy data generation are also known as well as $I_{\max}$, $\tau_E$ and $p_E$. The other parameters are estimated.
At step 1 of our methodology, for each simulated dataset and for each animal species, we fit the mixed-effects models using the stochastic approximation expectation-maximization (SAEM) algorithm~\citep*{delyon_1999} for the hybrid approach. For the dog and the rat, random effects (intraindividual variability - IIV) are also estimated on $IC_{50}$ and $k_e$, therefore $\mathbf{y}_2 = \mathbf{y}_3 = (\tilde{c}, \tilde{i})$, the concentration and the percentage of inhibition measured for each subject, and $\mathbf{f}_k(.) = \{ C(t, \boldsymbol{\theta}_k = \{ k_{a,k}, \mu_{CL,k}, \mu_{V,k}, \omega_{CL,k}, \omega_{V,k}\}) , I(t, \boldsymbol{\theta}_k = \{ k_{a,k}, \mu_{CL,k}, \mu_{V,k}, \omega_{CL,k}, \omega_{V,k}, \mu_{IC_{50}, k}, \mu_{k_e, k},$ $ \omega_{IC_{50}, k}, \omega_{k_e, k}\})\}$ for $k=2,3$, while for the mouse, random effects are only estimated on $CL$ (that is $\mathbf{f}_1(.) = \{C(t, \boldsymbol{\theta}_1 = \{ k_{a,1}, \mu_{CL,1}, \mu_{V,1}, \omega_{CL,1}, \}), I(t, \boldsymbol{\theta}_1 = \{ k_{a,1}, \mu_{CL,1}, \mu_{V,1}, \omega_{CL,1}, \mu_{IC_{50}, 1},$ $ \mu_{k_e, 1}\}) \}$ and $\mathbf{y}_1 = (\tilde{c}, \tilde{i})$).

For the next steps of the analysis based on mouse data, we then make the assumption that $\omega_{V} = \omega_{IC_{50}} = \omega_{k_e} = 0.7$ for mouse. As sensitivity analysis, we also consider the assumptions that $\omega_{V} = \omega_{IC_{50}} = \omega_{k_e} = 0.4$ or $\omega_{V} = \omega_{IC_{50}} = \omega_{k_e} = 
1$ for mouse. 

For the hybrid approach, we run 5 chains in parallel and we keep the average of the estimates. Then, we simulate $L = 1000$ different values for each parameter of the models from log-normal distributions using the estimates of the hyperparameters, that are used for steps 2 to 4 of the methodology.

In step 3, to compute the accuracy, for each scenario, the animal species that should be kept for the final step (because they are consistent) have to be defined. 
The extrapolated MTDs to humans from Table \ref{tab:simu_parameters_and_extrapolated_therapeutic_window}
are approximately equal to 502 mg (the true MTD value for humans) for all animal species (mouse, rat and dog) in scenarios 1 and 3, and only for mouse and dog in scenarios 2 and 4. The values of extrapolated MTDs for rat in scenario 2 and 4 are greater and near to 2000. So, for the MTD, the binary variable "true response" is set to 1 for all comparisons between animal species in scenarios 1 and 3, and only for the comparison between mouse and dog in scenarios 2 and 4; Otherwise, for all comparisons including rat in scenarios 2 and 4, it is set to 0. Note that, in the case of MTD, all scenarios are considered as relevant for choosing the Hellinger distance threshold.
On the other hand, the extrapolated MEDs to humans are approximately equal to 89 mg (the true MED value for humans) for all animal species in scenario 1, and only for mouse and dog in scenarios 3 and 4. So, for the MED, the variable ``true response" is set to 1 for all animal species comparisons in scenario 1 and only for mouse and dog in scenarios 3 and 4. Scenario 2 is not considered to be a relevant scenario to compute the accuracy for the MED since a clear decision cannot be taken \textit{a priori} (see \ref{sec:add_discussion}). For all other comparisons, the accuracy response is set to 0.

All analyses are performed using R software version 4.04 with Monolix software \citep{monolix_2019} and 
Stan \citep{stan_2021} package \textit{rstan} version 2.21.2. In \textit{rstan}, 3 chains, a burn-in of 3000 and 6000 other iterations are used and, as a convergence criterion, \cite{gelman_1992}'s potential scale reduction factor Rhat = 1.

\section{Web Appendix C: Additional Simulation Results}\label{sec:additional_simulation_results}

\subsection{Complement to Illustration}\label{subsec:compl_illustration}

\begin{table*}
\small
\caption{Mean estimates of the model parameters for one simulated dataset in scenarios 1 and 2 for the Bayesian approach (under the assumption that $\omega_{V} = 0.7$ for mouse). Standard deviations are shown in parentheses. \textbf{T}: True values; \textbf{B}: Bayesian results; $k_a$: Absorption rate constant for oral administration; $(\mu_{CL}, \mu_V)$: Exponential of the respective means of the logarithms of the distributions of the clearance elimination and the volume of distribution;  $(\omega_{CL}, \omega_V)$: Respective standard deviations of the logarithms of the distributions of the clearance elimination and the volume of distribution; $\sigma_C$: Standard deviations of the logarithms of the distributions of the measured concentration of the drug in blood plasma. \label{tab:parameters_estimates_Bayesian_tr4_for_sc1_and_sc2}}
\begin{center}
\begin{tabular}{lrrrrrr}
\hline \hline
\textbf{Parameters}&  & \textbf{Mouse} & \multicolumn{2}{c}{\textbf{Rat}} & \multicolumn{2}{c}{\textbf{Dog}}   \\  
& & & \textbf{Sc1} & \textbf{Sc2} & \textbf{Sc1} & \textbf{Sc2} \\
\hline 
$k_a$ (h$^{-1}$) & \textbf{T} &\multicolumn{5}{c}{2} \\
& \textbf{B} & 2.04 (0.0268) & 2.00 (0.00669) & 4.23 (1.61) & 1.98 (0.0176) & 1.98 (0.0174) \\ \hline
$\mu_{CL}$  & \textbf{T} &0.11 &0.40&1.59&\multicolumn{2}{c}{9.3} \\
(L.h$^{-1}$) & \textbf{B} & 0.0961 (0.00984) & 0.464 (0.0567) &1.45 (0.170) & 10.2 (1.48) & 10.2 (1.51) \\ \hline
$\mu_V$ (L) & \textbf{T}&0.04&\multicolumn{2}{@{}c|@{}}{0.21}&\multicolumn{2}{c}{14} \\
& \textbf{B} & 0.0508 (0.00552) & 0.216 (0.0250) & 0.581 (0.0261) & 14.9 (1.55) & 14.9 (1.51) \\ \hline
$\omega_{CL}$ & \textbf{T} &\multicolumn{5}{c}{0.7} \\ 
& \textbf{B} & 0.837 (0.0837)& 0.748 (0.0888) & 0.719 ( 0.0907) & 0.798 (0.114) & 0.794 (0.114) \\ \hline
$\omega_{V}$ & \textbf{T} &\multicolumn{5}{c}{0.7} \\ 
& \textbf{B} & - & 0.694 (0.0904) & 0.690 (0.0973) & 0.545 (0.0829) & 0.542 (0.0828) \\ \hline
$\sigma_{C}$ & \textbf{T} &\multicolumn{5}{c}{0.2} \\ 
& \textbf{B} & - & 0.192 (0.0123) & 0.304 (0.0761) & 0.222 (0.0166) & 0.222 (0.0169) \\ \hline
\end{tabular}
\end{center}
\end{table*}

\subsection{Parameters Estimation}\label{subsec:parameters_estimation}

\begin{landscape}
\begin{table*}
\caption{Mean estimates of the model parameters for all scenarios for the Bayesian approach over the 500 replications (under the assumption that $\omega_{V} = \omega_{IC_{50}} = \omega_{k_e} = 0.7$ for mouse). Standard deviations are shown in parentheses. \textbf{T}: True values; \textbf{B}: Bayesian results.\label{tab:parameters_estimates_Bayesian}}
\begin{center}
\footnotesize
\begin{tabular}{lrrrrrrrrrr}
\hline \hline
\textbf{Parameters}&  & \textbf{Mouse} & \multicolumn{4}{@{}c|@{}}{\textbf{Rat}} & \multicolumn{4}{c}{\textbf{Dog}}   \\  
& & & \textbf{Sc1} & \textbf{Sc2} & \textbf{Sc3} & \textbf{Sc4} & \textbf{Sc1} & \textbf{Sc2} & \textbf{Sc3} & \textbf{Sc4} \\
\hline  \hline
$k_a$ (h$^{-1}$) & \textbf{T} &\multicolumn{9}{c}{2} \\
& \textbf{B} &2.05 (0.673)&2.02 (0.0851)&2.40 (0.767)&2.01 (0.0812)&2.43(0.773)&2.00 (0.0215)&2.00 (0.0190)&2.00 (0.0226)&2.00 (0.0226) \\ \hline
$\mu_{CL}$  & \textbf{T} &0.11 &0.40&1.59&0.40&1.59&\multicolumn{4}{c}{9.3} \\
(L.h$^{-1}$) & \textbf{B} &0.128 (0.0479)&0.406 (0.0464)&1.61 (0.177)&0.404 (0.0425)&1.60 (0.170)&9.34 (1.28)&9.38 (1.28)&9.34 (1.28)&9.37 (1.28) \\ \hline
$\mu_V$ (L) & \textbf{T}&0.04&\multicolumn{4}{@{}c|@{}}{0.21}&\multicolumn{4}{c}{14} \\
& \textbf{B} &0.0598 (0.0499)&0.214 (0.0261)&0.270 (0.115)&0.214 (0.0249)&0.273 (0.118)&14.1 (1.88)&14.1 (1.89)&14.1 (1.87)&14.1 (1.87) \\ \hline
$\omega_{CL}$ & \textbf{T} &\multicolumn{9}{c}{0.7} \\ 
& \textbf{B} &0.725 (0.0733) &0.715 (0.0836) &0.697 (0.113)&0.715 (0.0809)&0.704 (0.110)&0.721 (0.0942)&0.721 (0.0949)&0.721 (0.0944)&0.721 (0.0945) \\ \hline
$\omega_{V}$ & \textbf{T} &\multicolumn{9}{c}{0.7} \\ 
& \textbf{B} &-&0.721 (0.0880)&0.711 (0.148)&0.720 (0.0924)&0.715 (0.140)&0.725 (0.0993)&0.726 (0.100)&0.725 (0.0989)&0.725 (0.0990) \\ \hline
$\sigma_C$ & \textbf{T} &\multicolumn{9}{c}{0.2} \\ 
& \textbf{B} &-&0.203 (0.0154)&0.227 (0.0596)&0.202 (0.0152)&0.228 (0.0607)&0.204 (0.0381)&0.205 (0.0484)&0.205 (0.0472)&0.202 (0.0153)\\ \hline
$\mu_{IC_{50}}$ & \textbf{T} &0.32 & 0.32 & 0.32 & 2.9 & 2.9 & \multicolumn{4}{c}{0.32} \\
(mg.L$^{-1}$)& \textbf{B} &0.341 (0.0577)&0.325 (0.0360)&0.325 (0.0376)&2.95 (0.352)&2.89 (0.339)&0.322 (0.0427)&0.322 (0.0426)&0.324 (0.0428)&0.324 (0.0427) \\ \hline
$\mu_{k_e}$ (h$^{-1}$) & \textbf{T} &\multicolumn{9}{c}{1.6}\\
& \textbf{B} &1.54 (0.311)&1.64 (0.183)&1.63 (0.184)&1.63 (0.191)&1.62 (0.177)&1.62 (0.213)&1.62 (0.213)&1.62 (0.213)&1.62 (0.213) \\ \hline
$\omega_{IC_{50}}$ & \textbf{T} &\multicolumn{9}{c}{0.7} \\
& \textbf{B} &-&0.718 (0.0854)&0.722 (0.0859)&0.716 (0.0809)&0.722 (0.0876)&0.723 (0.102)&0.723 (0.102)&0.723 (0.102)&0.723 (0.102) \\ \hline
$\omega_{k_e}$  &\textbf{T} &\multicolumn{9}{c}{0.7} \\
& \textbf{B} &-&0.723 (0.0880)&0.729 (0.0873)&0.722 (0.0929)&0.723 (0.0899)&0.728 (0.102)&0.727 (0.102)&0.727 (0.101)&0.727 (0.101) \\ \hline
$\sigma_I$ & \textbf{T} &\multicolumn{9}{c}{0.2} \\ 
& \textbf{B} &-&0.203 (0.0154)&0.209 (0.0208)&0.202 (0.0141)&0.211 (0.0216)&0.202 (0.0163)&0.203 (0.0316)&0.202 (0.0163)&0.202 (0.0163) \\ \hline
\end{tabular}
\end{center}
\end{table*}
\end{landscape}

\begin{landscape}
\begin{table*}
\caption{Mean estimates of the model parameters for all scenarios for the hybrid approach over the 500 replications (under the assumption that $\omega_{V} = \omega_{IC_{50}} = \omega_{k_e} = 0.7$ for mouse). Standard deviations are shown in parentheses. \textbf{T}: True values; \textbf{H}: Hybrid results.\label{tab:parameters_estimates_hybrid}}
\begin{center}
\begin{tabular}{lrrrrrrr}
\hline \hline
\textbf{Parameters}&  & \textbf{Mouse} & \multicolumn{4}{c}{\textbf{Rat}} & \textbf{Dog}   \\  
& & & \textbf{Sc1} & \textbf{Sc2} & \textbf{Sc3} & \textbf{Sc4} &  \\
\hline  \hline
$k_a$ (h$^{-1}$) & \textbf{T} &\multicolumn{6}{c}{2} \\
& \textbf{H} &2.11 (0.201)& 2.01 (0.0508)& 2.01 (0.0359)& 2.01 (0.0457)& 2.01 (0.0293) & 2.01 (0.0988) \\ \hline
$\mu_{CL}$  & \textbf{T} &0.11 &0.40&1.59&0.40&1.59&9.3 \\
(L.h$^{-1}$) & \textbf{H} &0.101 (0.0120)& 0.402 (0.0461)& 1.59 (0.0179)& 0.400 (0.0425)& 1.59 (0.171)& 9.51 (1.29) \\ \hline
$\mu_V$ (L) & \textbf{T}&0.04&\multicolumn{4}{c}{0.21}&14 \\
& \textbf{H} & 0.0476 (0.00765)& 0.208 (0.0252)& 0.209 (0.0309)& 0.210 (0.0251)& 0.208 (0.0328)& 14.4 (1.93) \\ \hline
$\omega_{CL}$ & \textbf{T} &\multicolumn{6}{c}{0.7} \\ 
& \textbf{H} &0.715 (0.0966)& 0.674 (0.0816)& 0.690 (0.0804)& 0.672 (0.0788)& 0.697 (0.0763)& 0.654 (0.0953) \\ \hline
$\omega_{V}$ & \textbf{T} &\multicolumn{6}{c}{0.7} \\ 
& \textbf{H} &-&0.676 (0.0912)&0.708 (0.127)&0.674 (0.0933)&0.712 (0.120)&0.664 (0.102) \\ \hline
$\sigma_C$ & \textbf{T} &\multicolumn{6}{c}{0.2} \\ 
& \textbf{H} &-&0.164 (0.0402)&0.137 (0.0388)&0.163 (0.0415)&0.138 (0.0333)&0.243 (0.0618) \\ \hline
$\mu_{IC_{50}}$ & \textbf{T} &0.32 & 0.32 & 0.32 & 2.9 & 2.9 & 0.32 \\
(mg.L$^{-1}$)& \textbf{H} &0.292 (0.0409)&0.324 (0.0365)&0.325 (0.0387)&2.96 (0.360)&2.90 (0.345)&0.311 (0.0429) \\ \hline
$\mu_{k_e}$ (h$^{-1}$) & \textbf{T} &\multicolumn{6}{c}{1.6}\\
& \textbf{H} &1.99 (0.195)&1.60 (0.179)&1.60 (0.184)&1.60 (0.192))&1.60 (0.181)&1.61 (0.218) \\ \hline
$\omega_{IC_{50}}$ & \textbf{T} &\multicolumn{6}{c}{0.7} \\
& \textbf{H} &-&0.690 (0.0847)&0.696 (0.0825)& 0.687 (0.0806)& 0.693 (0.0837)& 0.698 (0.106) \\ \hline
$\omega_{k_e}$  &\textbf{T} &\multicolumn{6}{c}{0.7} \\
& \textbf{H} &-&0.689 (0.0857)&0.689 (0.0873)&0.689 (0.0901)&0.686 (0.0891)&0.688 (0.0970) \\ \hline
$\sigma_I$ & \textbf{T} &\multicolumn{6}{c}{0.2} \\ 
& \textbf{H} &-&0.210 (0.0184)&0.212 (0.0239)&0.210 (0.0182)&0.212 (0.0246)&0.203 (0.0185) \\
\hline
\end{tabular}
\end{center}
\end{table*}
\end{landscape}

\subsection{Hellinger Distance Results for the Hybrid Approach, MED Estimation and Additional Scenarios}\label{subsec:add_commensurability_checking_res}

\begin{figure}
\centerline{
\subfigure[]{\includegraphics[scale=0.25]{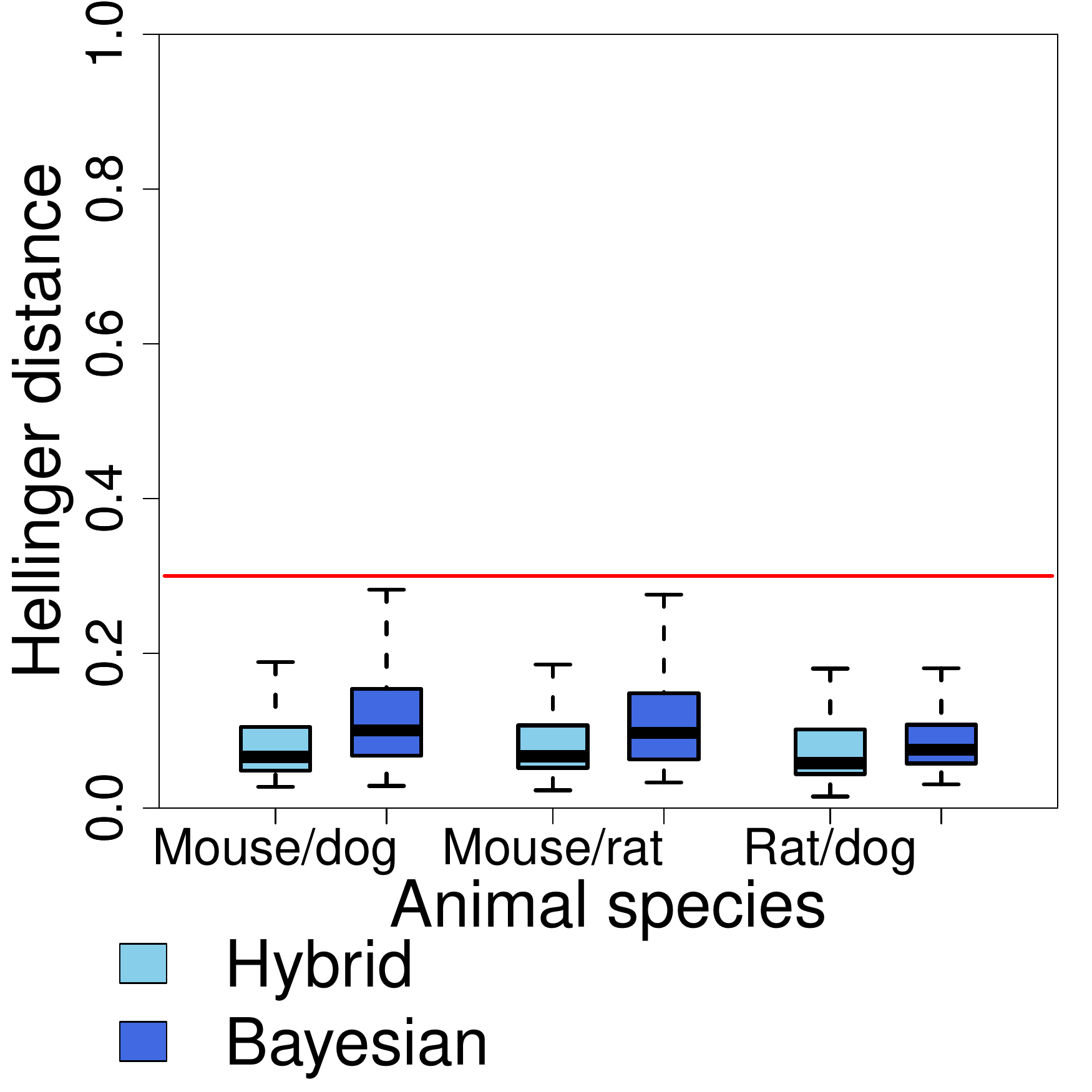}\label{fig:hellinger_dist_PD_eff_model_sc1_omega1}}
\subfigure[]{\includegraphics[scale=0.25]{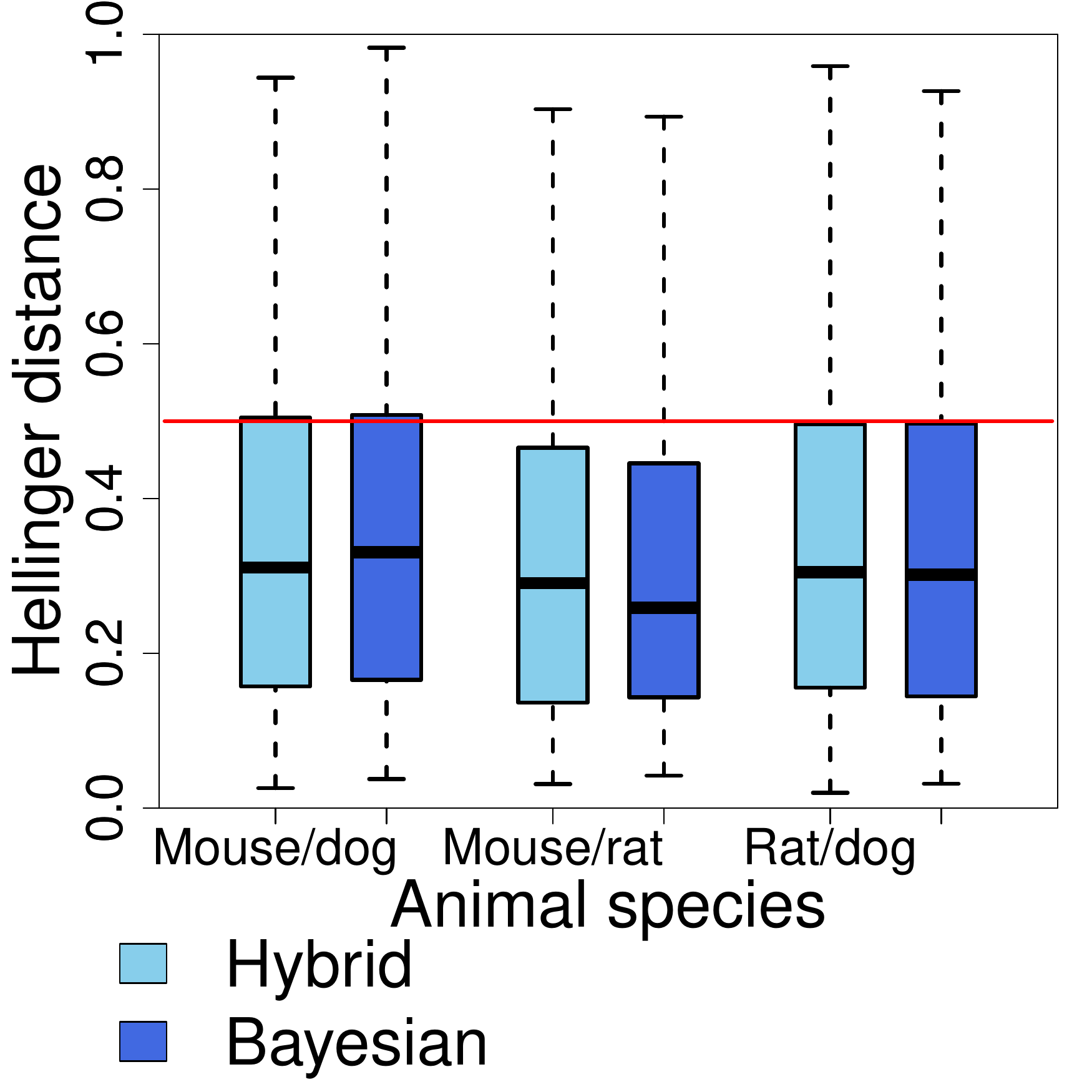}\label{fig:hellinger_dist_PK_model_sc1}}}
\centerline{
\subfigure[]{\includegraphics[scale=0.25]{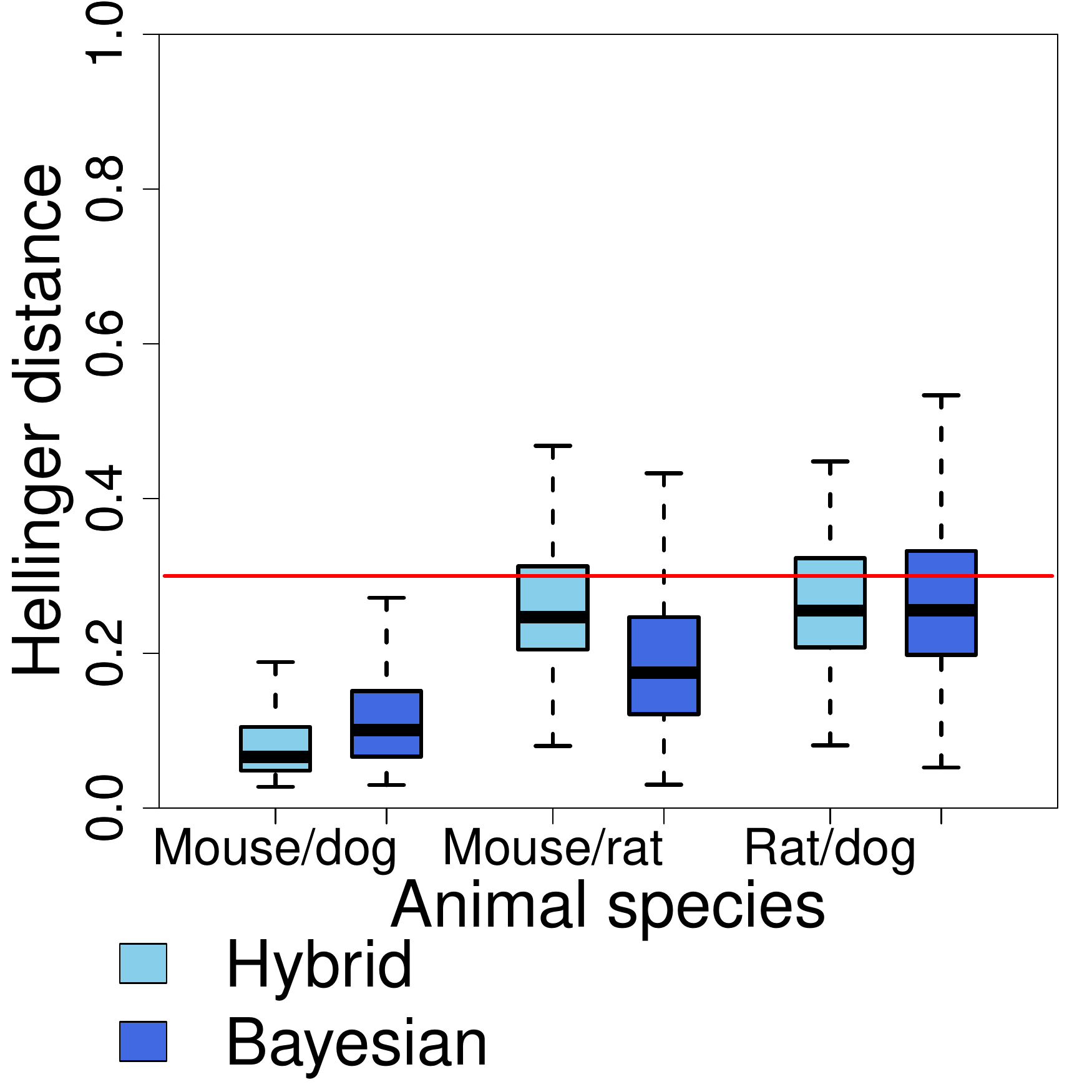}\label{fig:hellinger_dist_PD_eff_model_sc2_omega1}}
\subfigure[]{\includegraphics[scale=0.25]{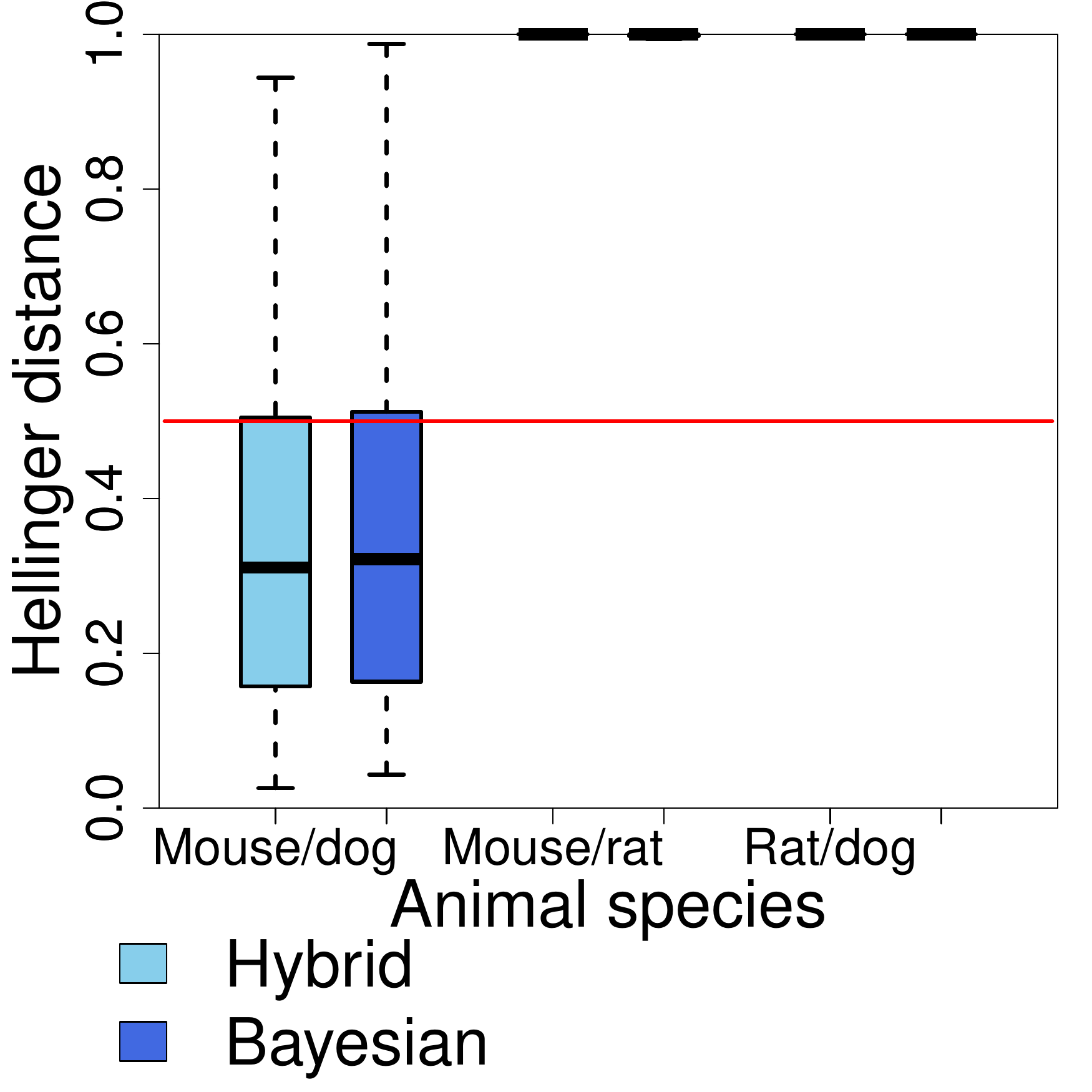}\label{fig:hellinger_dist_PK_model_sc2}}}
\centerline{
\subfigure[]{\includegraphics[scale=0.25]{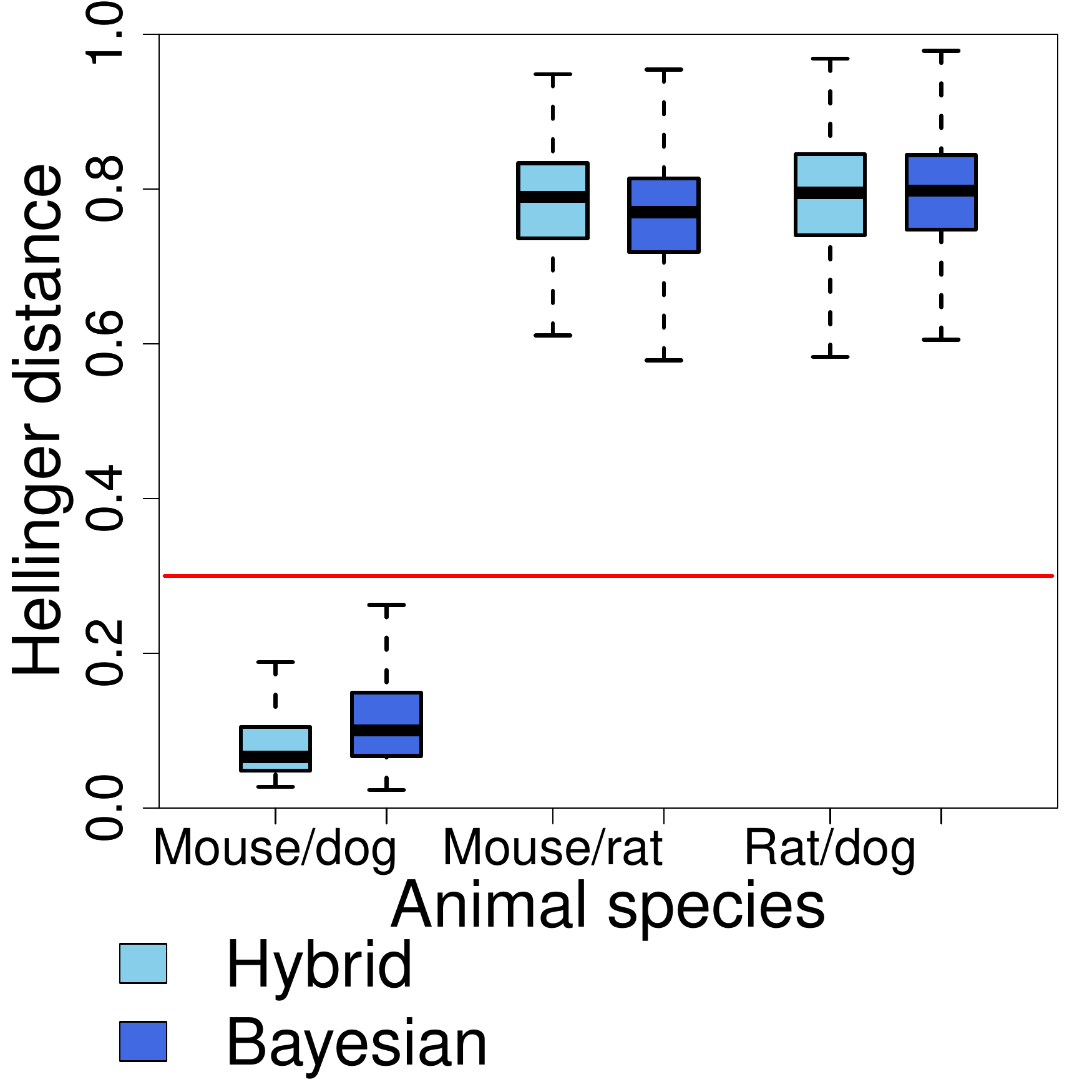}\label{fig:hellinger_dist_PD_eff_model_sc3_omega1}}
\subfigure[]{\includegraphics[scale=0.25]{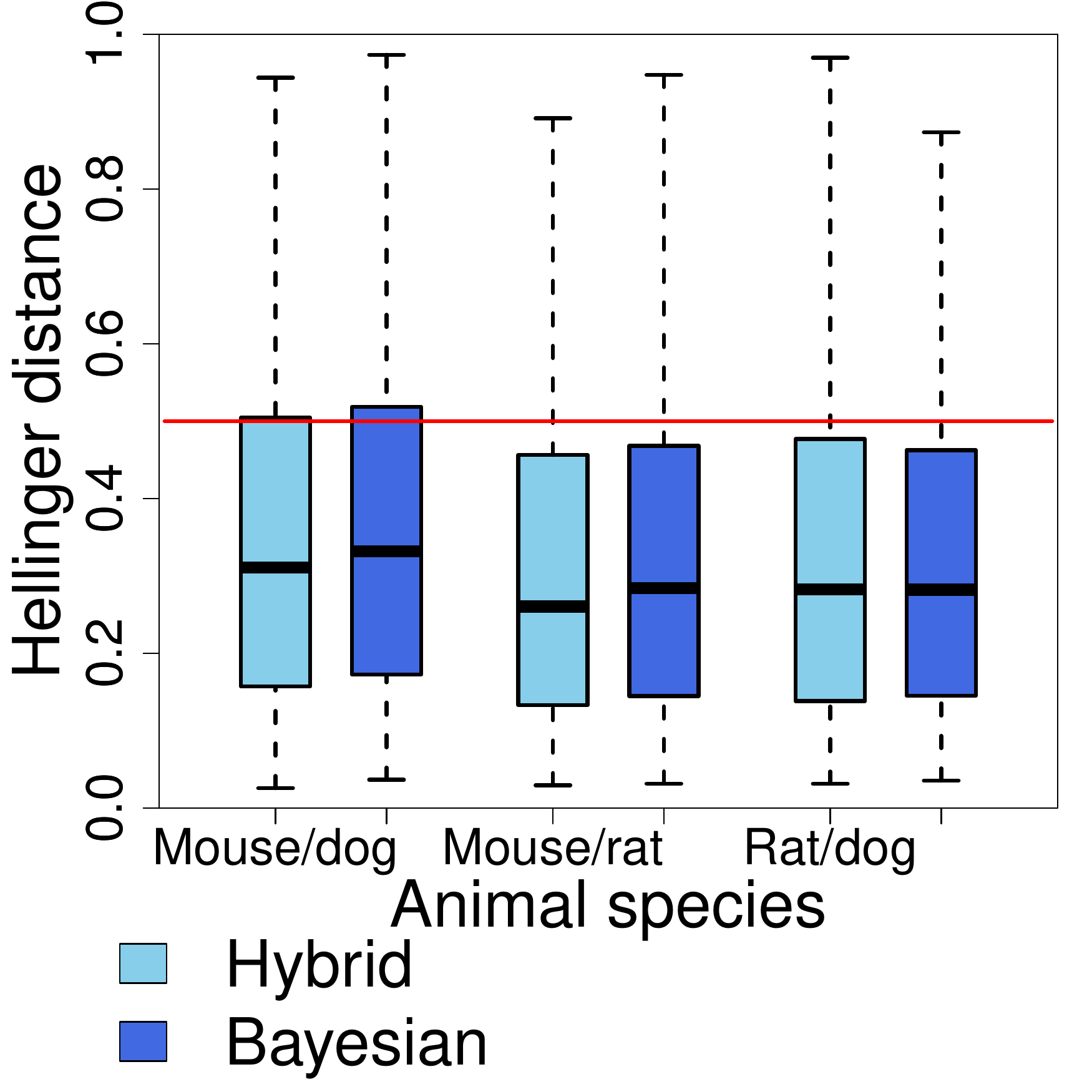}\label{fig:hellinger_dist_PK_model_sc3}}}
\centerline{
\subfigure[]{\includegraphics[scale=0.25]{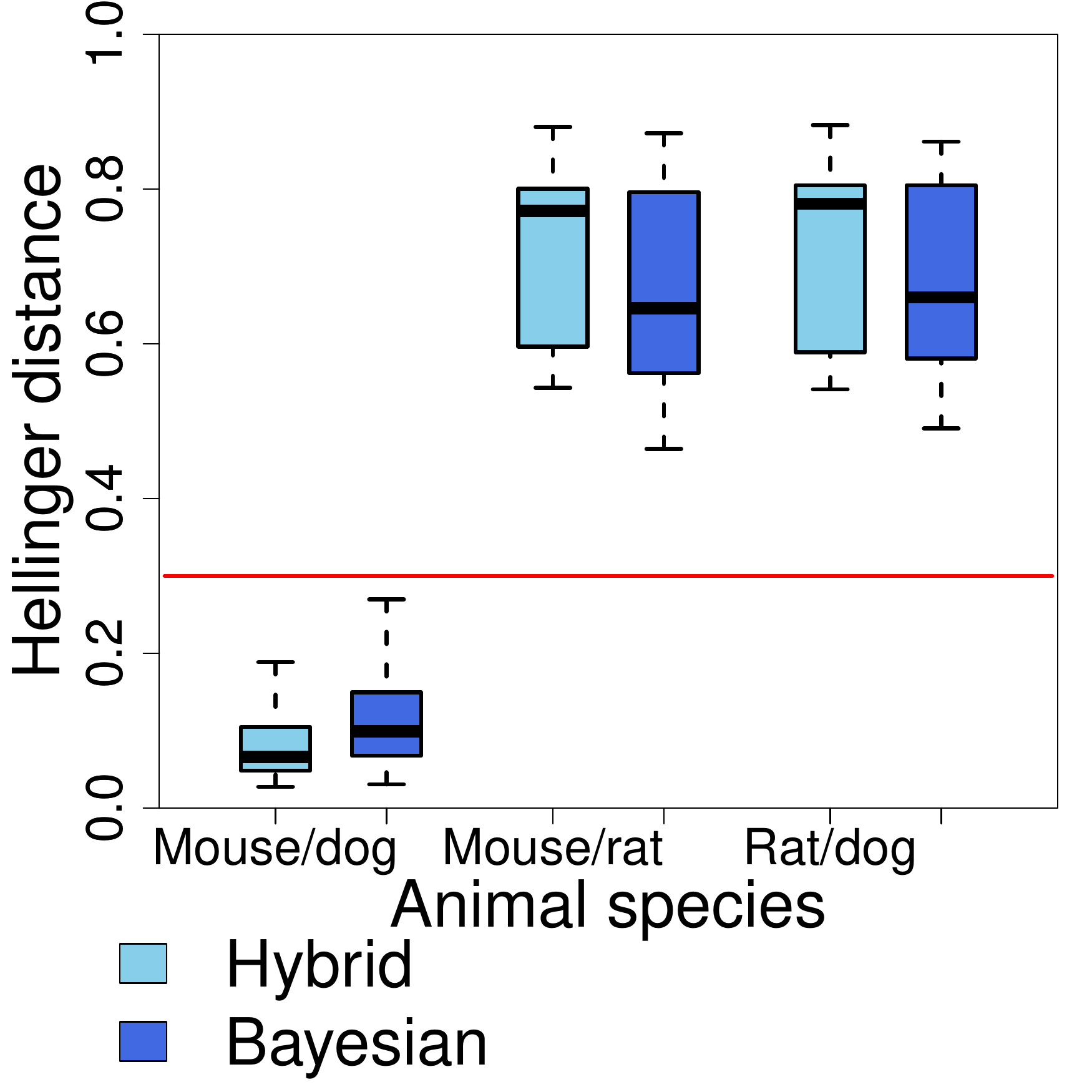}\label{fig:hellinger_dist_PD_eff_model_sc4_omega1}}
\subfigure[]{\includegraphics[scale=0.25]{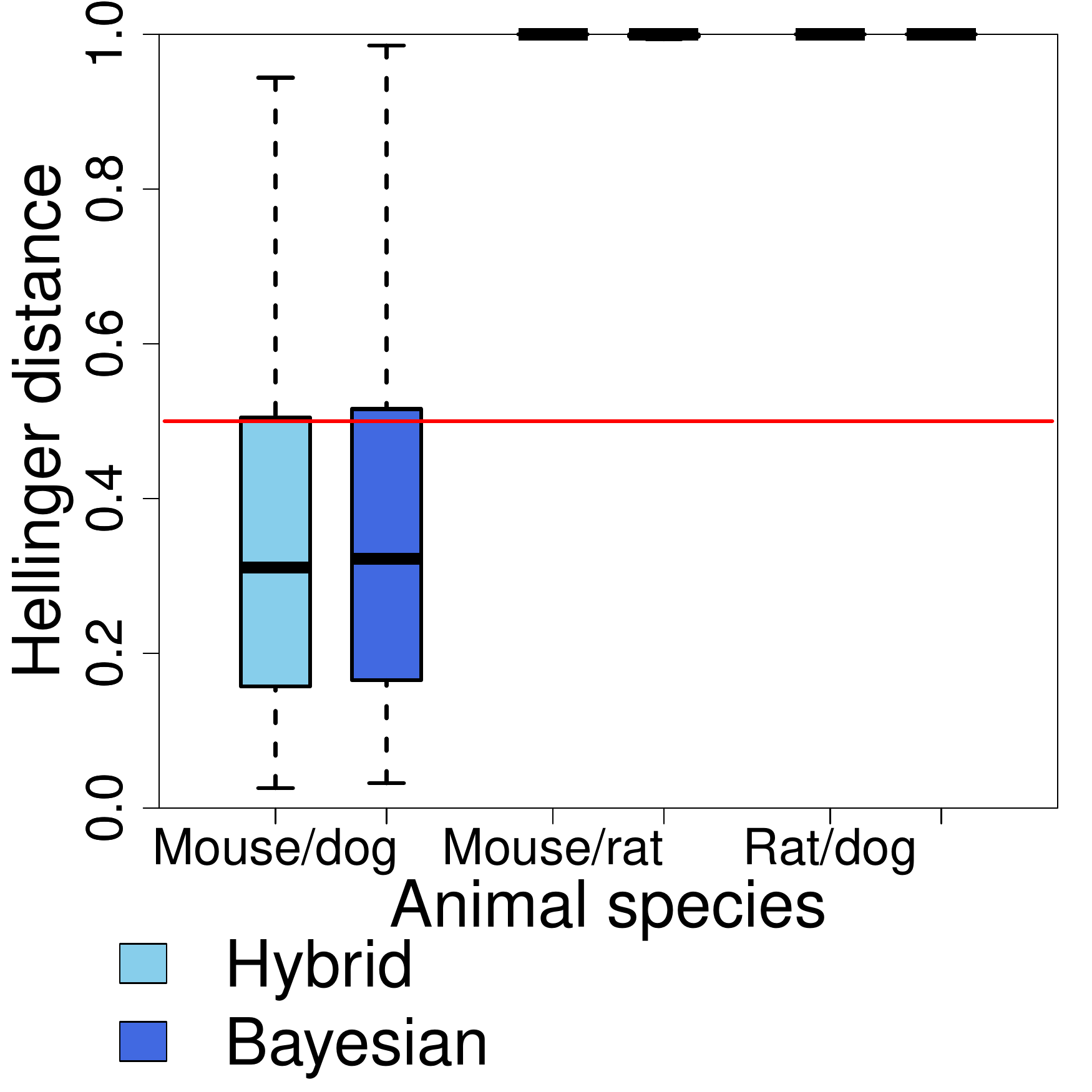}\label{fig:hellinger_dist_PK_model_sc4}}}
\caption{Hellinger distance of the transformed predicted MED (first column) and MTD (second column) distributions in humans between animal species for scenario 1 (a, b), scenario 2 (c, d), scenario 3 (e, f) and scenario 4 (g, h) for both approaches over 500 replications (under the assumption that $\omega_{V} = \omega_{IC_{50}} = \omega_{k_e} = 0.7$ for mouse). MED: Minimum effective dose ; MTD: Maximum tolerated dose.}
\label{fig:hellinger_dist_omega1}
\end{figure}

We then applied our methodology to the four scenarios with 500 replications each. Figure \ref{fig:hellinger_dist_omega1} shows boxplots of the Hellinger distances between two species for both the hybrid and the Bayesian approaches in the different scenarios, for the MED and the MTD. As seen in this figure, the Hellinger distance easily distinguishes animal species between those that show similarities in extrapolated toxicity and efficacy and those that show dissimilarities.

Indeed, for scenario 1 for which the extrapolation is correct from all animal species to human, the Hellinger distances between animal species for the MED distributions are relatively small (less than 0.3 in more than 97\% of cases). This indicates consistent results across all animal species and therefore the possibility of using them all to derive human MEDs. 
Hellinger distances for these distributions increase for scenario 2, for which the true values of MED differ slightly (see Table \ref{tab:simu_parameters_and_extrapolated_therapeutic_window}).
In scenarios 3 and 4, that are based on an inaccurate extrapolation of efficacy model parameters for rat, the Hellinger distances for the MED distributions are large for rat versus mouse and versus dog (more than 0.3 in 100\% of cases), but again small for mouse versus dog (less than 0.3 in more than 97\% of cases). This indicates MED distributions are consistent between mouse and dog, but not for rat.

In the case of MTD, Hellinger distances are larger (less than 0.5 in less than 83\% of cases). As a result, animal species will be less often selected whatever the threshold chosen. Nevertheless, for scenario 2 that uses an inaccurate extrapolation of toxicity model parameters for rat, the Hellinger distance between mouse and rat, and between rat and dog is close to 1. Similarly, scenario 4 combines inaccurate extrapolations for both toxicity and efficacy, and thus all Hellinger distances involving rat are close to 1. This means that the rat will be never selected for the final MTD calculation for these two scenarios.

\begin{figure}
\centerline{
\includegraphics[scale=0.25]{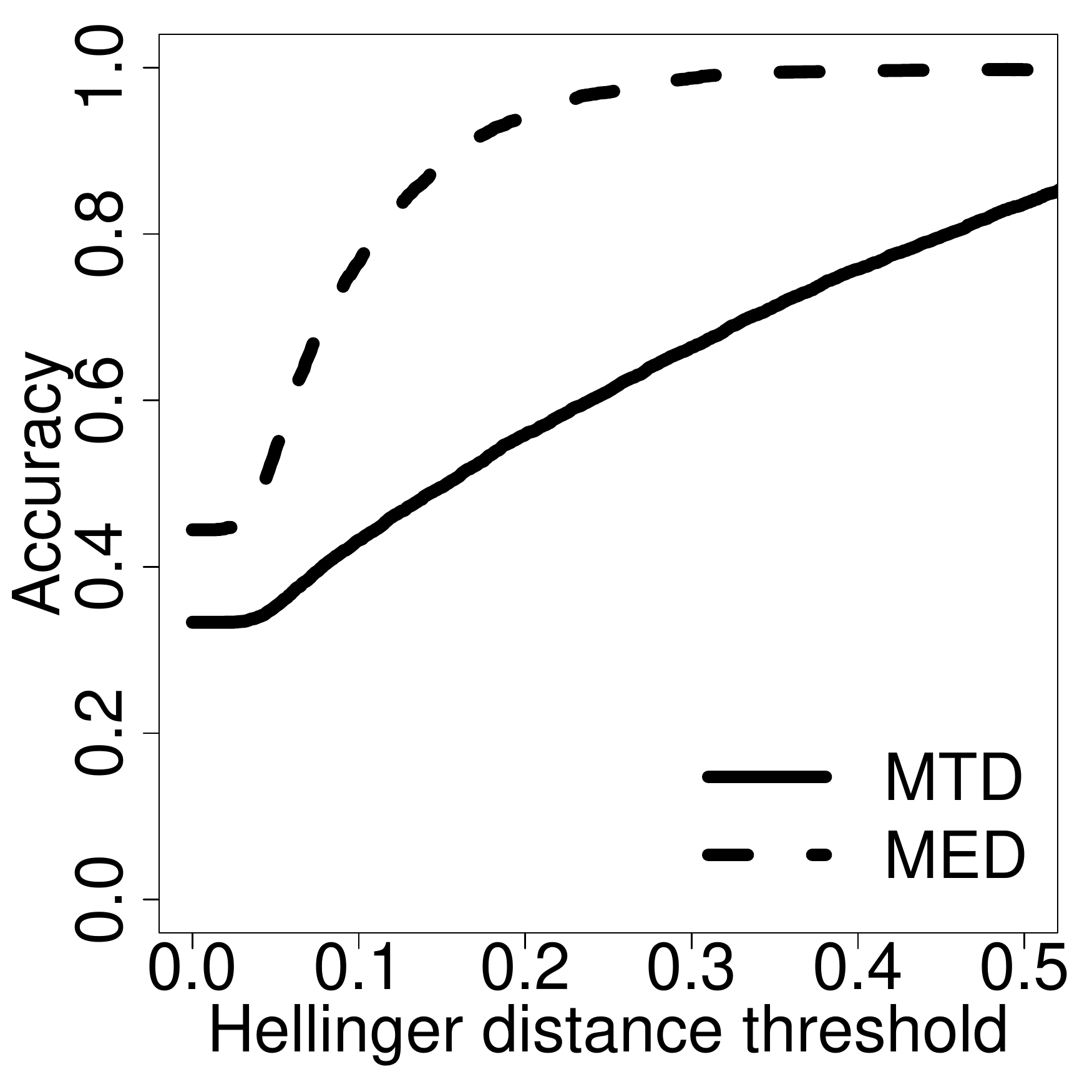}}
\caption{Accuracy by Hellinger distance threshold for MTD and MED for all scenarios for both methods over 500 replications. MED: Minimum effective dose ; MTD: Maximum tolerated dose.}
\label{fig:acc_by_thres} 
\end{figure}

As shown in Figure \ref{fig:acc_by_thres}, as Hellinger distance threshold increases, so does accuracy for both MTD and MED. Nevertheless, for MED, accuracy stabilizes at almost 1 from a threshold of 0.3, threshold we choose for the following step. For MTD, we set the threshold at 0.5.

\subsection{Final Estimations of MED and MTD in Humans for the Hybrid Approach and Additional Scenarios}

\begin{figure}
\centerline{
\subfigure[]{\includegraphics[scale=0.25]{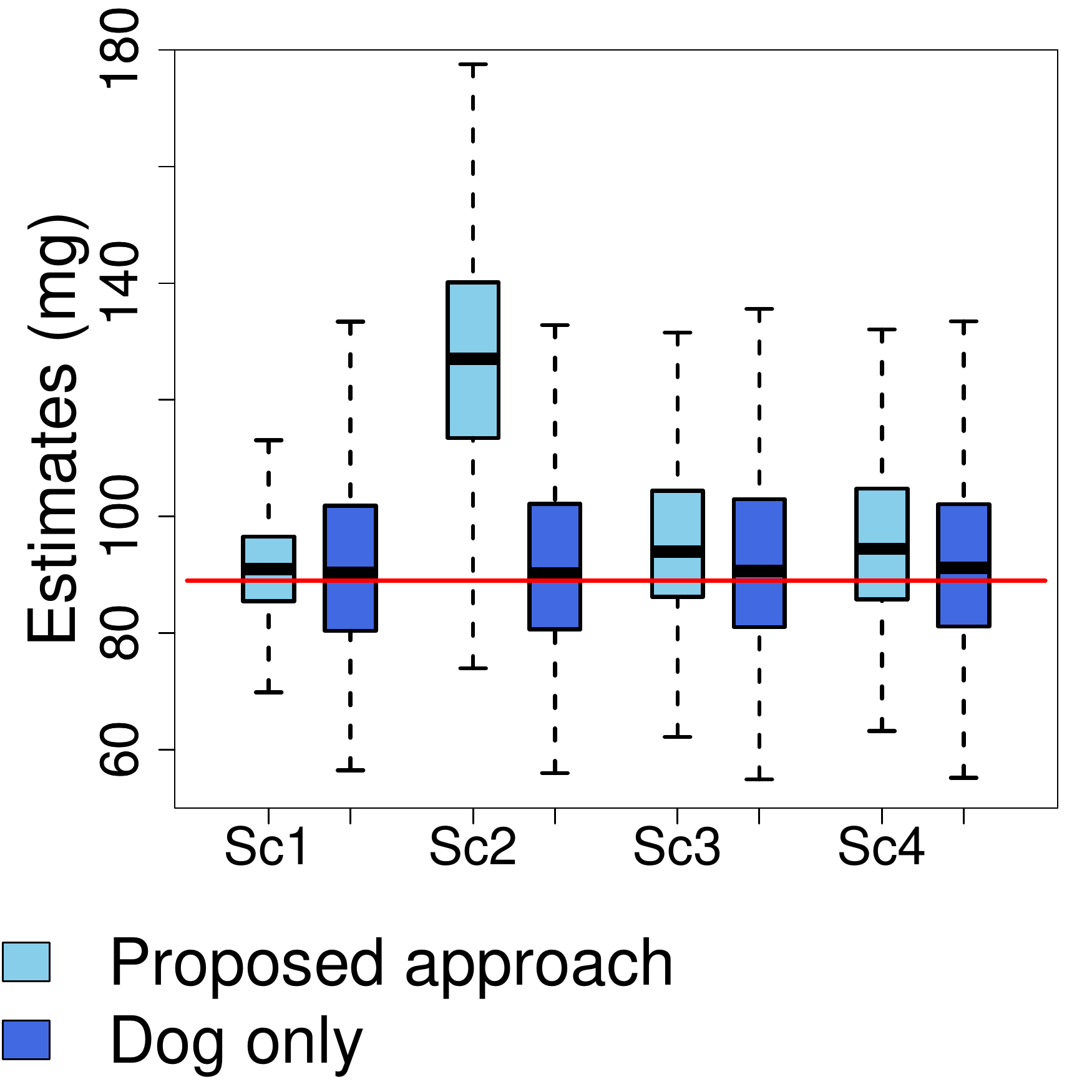}\label{fig:MED_posterior_mean_Bayesian_omega1}}
\subfigure[]{\includegraphics[scale=0.25]{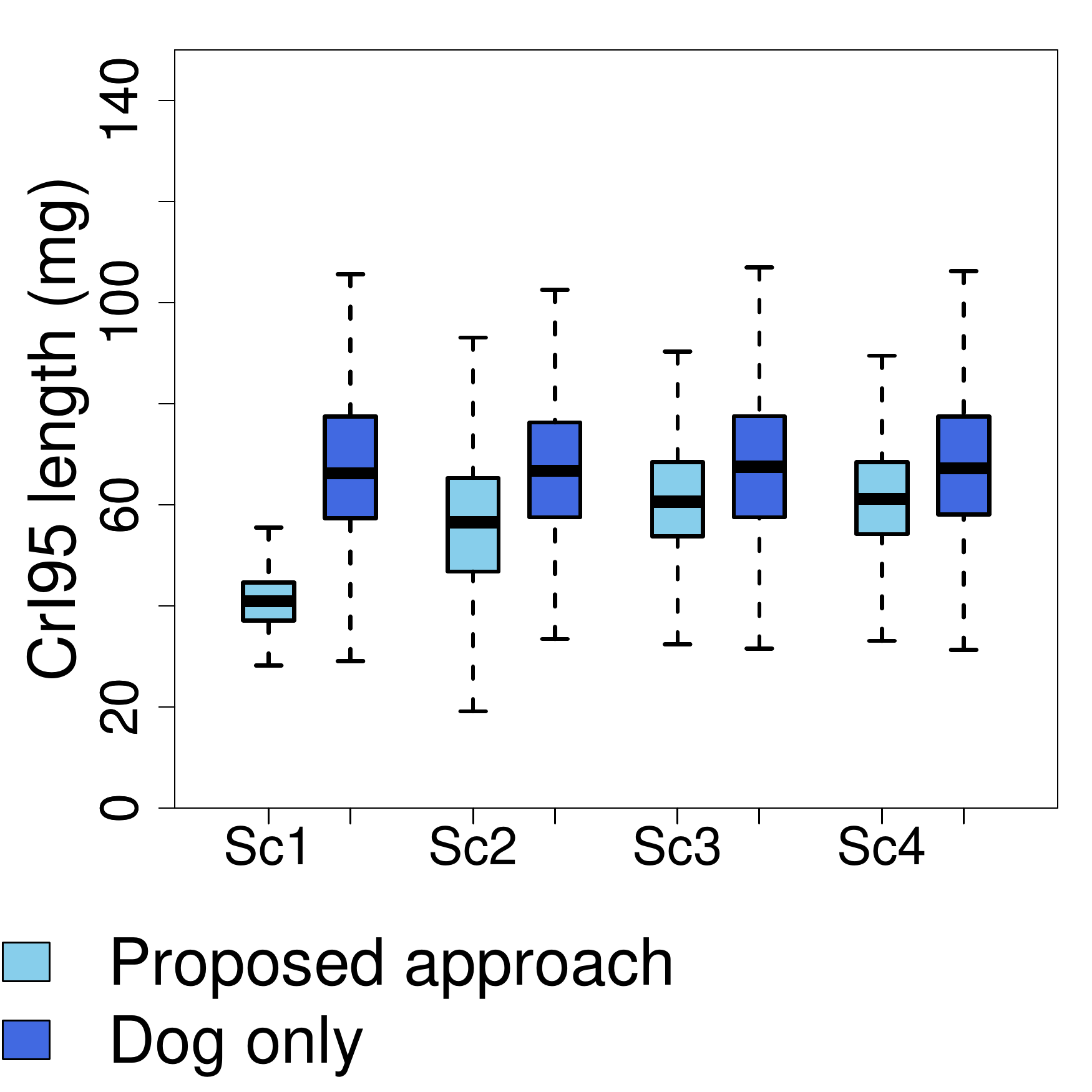}\label{fig:MED_posterior_IC95_length_Bayesian_omega1}}}
\centerline{
\subfigure[]{\includegraphics[scale=0.25]{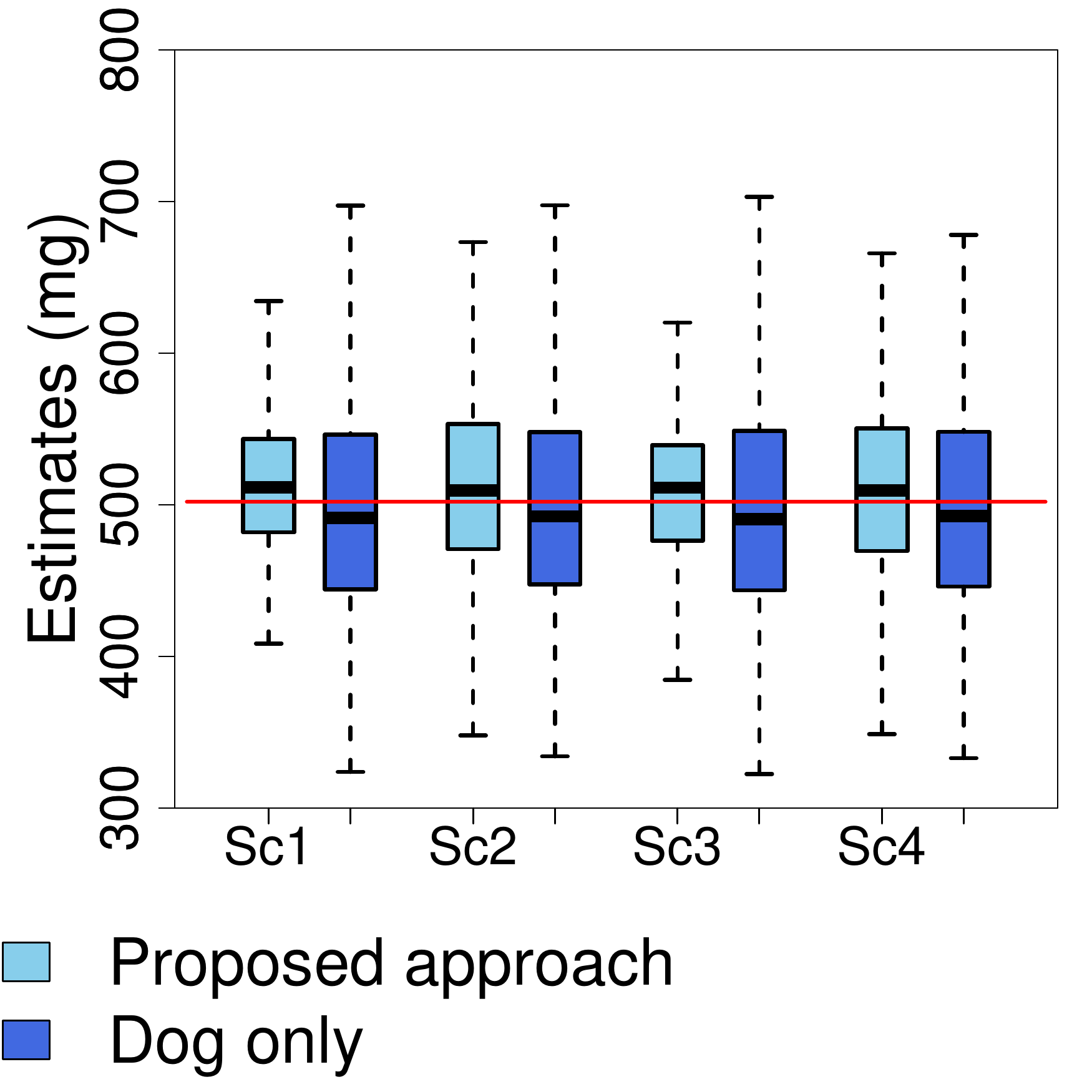}\label{fig:MTD_posterior_mean_Bayesian}} 
\subfigure[]{\includegraphics[scale=0.25]{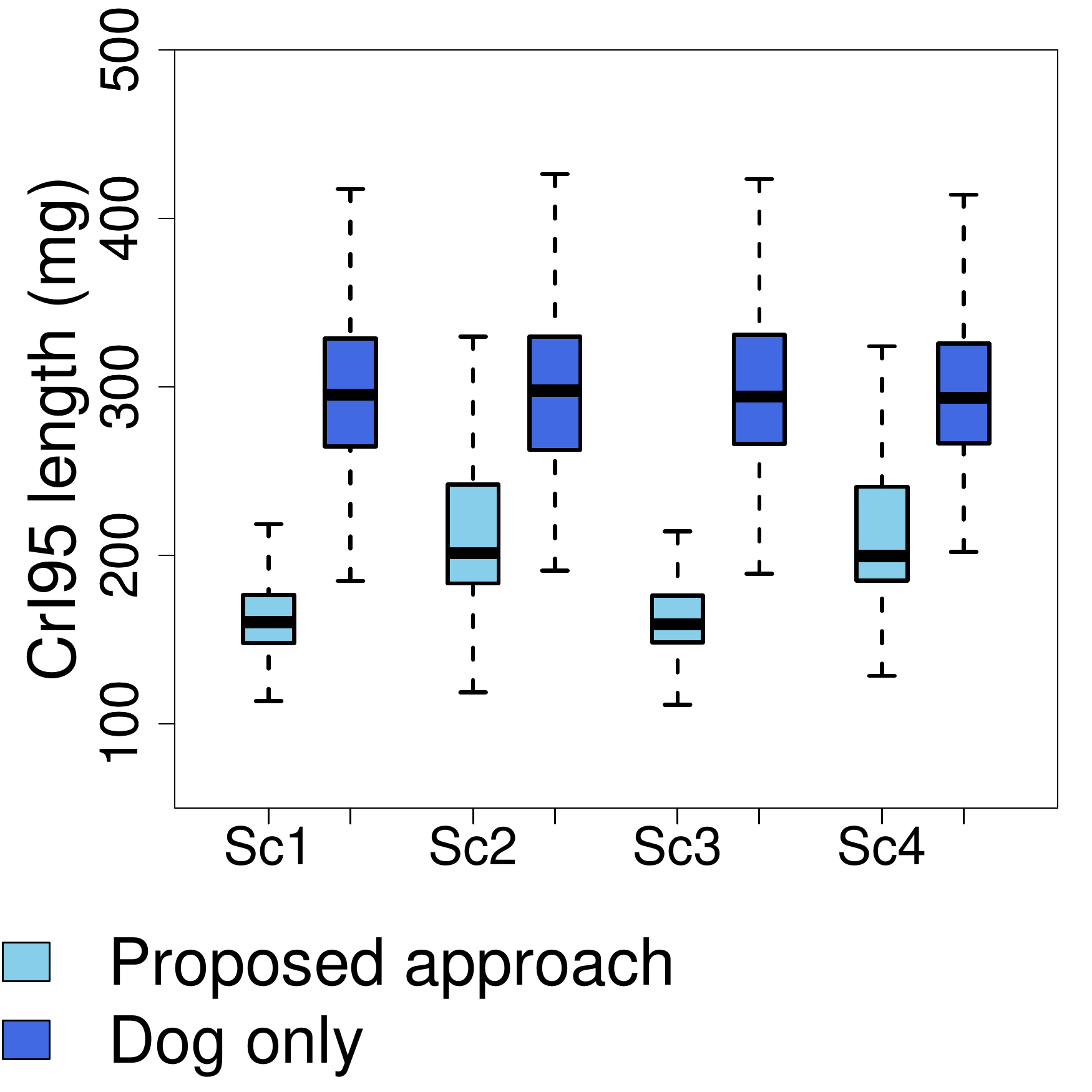}\label{fig:MTD_posterior_IC95_length_Bayesian}}} 
\caption{Estimated MED (a) and MTD (c) in humans and the length of the corresponding 95\% credibility interval (CrI95) (b, d) for all scenarios for the Bayesian approach and using the standard approach (that is, only dog data) over 500 replications, under the assumption that $\omega_{V} = \omega_{IC_{50}} = \omega_{k_e} = 0.7$ for mouse. MED: Minimum effective dose; MTD: Maximum tolerated dose.}
\label{fig:dose_posterior_mean_and_IC95_length_Bayesian_omega1}
\end{figure}

\begin{figure}
\centerline{
\subfigure[]{\includegraphics[scale=0.25]{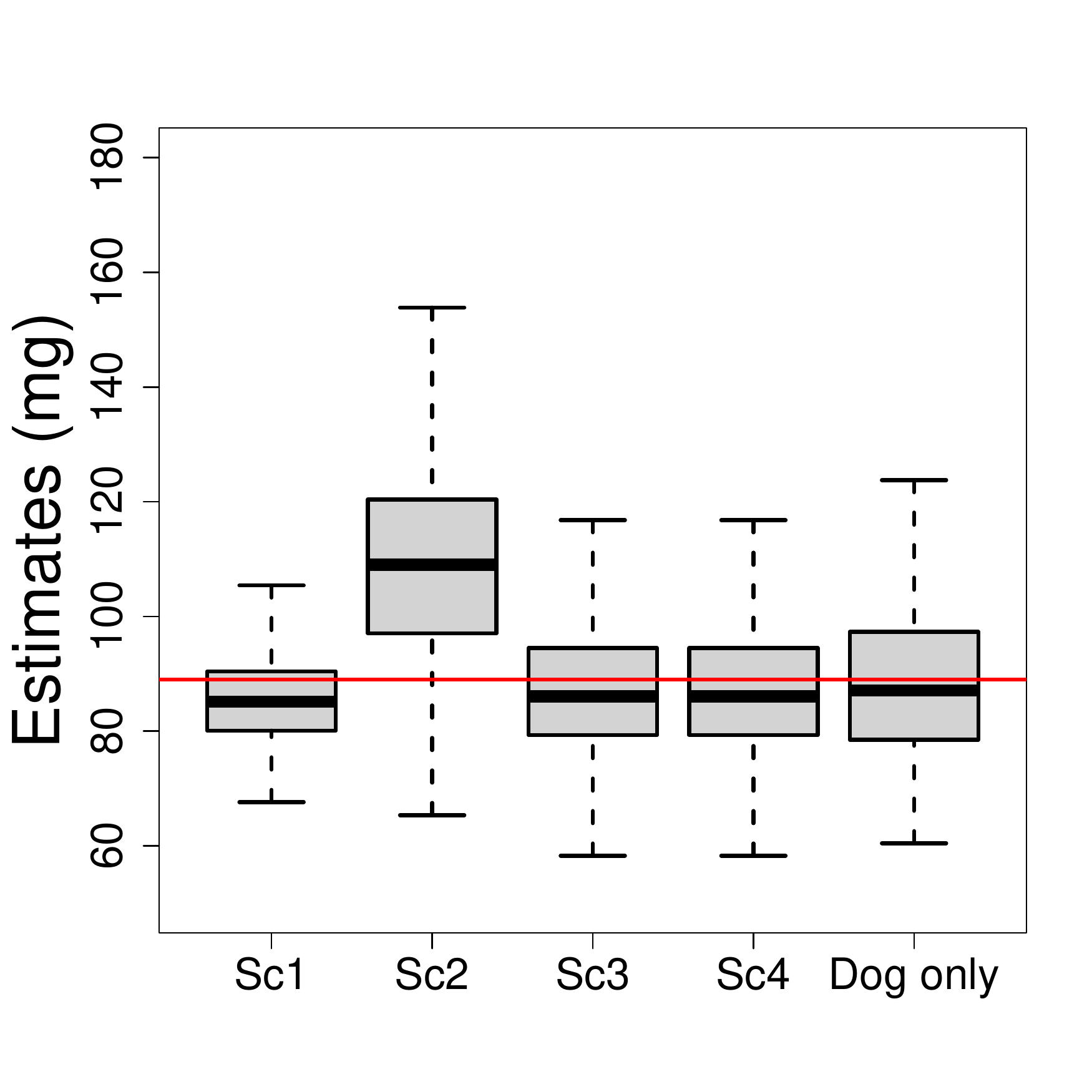}\label{fig:MED_posterior_mean_hybrid_omega1}} 
\subfigure[]{\includegraphics[scale=0.25]{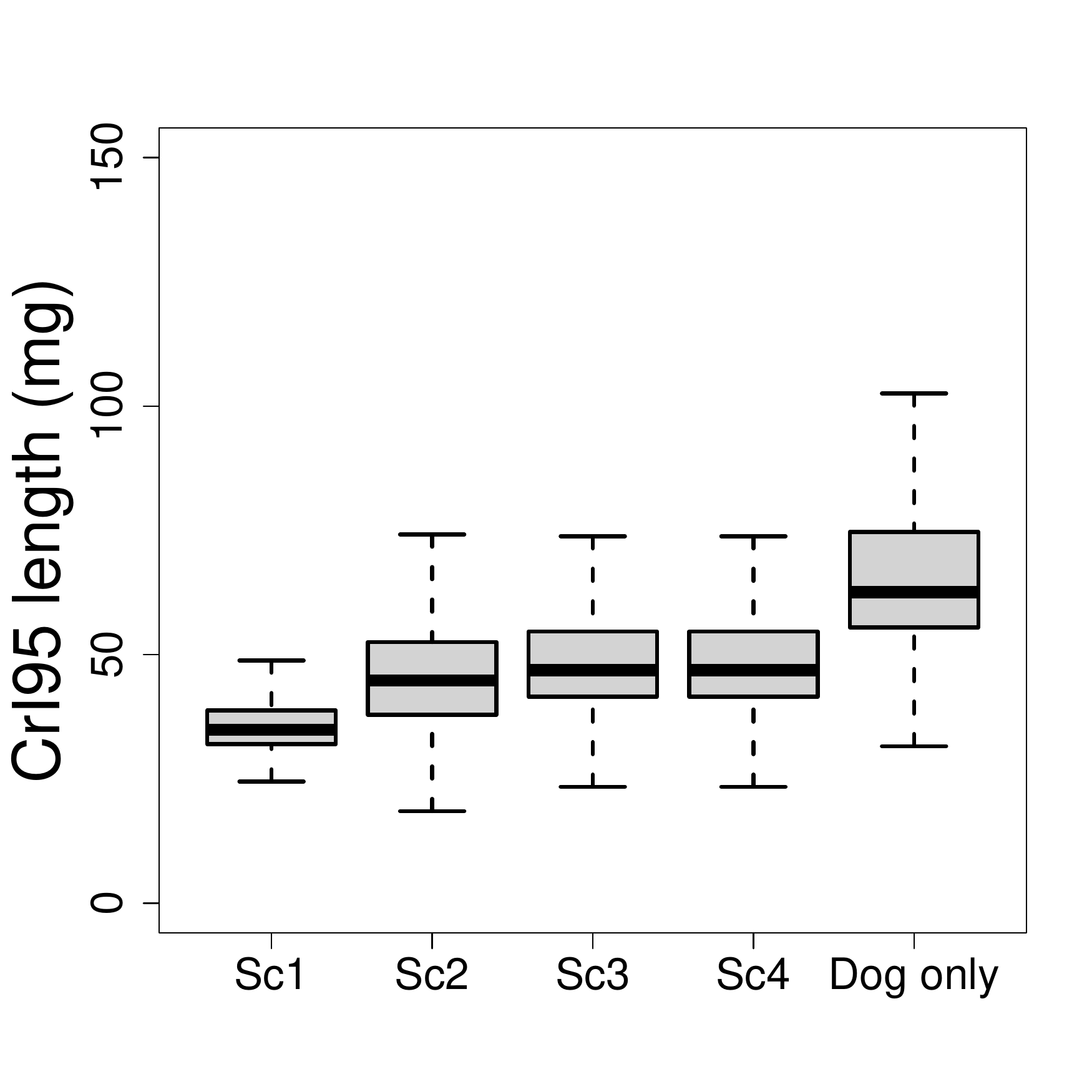}\label{fig:MED_posterior_IC95_length_hybrid_omega1}}}
\centerline{ 
\subfigure[]{\includegraphics[scale=0.25]{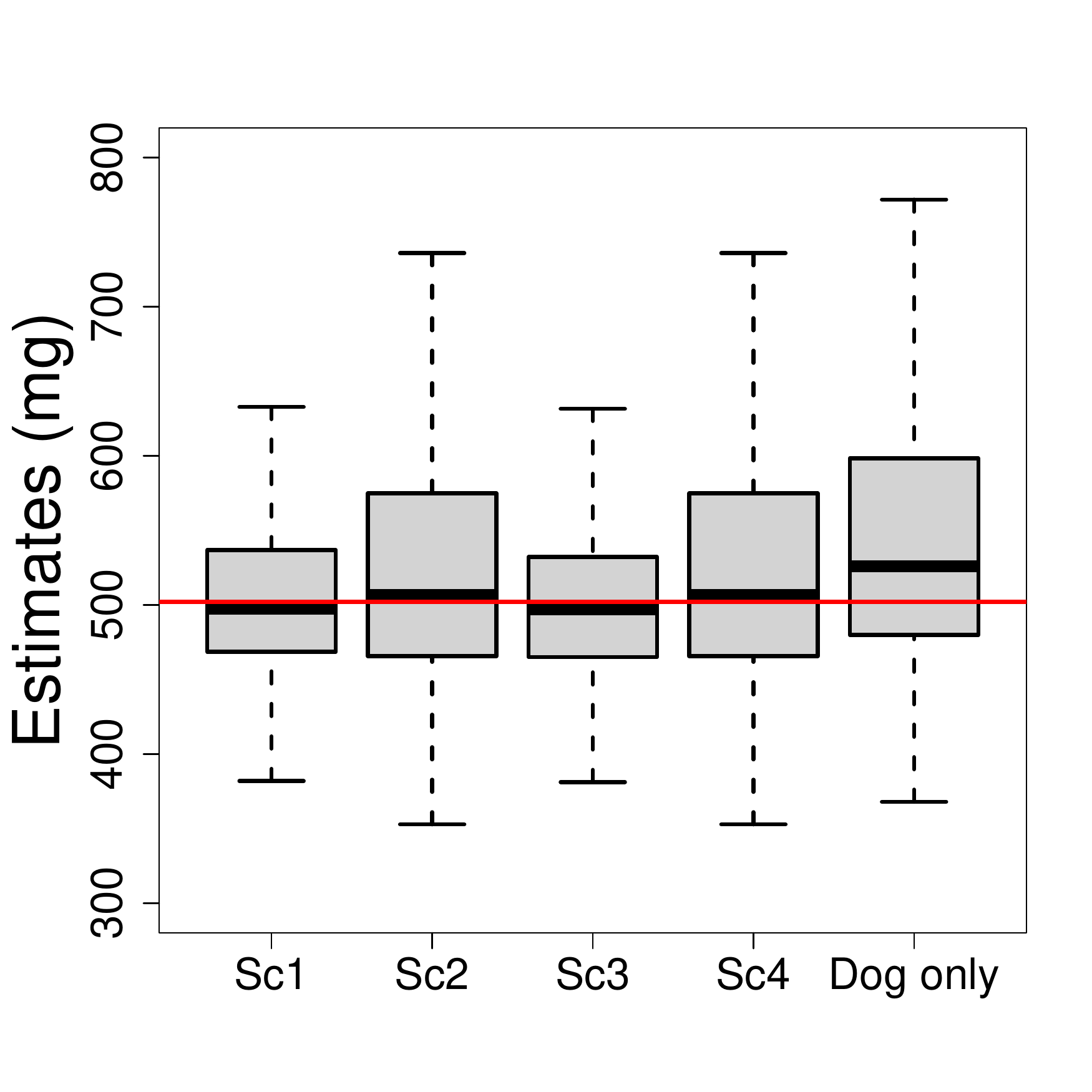}\label{fig:MTD_posterior_mean_hybrid}} 
\subfigure[]{\includegraphics[scale=0.25]{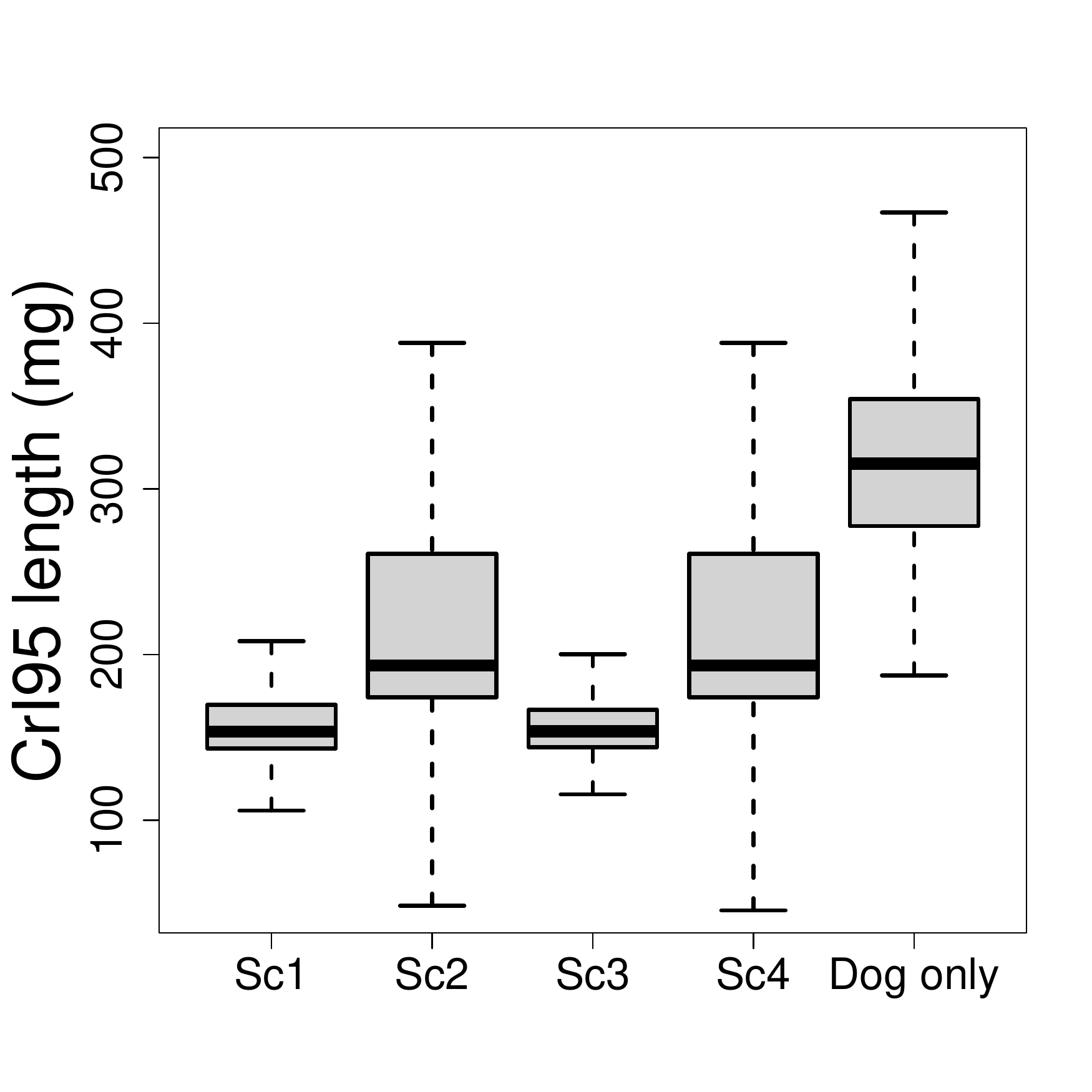}\label{fig:MTD_posterior_IC95_length_hybrid}}} 
\caption{Estimated MED (a) and MTD (c) in humans and the length of the corresponding 95\% credibility interval (CrI95) (b, d) for all scenarios for the hybrid approach and using the standard approach (that is, only dog data) over 500 replications, under the assumption that $\omega_{V} = \omega_{IC_{50}} = \omega_{k_e} = 0.7$ for mouse. MED: Minimum effective dose; MTD: Maximum tolerated dose.}
\label{fig:dose_posterior_mean_and_IC95_length_hybrid_omega1}
\end{figure}

As shown in Figures \ref{fig:MED_posterior_mean_Bayesian_omega1} and \ref{fig:MED_posterior_mean_hybrid_omega1}, for scenario 1 for which data from all animal species are used to estimate the final MED in almost all cases, the MED is correctly estimated to 91 mg by Bayesian approach and to 86 mg by the approach using the hybrid (close to the true value of 89 mg). Comparable results are obtained with the hybrid approach for scenarios 3 and 4 for which, in most cases, only mouse and dog data are used. For these scenarios, the estimate of MED is slightly greater (96 mg) for the Bayesian approach. 
However, for scenario 2, the MED is overestimated for both approaches. Indeed, for this scenario, the theoretical MED of the rat extrapolated to humans is higher than that of the mouse and dog, but lower than in scenarios 3 and 4 (see Table \ref{tab:simu_parameters_and_extrapolated_therapeutic_window}). As a result, the Hellinger distance between the rat and other animal species is smaller than for scenarios 3 and 4, and a threshold set at 0.3 struggles to differentiate the MED distributions.

The MTD is estimated between 510 and 561 mg for the Bayesian approach (see Figure \ref{fig:MTD_posterior_mean_Bayesian}), and between 502 and 526 mg for the hybrid approach (see Figure \ref{fig:MTD_posterior_mean_hybrid}), again close to the true value of 502 mg. Estimates of 561 mg, 553 mg, 526 mg and 526 mg respectively are obtained for scenarios 2 and 4 with Bayesian approach and scenarios 2 and 4 with hybrid approach.  

The length of the 95\% credibility interval (CrI95) is greater when only the dog results are used (that is the standard approach) to calculate the MED and the MTD than when using the proposed methods (see Figures \ref{fig:MED_posterior_IC95_length_Bayesian_omega1}, \ref{fig:MED_posterior_IC95_length_hybrid_omega1}, \ref{fig:MTD_posterior_IC95_length_Bayesian} and \ref{fig:MTD_posterior_IC95_length_hybrid}). 

\section{Web Appendix D: Sensitivity Analyses}

\begin{figure}
\centerline{
\subfigure[]{\includegraphics[scale=0.25]{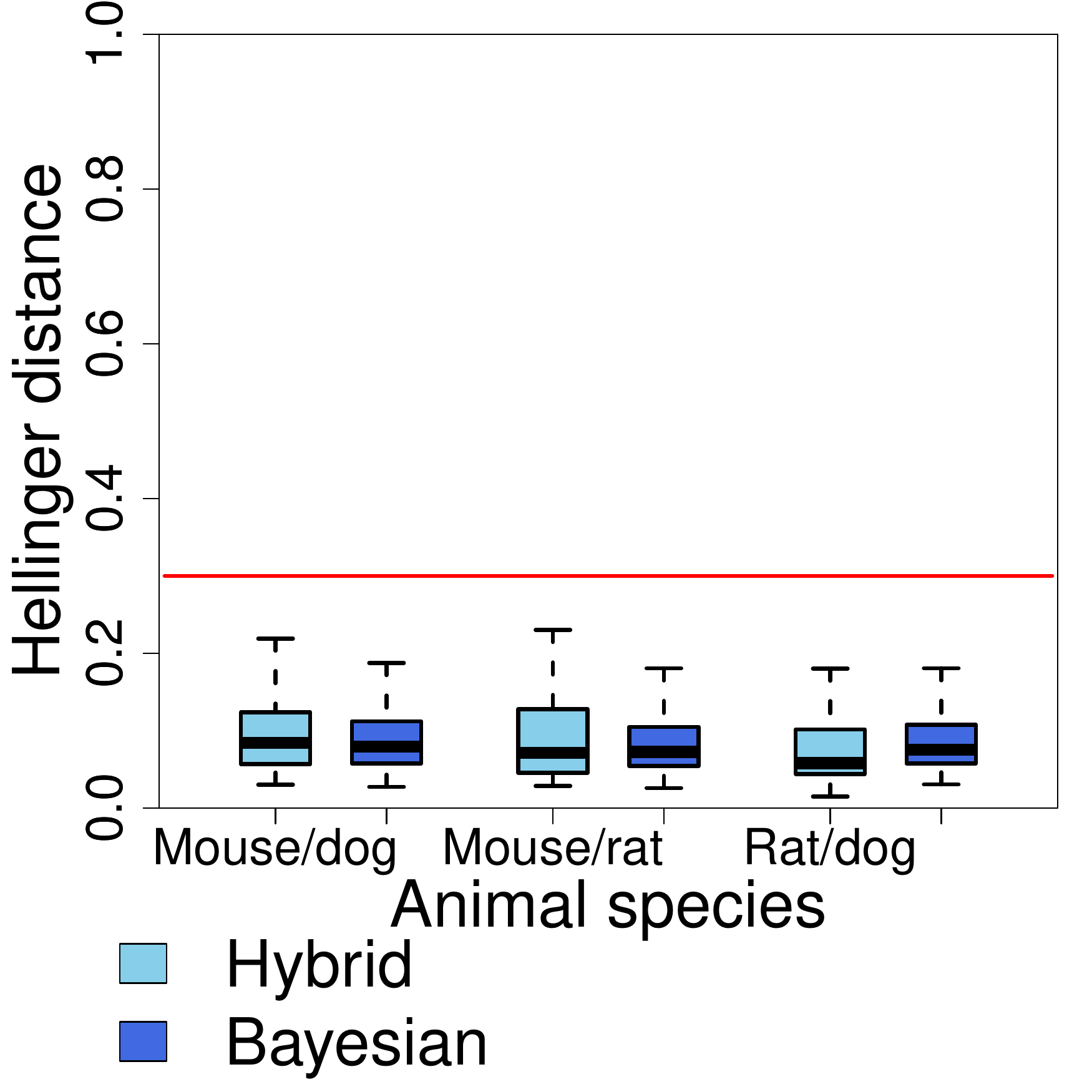}}
\subfigure[]{\includegraphics[scale=0.25]{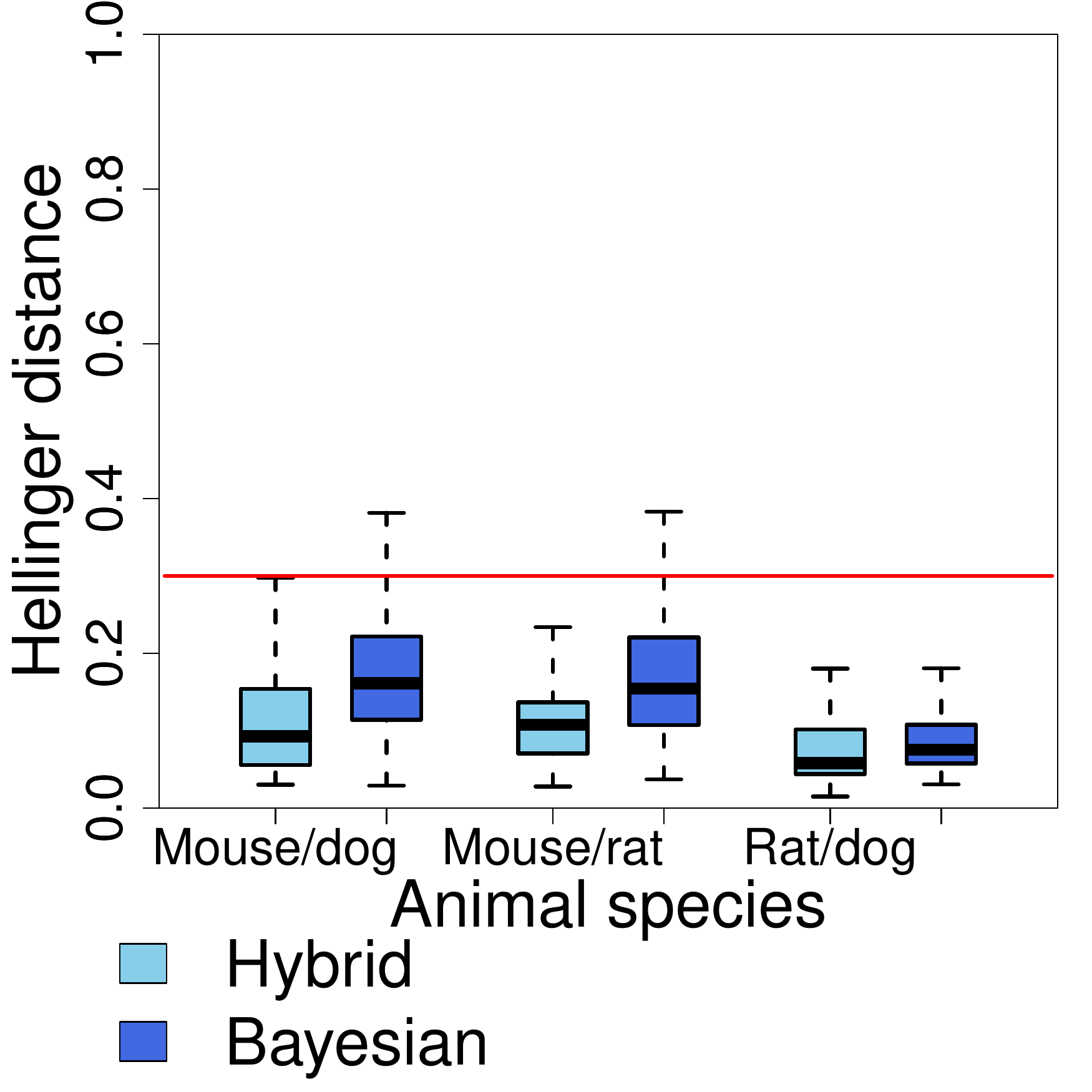}}}
\centerline{
\subfigure[]{\includegraphics[scale=0.25]{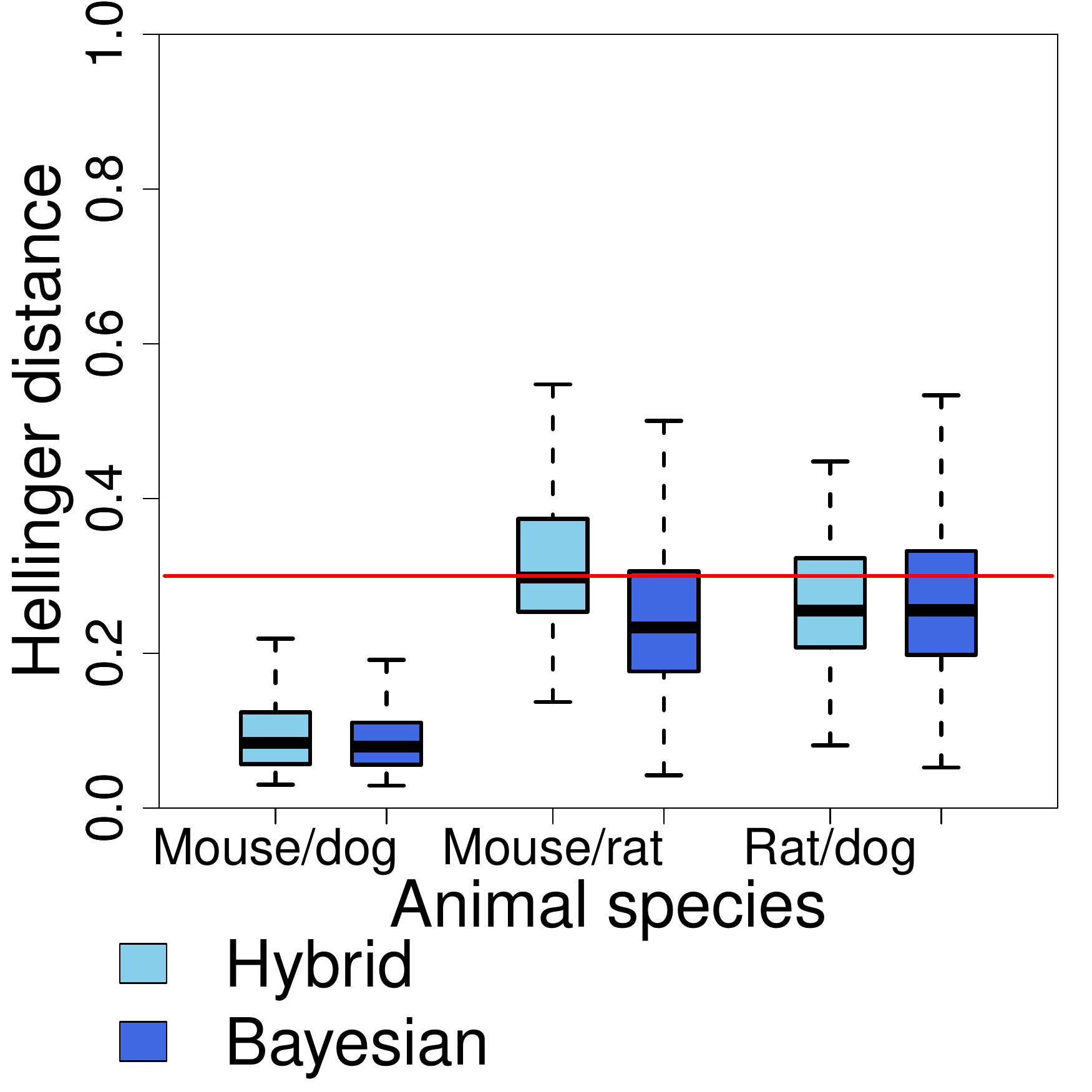}}
\subfigure[]{\includegraphics[scale=0.25]{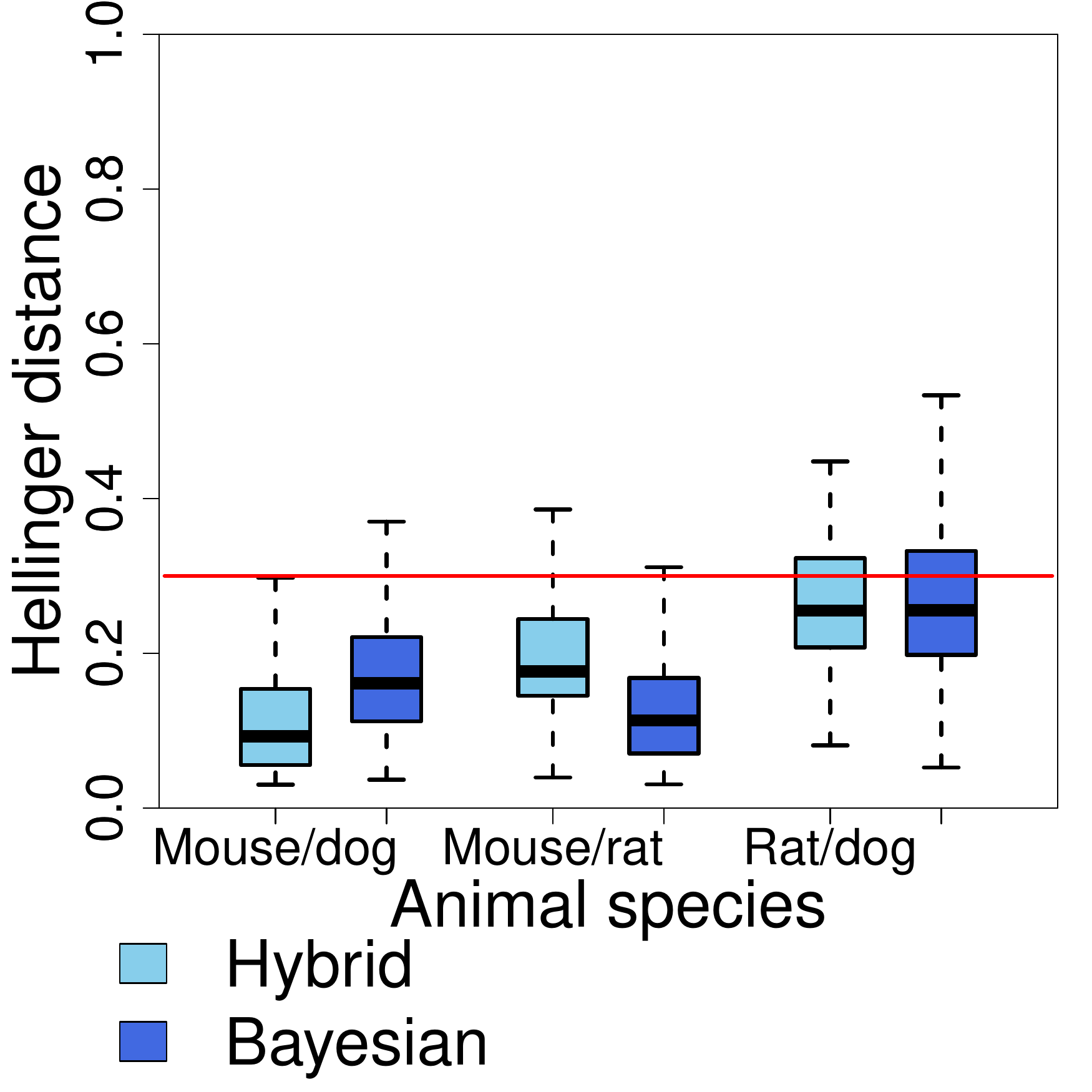}}}
\centerline{
\subfigure[]{\includegraphics[scale=0.25]{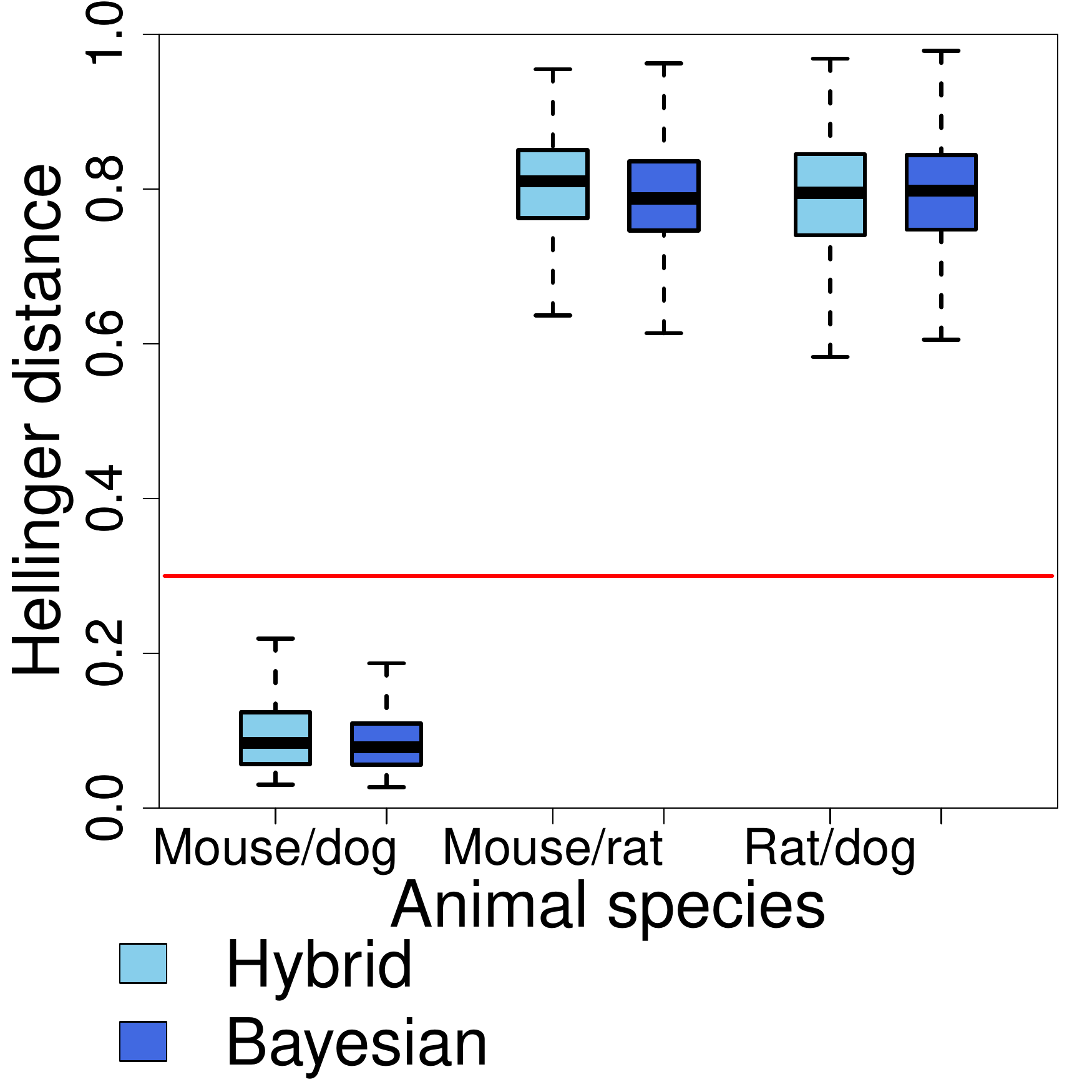}}
\subfigure[]{\includegraphics[scale=0.25]{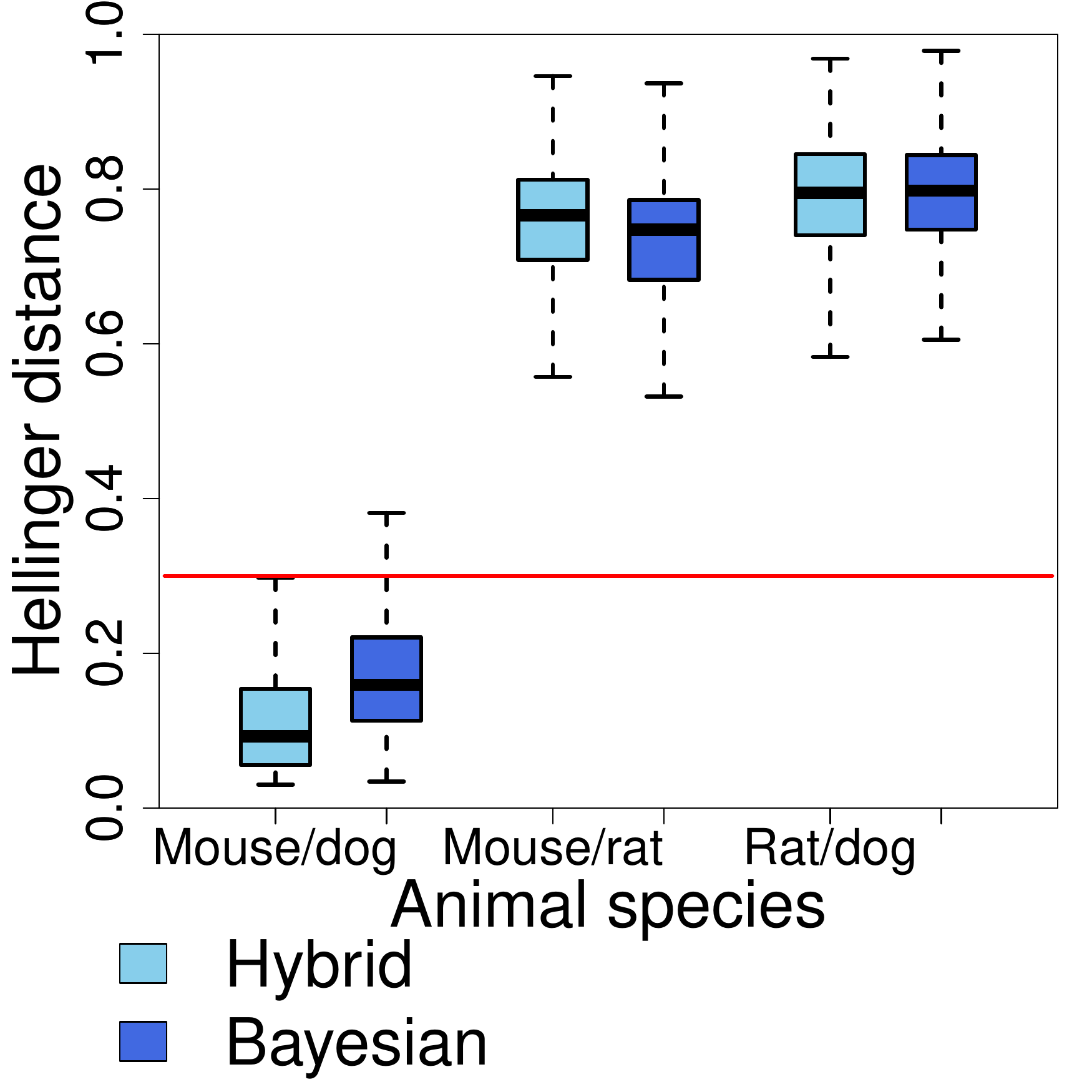}}}
\centerline{
\subfigure[]{\includegraphics[scale=0.25] {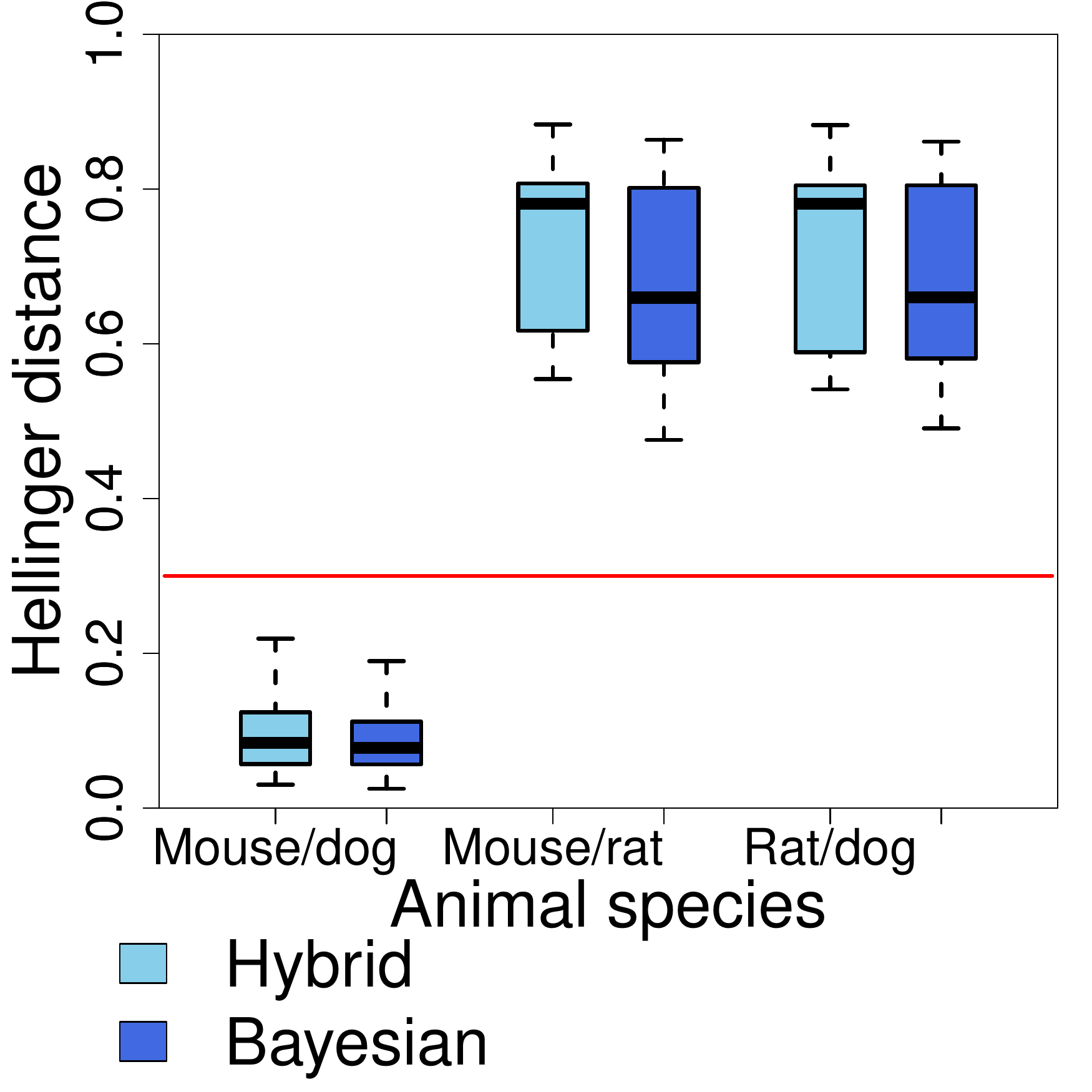}}
\subfigure[]{\includegraphics[scale=0.25] {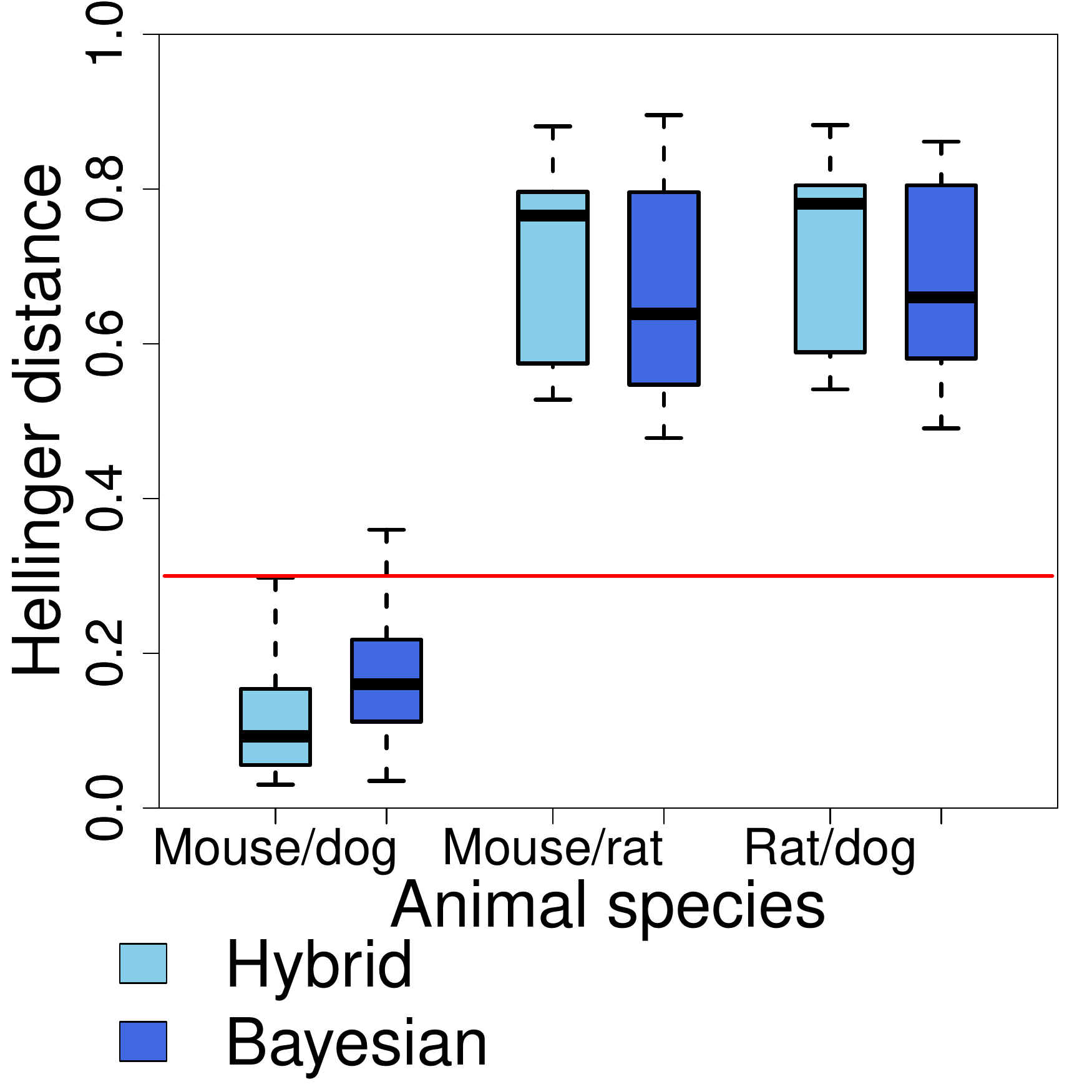}}} 
\caption{Hellinger distance of the transformed predicted MED distributions in humans between animal species over 500  replications, under the assumption that $\omega_{V} = \omega_{IC_{50}} = \omega_{k_e} = 0.4$ (first column) and $\omega_{V} = \omega_{IC_{50}} = \omega_{k_e} = 1$ (second column) for mouse in scenario 1 (a, b), scenario 2 (c, d), scenario 3 (e, f) and scenario 4 (g, h). MED: Minimum effective dose.}
\label{fig:hellinger_dist_MED_omega2and3}
\end{figure}

\begin{figure}
\centerline{
\subfigure[]{\includegraphics[scale=0.25]{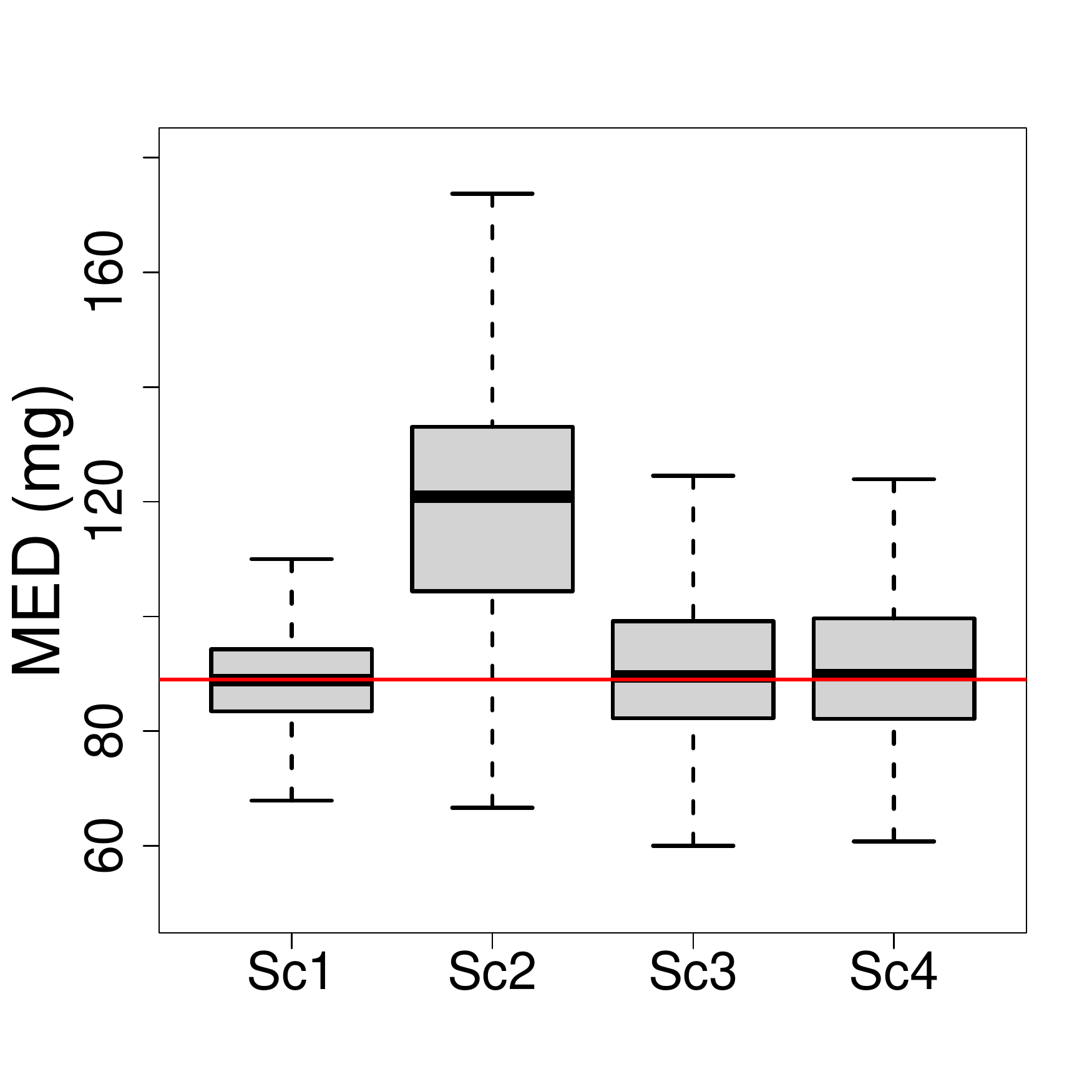}} 
\subfigure[]{\includegraphics[scale=0.25]{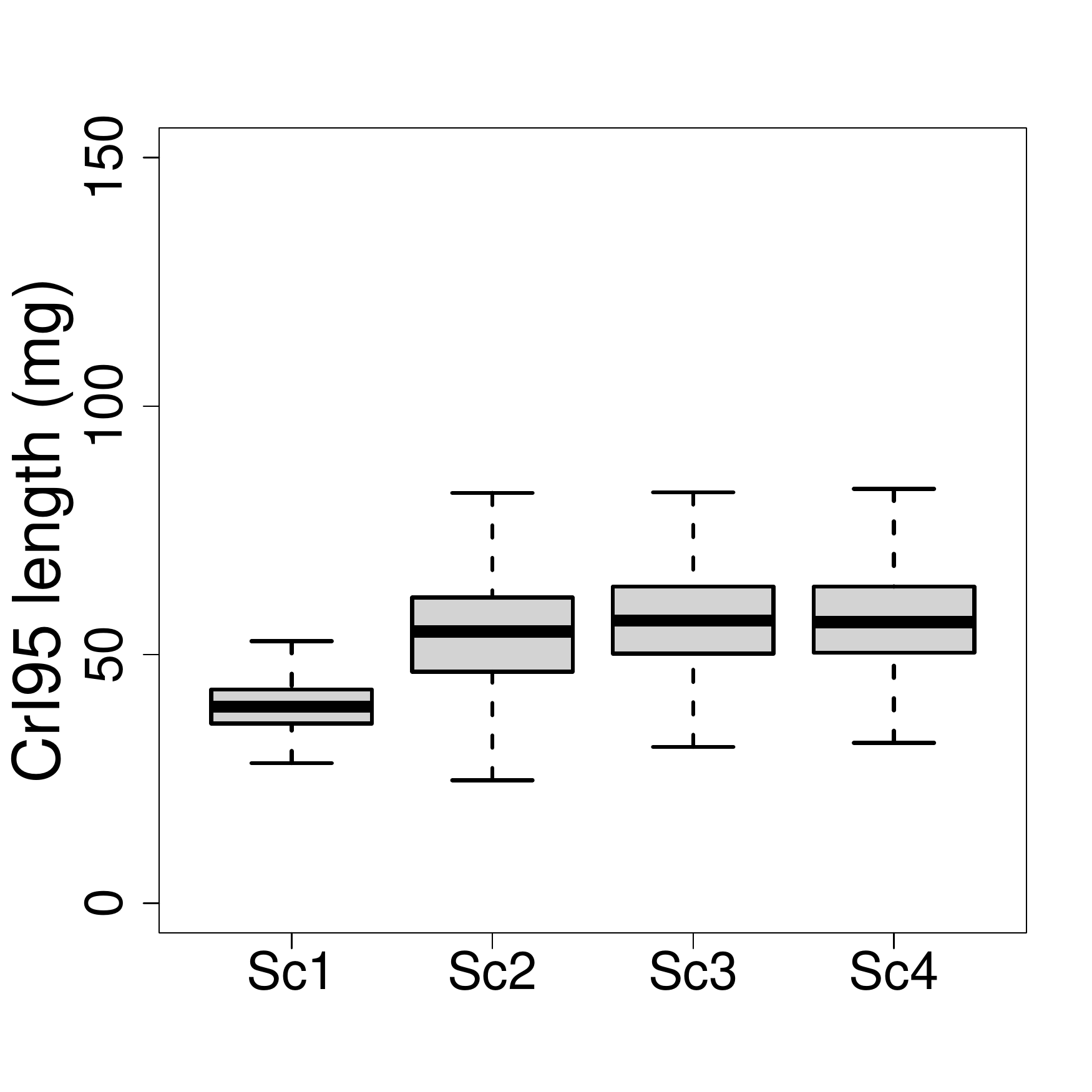}}}
\centerline{ 
\subfigure[]{\includegraphics[scale=0.25]{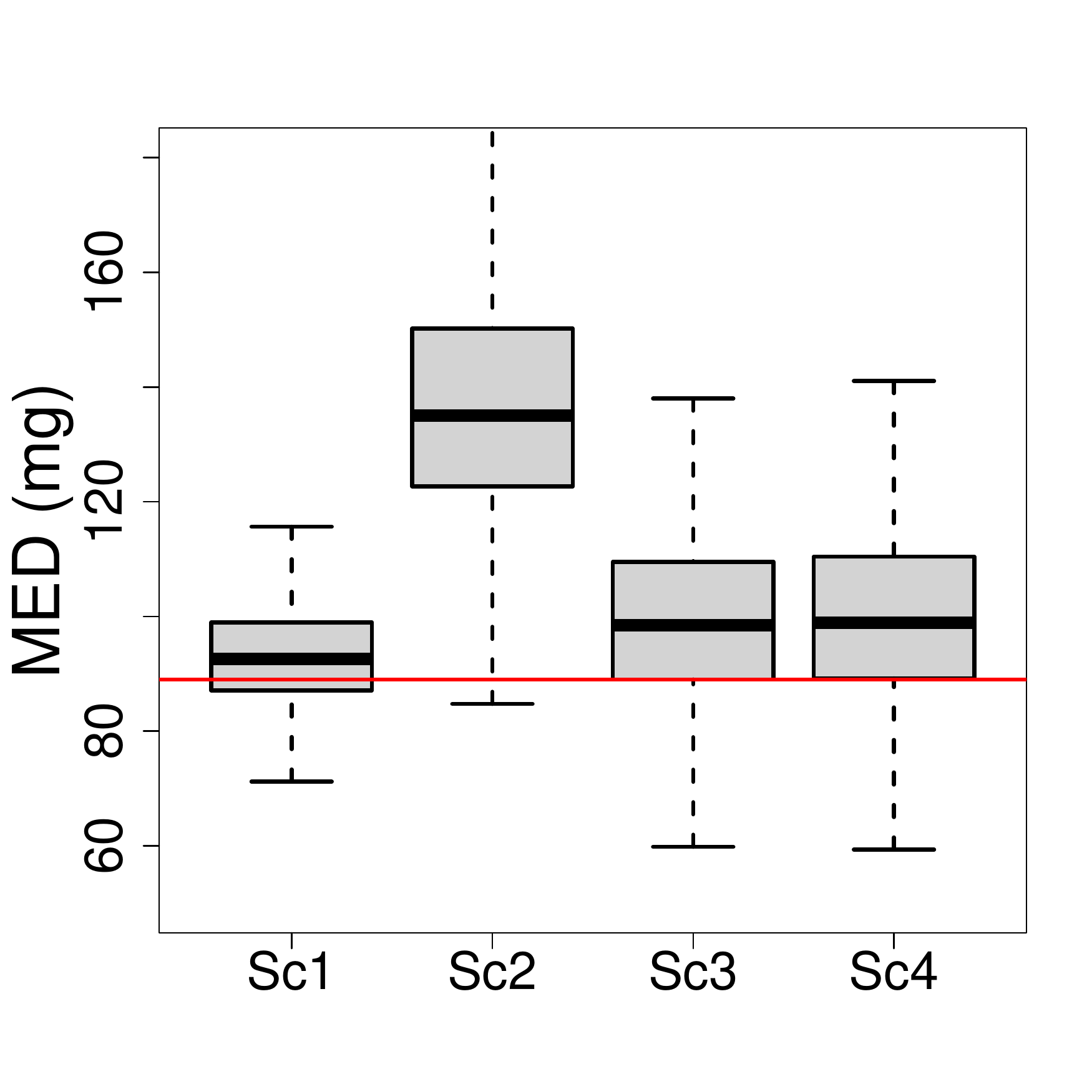}} 
\subfigure[]{\includegraphics[scale=0.25]{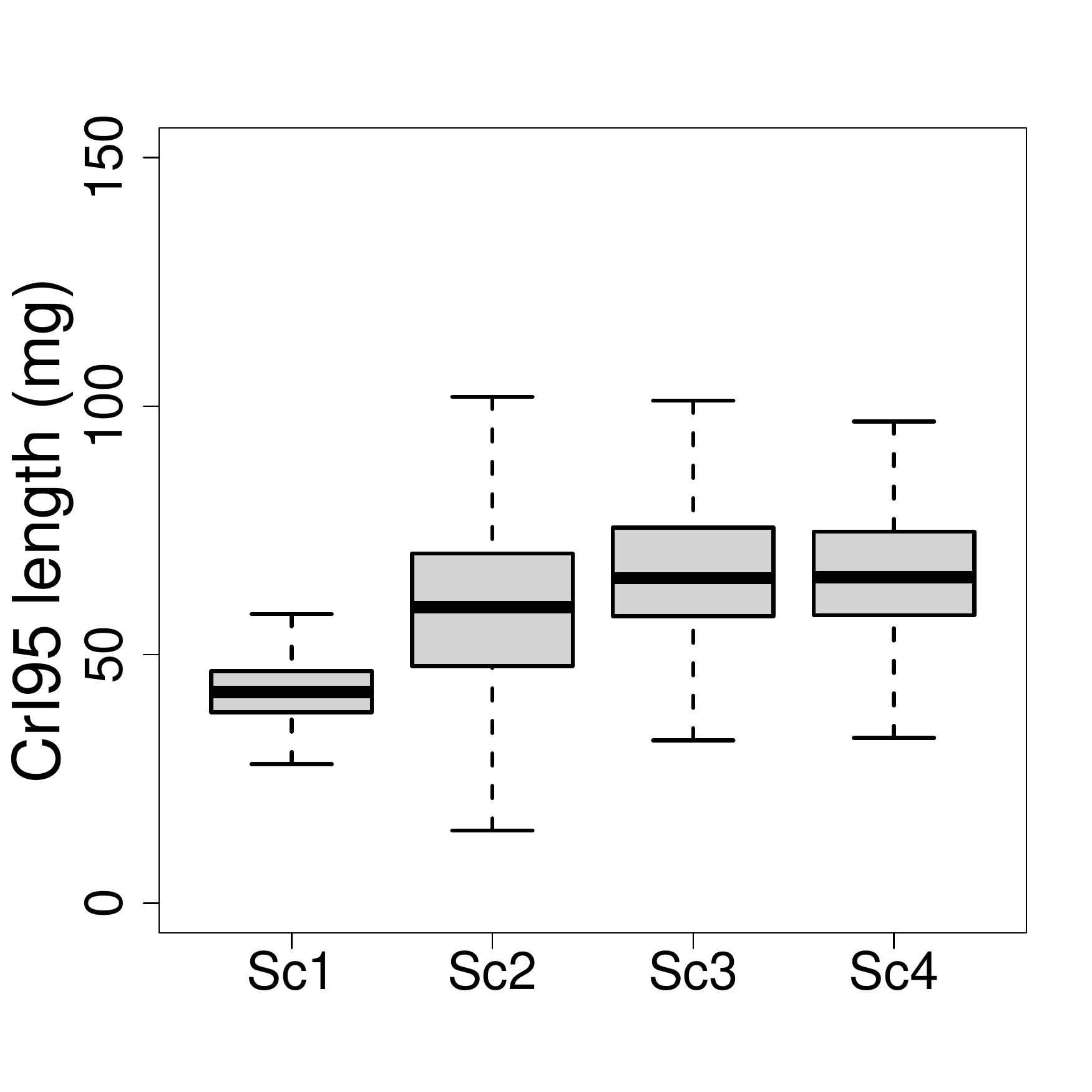}}}  
\caption{Estimated MED (a, c) in humans and the length of the corresponding 95\% credibility interval (CrI95) (b, d) for all scenarios for the Bayesian approach over 500 replications,  under the assumption that $\omega_{V} = \omega_{IC_{50}} = \omega_{k_e} = 0.4$ (first line) and $\omega_{V} = \omega_{IC_{50}} = \omega_{k_e} = 1$ (second line) for mouse. MED: Minimum effective dose.}
\label{fig:MED_posterior_mean_and_IC95_length_Bayesian_omega2and3}
\end{figure}

\begin{figure}
\centerline{
\subfigure[]{\includegraphics[scale=0.25]{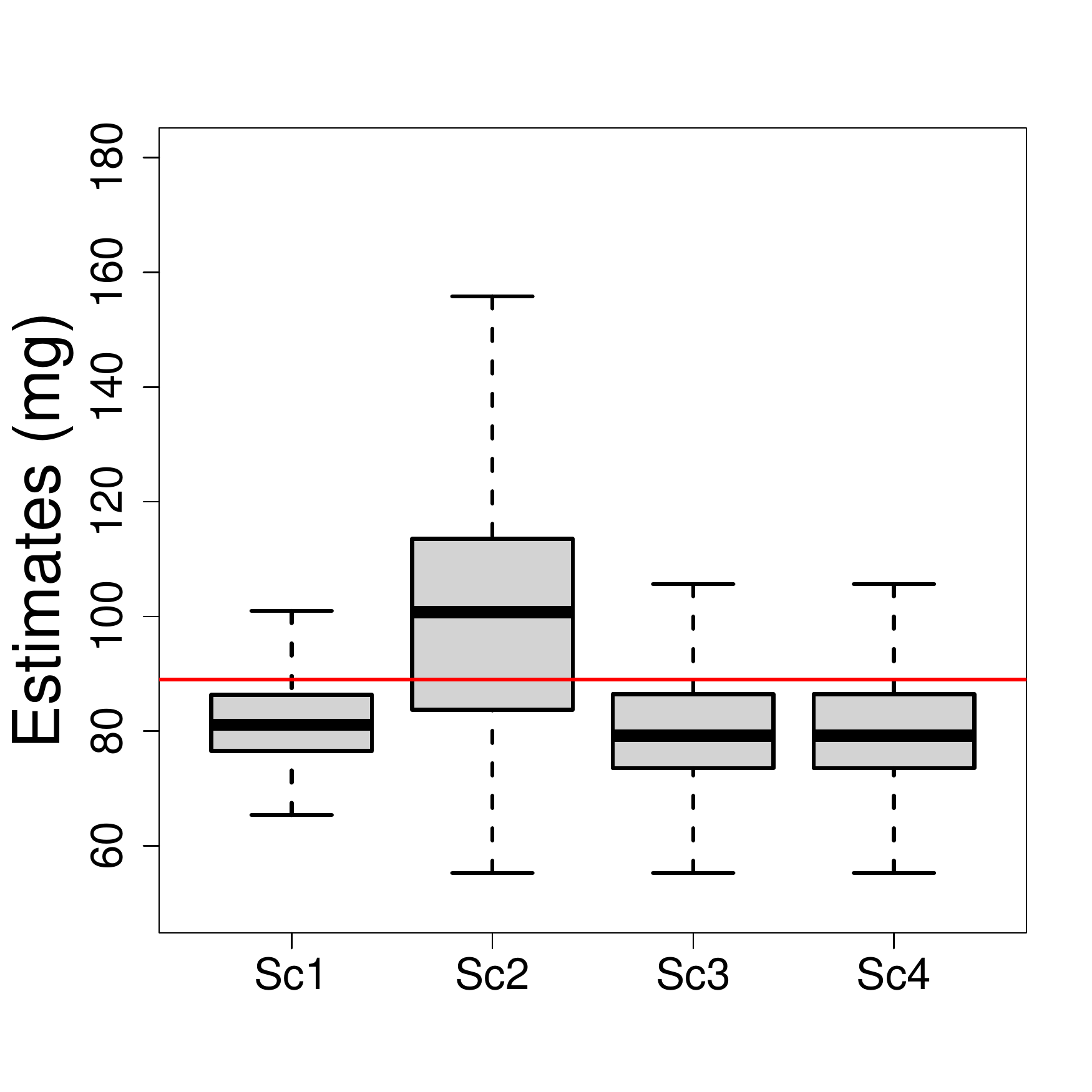}} 
\subfigure[]{\includegraphics[scale=0.25]{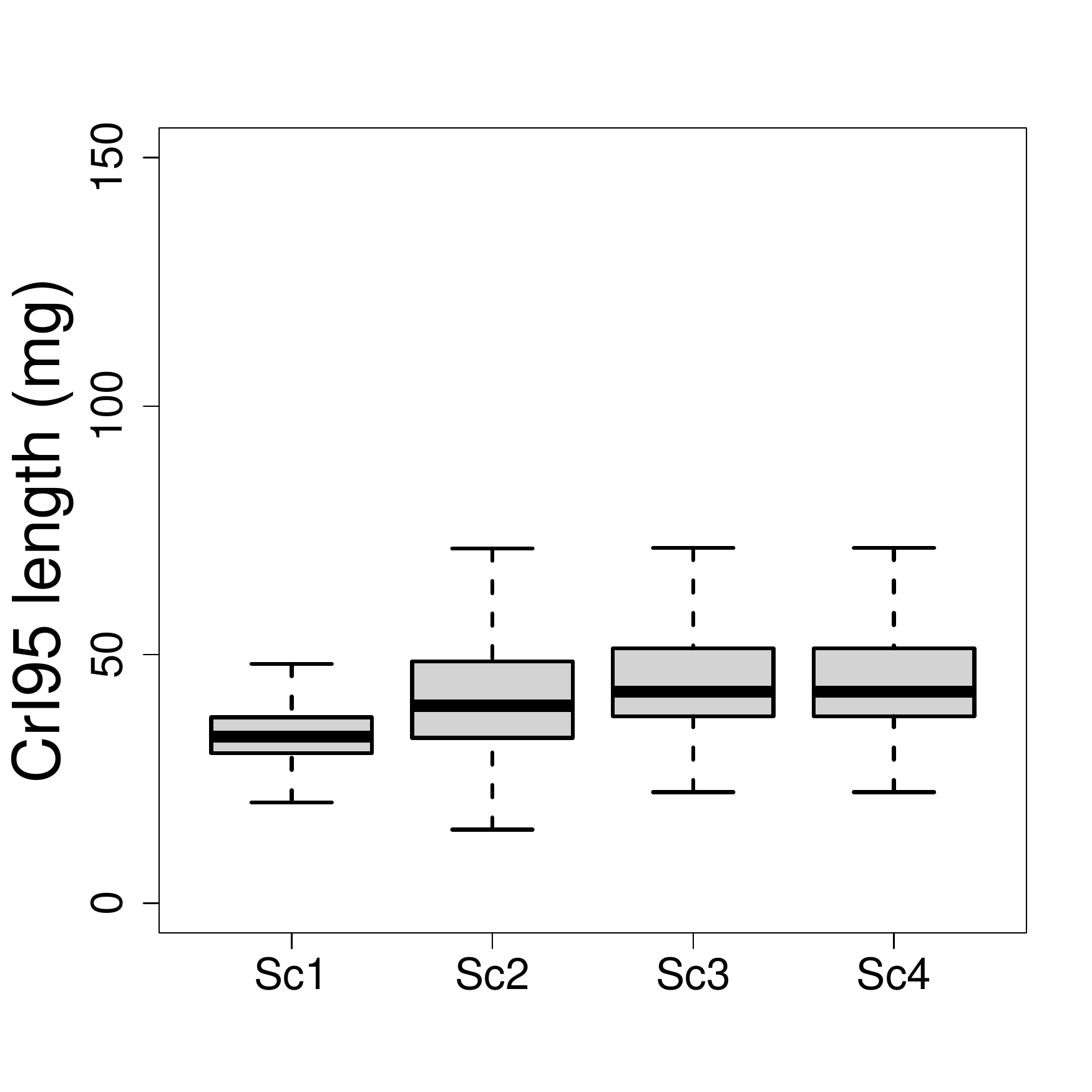}}}
\centerline{ 
\subfigure[]{\includegraphics[scale=0.25]{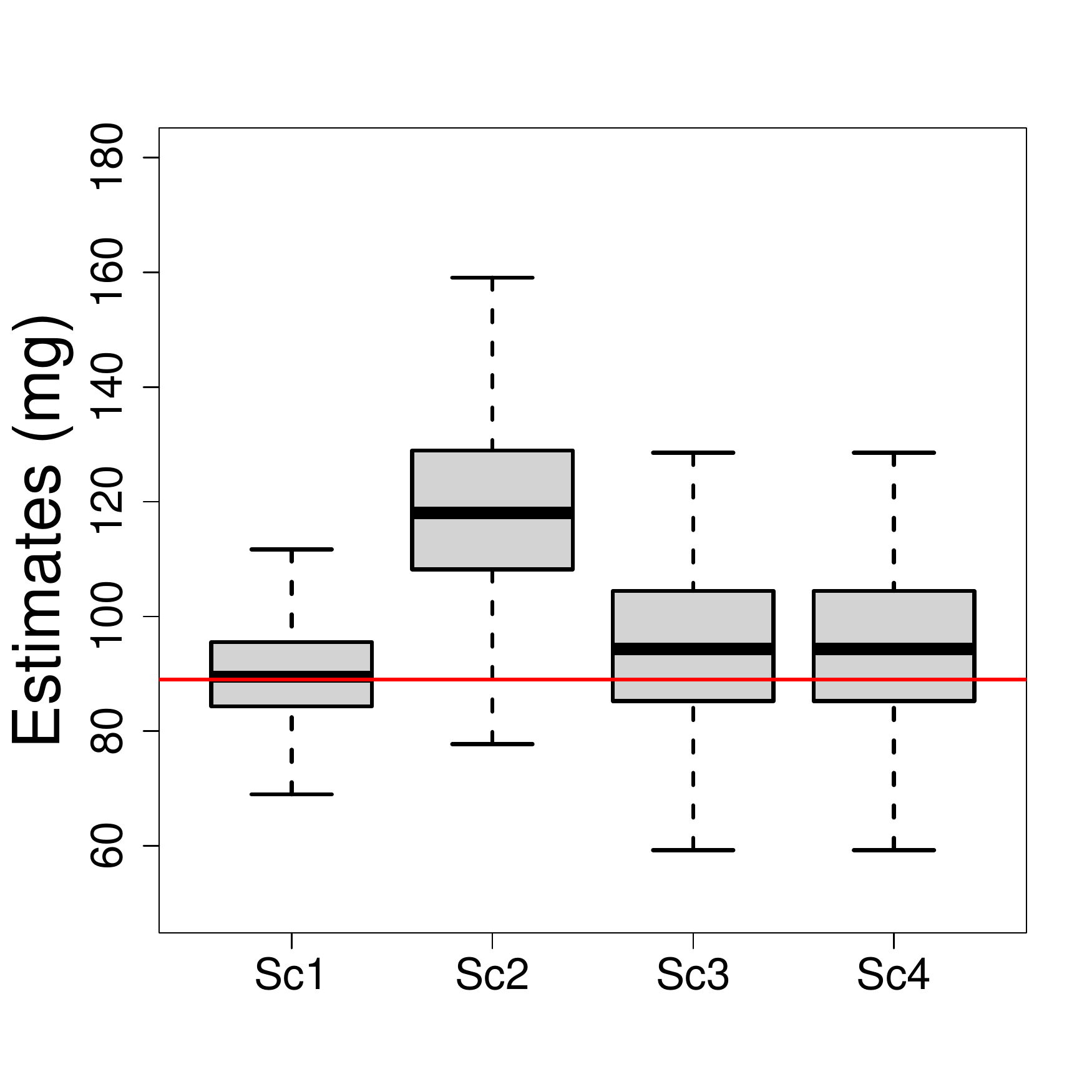}} 
\subfigure[]{\includegraphics[scale=0.25]{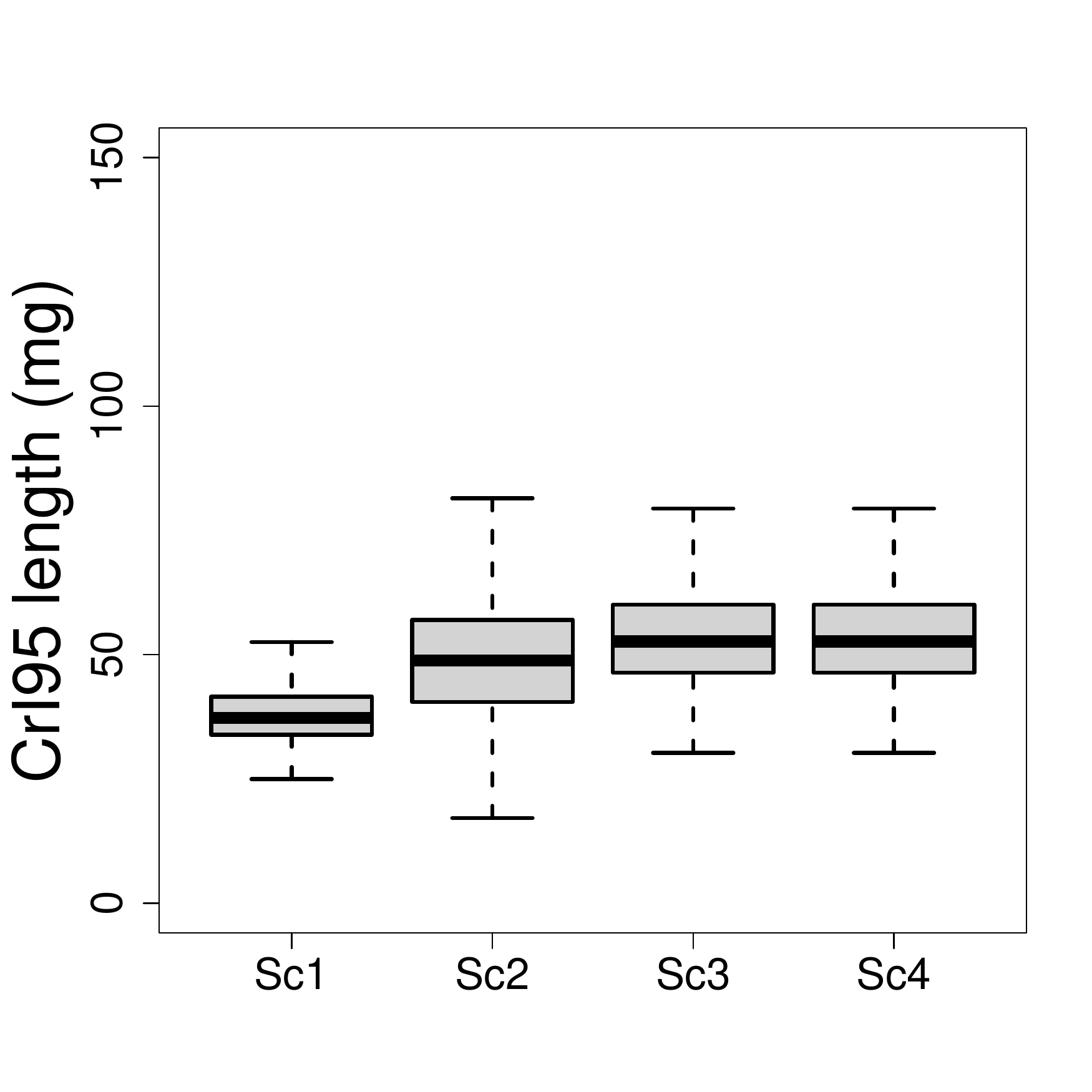}}}  
\caption{Estimated MED (a, c) in humans and the length of the corresponding 95\% credibility interval (CrI95) (b, d) for all scenarios for the hybrid approach over 500 replications,  under the assumption that $\omega_{V} = \omega_{IC_{50}} = \omega_{k_e} = 0.4$ (first line) and $\omega_{V} = \omega_{IC_{50}} = \omega_{k_e} = 1$ (second line) for mouse. MED: Minimum effective dose.}
\label{fig:MED_posterior_mean_and_IC95_length_hybrid_omega2and3}
\end{figure}

The results of the sensitivity analysis for $\omega_{V} = \omega_{IC_{50}} = \omega_{k_e} = 0.4$ and $\omega_{V} = \omega_{IC_{50}} = \omega_{k_e} = 
1$ for mouse do not differ from those obtained for $\omega_{V} = \omega_{IC_{50}} = \omega_{k_e} = 0.7$ as shown in Figures \ref{fig:hellinger_dist_MED_omega2and3}, \ref{fig:MED_posterior_mean_and_IC95_length_Bayesian_omega2and3} and \ref{fig:MED_posterior_mean_and_IC95_length_hybrid_omega2and3}.

\section{Web Appendix E: Additional Elements to Discussion}\label{sec:add_discussion}

In our simulation setting, the theoretical dose ranges (i.e. the dose ranges calculated from the true values of the model parameters)  extrapolated to humans shown in Table \ref{tab:simu_parameters_and_extrapolated_therapeutic_window} 
are not exactly the same between animal species, even for scenario 1, but, in general, very close each other. Nevertheless, the mouse extrapolated MED of 41 mg cannot be easily approached to 89 mg (true human value) because there is no extrapolation on PD parameters and the PD outcome relies on the PK results. Indeed, the accuracy of the extrapolation of the toxicity model from rats to human depends on the clearance parameter while the accuracy of the extrapolation of the efficacy model depends on PK parameters and $IC_{50}$. 
Therefore, this imperfection of the simulation scenarios should be kept in mind when interpreting the results. However, in real life, the extrapolation of a dose between two species is always subject to a margin of error.

Furthermore, Hellinger distance threshold depends on the outcome scale and variability. As $IC_{50}$ scale value is lower than the plasma concentration (0.32 or 2.9 mg.L$^{-1}$ vs greater to 5 mg.L$^{-1}$ concentration for hours), the impact of changing $IC_{50}$ is lower on the predicted dose than when changing PK parameters. 

As expected, parameter estimation at step 1 by MCMC methods is more costly in terms of time than using frequentist methods. Also, compared to the hybrid approach, the Bayesian approach seems to more frequently overestimate the proposed doses in our scenarios. In our example, it could be due to the overestimation of the IIV parameters (results shown in supplementary material), as well as the eventual \textit{posterior} distribution heavy tails. To note, in the Bayesian approach, \textit{posterior} distributions are not approximate by normal (or multivariate normal) distribution. Therefore a hybrid approach could be a viable option in practice when researcher face instability computation for complex models at the first Bayesian estimation step.

\bibliography{FAIR_art1_supporting_information_30-Nov-2022_arxiv_format.bib}